\documentclass[sigconf, nonacm]{acmart}
\newcommand{\eat}[1]{}

\usepackage{dsfont}
\usepackage{latexsym}
\usepackage{amsfonts}
\usepackage{amsmath}
\usepackage{xcolor}
\usepackage{colortbl}
\usepackage{epsfig}
\usepackage{xspace}
\usepackage{graphicx}
\usepackage{paralist}
\usepackage{enumerate}
\usepackage[color,matrix,arrow,all]{xy}
\usepackage{comment}
\usepackage{booktabs}
\usepackage{balance}
\usepackage{stmaryrd}
\usepackage{pifont}
\usepackage{hhline}
\usepackage{listings}
\usepackage{array}
\usepackage{float}
\usepackage[flushleft]{threeparttable}

\usepackage{mathrsfs}
\usepackage{makecell}
\usepackage{xparse}
\usepackage{wrapfig}

\usepackage{epsfig}
\usepackage{multirow}
\usepackage{url}

\usepackage{multirow}
\usepackage{natbib}
\usepackage{graphicx}

\usepackage{listings}
\usepackage{framed}
\usepackage{xcolor}
\usepackage{color}
\usepackage{geometry}
\usepackage{changepage}
\colorlet{shadecolor}{gray!20}
\usepackage{makecell}

\usepackage{rotating}

\usepackage{subcaption}

\usepackage{algorithm}
\usepackage{algorithmic}

\definecolor{shadecolor}{RGB}{220,220,220}

\definecolor{inputcolor}{RGB}{255,139,35}
\definecolor{outputcolor}{RGB}{120,212,252}
\definecolor{embedcolor}{RGB}{254,127,156}
\definecolor{maskcolor}{RGB}{122,128,255}
\definecolor{ecolor}{RGB}{58,149,54}

\definecolor{highcolor}{RGB}{255,153,153}
\definecolor{midcolor}{RGB}{255,204,204}
\definecolor{lowcolor}{RGB}{204,229,255}

\usepackage{tikz}
\usetikzlibrary{shapes,snakes}
\usetikzlibrary{calc}

\usepackage{algorithm} 
\usepackage{algorithmic}  
\usepackage[algo2e]{algorithm2e} 

\usepackage[export]{adjustbox}

\definecolor{green}{RGB}{0,128,0}

\definecolor{yellow}{RGB}{255,200,18}

\newcommand{\att}[1]{\textbf{\texttt{#1}}}

\newcommand{\stab}{\vspace{1.2ex}\noindent}

\newtheorem{example}{Example}

\newcommand{\bi}{\begin{itemize}}
\newcommand{\ei}{\end{itemize}}

\newcommand{\be}{\begin{enumerate}}
\newcommand{\ee}{\end{enumerate}}
\newcommand{\beqn}{\begin{eqnarray*}}
\newcommand{\eeqn}{\end{eqnarray*}}

\newcommand{\stitle}[1]{\stab\noindent{\bf #1}}
\newcommand{\etitle}[1]{\vspace{1mm}\noindent{\underline{\em #1}}}
\newcommand{\vs}{\textit{vs.} \xspace}
\newcommand{\ie}{\textit{i.e.,} \xspace}
\newcommand{\eg}{\textit{e.g.,} \xspace}

\newcommand{\eop}{\hspace*{\fill}\mbox{$\Box$}}     

\newcommand{\sys}{{\att {SuperSQL}}\xspace}

\newcommand{\testdb}{{\att {NL2SQL360}}\xspace}

\newcommand{\nlq}{{\sc nl}\xspace}
\newcommand{\sql}{{\sc sql}\xspace}
\newcommand{\sqls}{{\sc sqls}\xspace}
\newcommand{\nlsql}{{\sc nl2sql}\xspace}

\makeatletter
    \newcommand\figcaption{\def\@captype{figure}\caption}
    \newcommand\tabcaption{\def\@captype{table}\caption}
\makeatother

\tikzstyle{mybox} = [draw=black, fill=black!5, thick,
   rectangle, rounded corners, inner sep=0pt, inner ysep=6pt]
\tikzstyle{fancytitle} =[fill=black, text=white]

\newcommand{\rev}[1]{{#1}}

\NewDocumentCommand{\nan}{ mO{} }{\textcolor{blue}{\textsuperscript{\textit{Nan}}\textsf{\textbf{\small[#1]}}}}

\NewDocumentCommand{\yuyu}{ mO{} }{\textcolor{green}{\textsuperscript{\textit{Yuyu}}\textsf{\textbf{\small[#1]}}}}

\newcommand{\note}[1]{}

\usepackage{marginnote}
\let\oldmarginpar\marginpar
\renewcommand\marginpar[1]{\-\oldmarginpar[\raggedleft\footnotesize #1]%
	{\raggedright\footnotesize\color{blue} #1}} 
\usepackage{marginnote}
\let\oldmarginnote\marginnote
\renewcommand\marginnote[1]{\-\oldmarginnote[\raggedleft\footnotesize #1]%
	{\raggedright\footnotesize\color{blue} #1}} %

\usepackage{tcolorbox}
\tcbuselibrary{most}

\newtcolorbox[]{finding}[0]{colback=gray!10, colframe=black, width=\columnwidth, boxrule=0.4pt, left=0mm, right=0mm, top=0mm, bottom=0mm, before skip balanced=0pt, after skip balanced=0pt, sharp corners}

\newcommand{\myvspace}[1]{{\vspace{#1}}}

\newcommand\vldbdoi{10.14778/3681954.3682003}
\newcommand\vldbpages{3318 - 3331}
\newcommand\vldbvolume{17}
\newcommand\vldbissue{11}
\newcommand\vldbyear{2024}

\newcommand\vldbtitle{\shorttitle} 
\newcommand\vldbavailabilityurl{https://github.com/HKUSTDial/NL2SQL360}
\newcommand\vldbpagestyle{empty}

\begin{document}




\title{The Dawn of Natural Language to SQL: Are We Fully Ready?}


\author{Boyan Li}
\affiliation{%
	\institution{HKUST (GZ)}
}
\email{boyanli@hkust-gz.edu.cn}

\author{Yuyu Luo*}
\affiliation{%
	\institution{HKUST (GZ) / HKUST}
}
\email{yuyuluo@hkust-gz.edu.cn}

\author{Chengliang Chai}
\affiliation{%
	\institution{Beijing Institute of Technology}
}
\email{ccl@bit.edu.cn}

\author{Guoliang Li}
\affiliation{%
	\institution{Tsinghua University}
}
\email{liguoliang@tsinghua.edu.cn}

\author{Nan Tang}
\affiliation{%
	\institution{HKUST (GZ) / HKUST}
}
\email{nantang@hkust-gz.edu.cn}

\begin{abstract}
\rev{Translating users' natural language questions into SQL queries (\ie \nlsql) significantly lowers the barriers to accessing relational databases. 
The emergence of Large Language Models has introduced a novel paradigm in \nlsql tasks, enhancing capabilities dramatically. 
 However, this raises a critical question: 
\textit{Are we fully prepared to deploy \nlsql models in production?}}
 
To address the posed questions, we present a multi-angle \nlsql evaluation framework, \testdb, to 
facilitate the design and test of new \nlsql methods for researchers.
Through \testdb, we conduct a detailed comparison of leading \nlsql methods across a range of application scenarios, such as different data domains and \sql characteristics, offering valuable insights for selecting the most appropriate \nlsql methods for specific needs.
Moreover, we explore the \nlsql design space, leveraging \testdb to automate the identification of an optimal \nlsql solution tailored to user-specific needs. 
Specifically, \testdb identifies an effective \nlsql method, \sys, distinguished under the Spider dataset using the execution accuracy metric.
\rev{Remarkably, \sys achieves competitive performance with execution accuracy of \textbf{87\%} and \textbf{62.66\%} on the Spider and BIRD test sets, respectively.}

\end{abstract}

\maketitle

\pagestyle{\vldbpagestyle}
\begingroup\small\noindent\raggedright\textbf{PVLDB Reference Format:}\\
Boyan Li, Yuyu Luo, Chengliang Chai, Guoliang Li, Nan Tang. \vldbtitle. PVLDB, \vldbvolume(\vldbissue): \vldbpages, \vldbyear.\\
\href{https://doi.org/\vldbdoi}{doi:\vldbdoi}
\endgroup
\begingroup
\renewcommand\thefootnote{}\footnote{\noindent
*Yuyu Luo is the corresponding author. \\
This work is licensed under the Creative Commons BY-NC-ND 4.0 International License. Visit \url{https://creativecommons.org/licenses/by-nc-nd/4.0/} to view a copy of this license. For any use beyond those covered by this license, obtain permission by emailing \href{mailto:info@vldb.org}{info@vldb.org}. Copyright is held by the owner/author(s). Publication rights licensed to the VLDB Endowment. \\
\raggedright Proceedings of the VLDB Endowment, Vol. \vldbvolume, No. \vldbissue\ %
ISSN 2150-8097. \\
\href{https://doi.org/\vldbdoi}{doi:\vldbdoi} \\
}\addtocounter{footnote}{-1}\endgroup

\ifdefempty{\vldbavailabilityurl}{}{
\vspace{.3cm}
\begingroup\small\noindent\raggedright\textbf{PVLDB Artifact Availability:}\\
The source code, data, and/or other artifacts have been made available at \url{\vldbavailabilityurl}.
\endgroup
}

\setcounter{figure}{0}

\section{Introduction}
\label{sec:intro}

\begin{figure}[t!]
\hspace*{-.5em}
	\includegraphics[width=1.05\columnwidth]{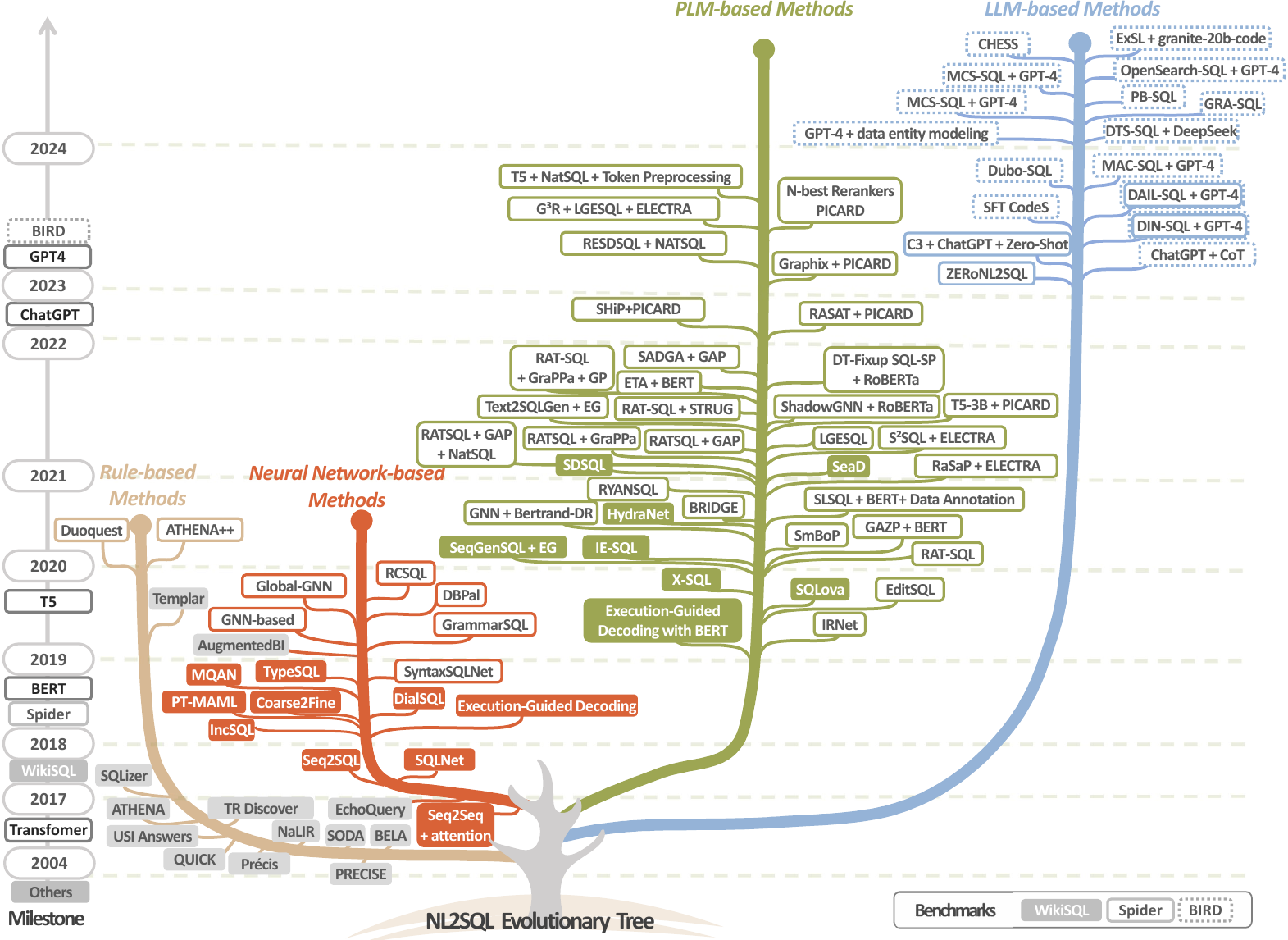}
    \myvspace{-2em}
	\caption{An Overview of \nlsql Methods.}
	\label{fig:tree}
    \vspace{-1em}
\end{figure}

\begin{figure}[t!]
	\includegraphics[width=\columnwidth]{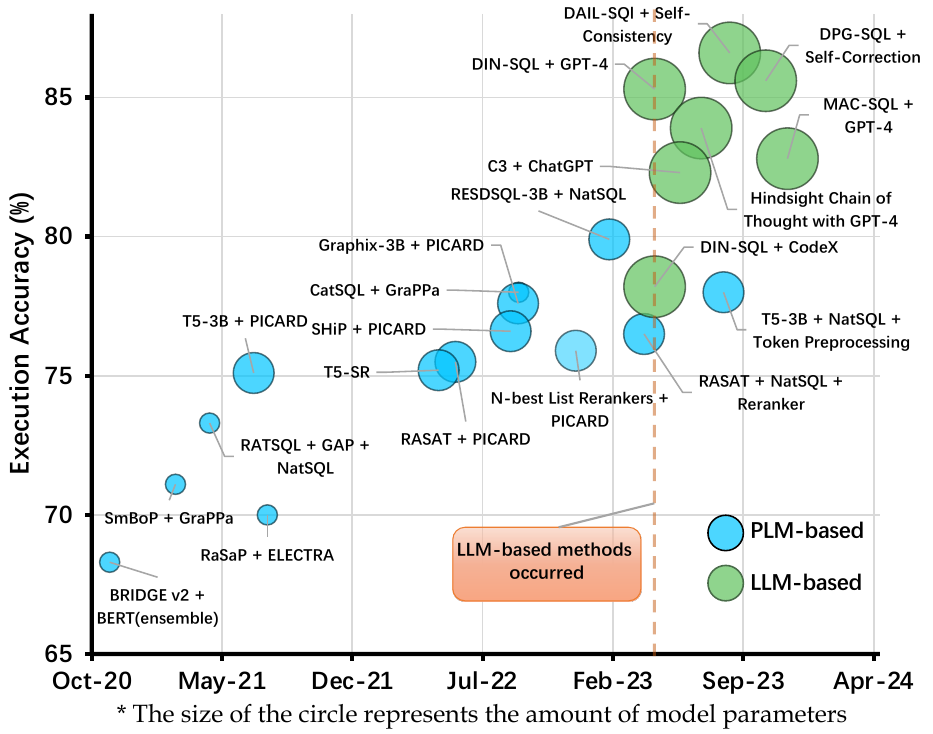}
    \myvspace{-2em}
	\caption{Evolution of PLM- and LLM-based \nlsql Models.}
	\label{fig:trend}
    \vspace{-1.75em}
\end{figure}

\begin{figure*}[t!]
	\centering
	\includegraphics[width=\textwidth]{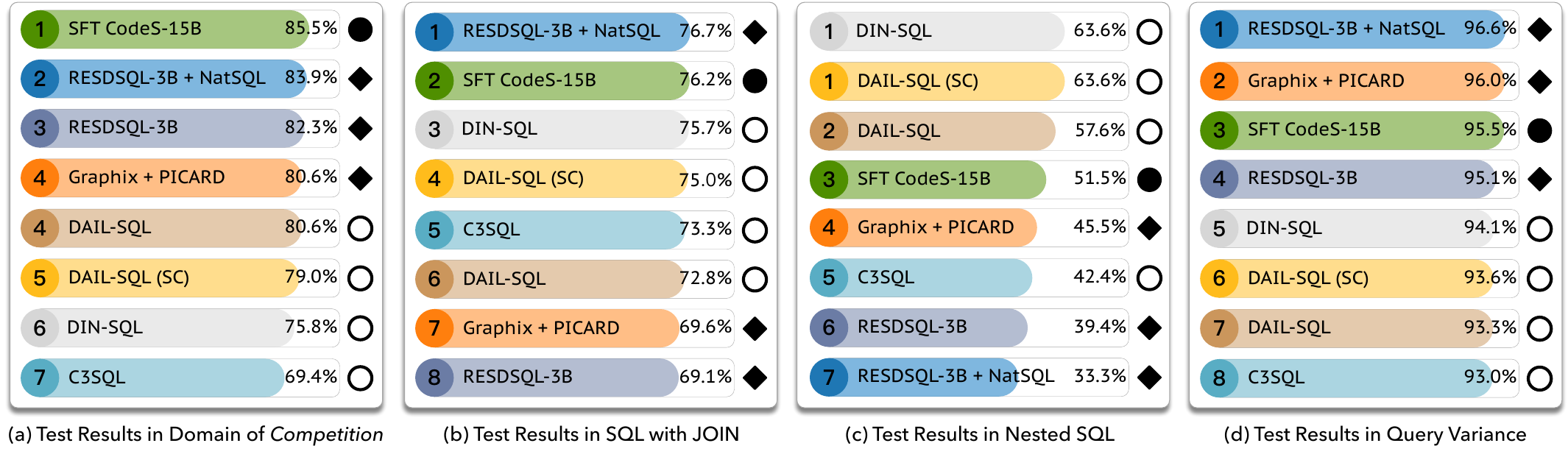}
    \myvspace{-2.em}
	\caption{NL2SQL Models on Spider from Different Angles {\small ($\bigcirc$: Prompting LLM, \ding{108}: Fine-tuning LLM, \ding{117}: Fine-tuning PLM)}.}
	\label{fig:leaderboard}
    \myvspace{-1em}
\end{figure*}

Natural Language to SQL (\nlsql), which converts a natural language query (\nlq) into an SQL query (\sql), can significantly lower the barrier for both lay users and expert users in accessing massive datasets and deriving insights~\cite{DBLP:journals/pacmmod/GuF00JM023, DBLP:conf/acl/ChenCWMPSS023, DBLP:conf/kdd/WangQHLYWLSHSL22,DBLP:conf/kdd/LiuH0W22,pourreza2023din,gao2023text,li2023resdsql, DBLP:conf/sigmod/LuoQ00W18, DBLP:conf/icde/LuoQ0018, DBLP:conf/sigmod/Luo00CLQ21,
	DBLP:journals/tvcg/LuoTLTCQ22,
	DBLP:conf/sigmod/TangLOLC22, xie2024haicharthumanaipaired, DBLP:journals/tkde/LuoQCTLL22, DBLP:journals/corr/abs-2109-03506, nl2sqlsurvey}.
Especially being empowered by the recent advances of large language models, the performance of \nlsql solutions has been significantly improved. The trend of providing \nlsql solutions by database vendors has shifted from a myth to a must-go. 

\rev{Despite all these efforts in tackling \nlsql, there are still many important questions, from where we are now, what \nlsql research topic should be studied next for researchers, to which method one should apply to a specific application for practitioners -- this paper systematically examines and answers these questions.}






\stitle{Q1: Where Are We Now?}
Figure~\ref{fig:tree} shows the evolution of \nlsql methods in the last two decades, from rule-based methods, deep neural network-based methods, tunable pre-trained language models (PLMs), to giant large language models (LLMs), alongside the development of benchmarks like Spider~\cite{DBLP:conf/emnlp/YuZYYWLMLYRZR18} and BIRD~\cite{bird}. 
Note that LLMs (\eg GPT-4~\cite{gpt4} and Llama2~\cite{DBLP:journals/corr/abs-2307-09288}) are larger language models compared to PLMs (e.g., GPT-2~\cite{gpt2} and BART~\cite{lewis2019bart}) and exhibit advanced language understanding and emergent abilities~\cite{minaee2024large, DBLP:conf/cidr/0001YF0LH24, YE202443, DBLP:journals/corr/abs-2406-07815}. Employing PLMs for the \nlsql task requires fine-tuning on task-specific datasets, while harnessing LLMs for this task can be done through prompts (in-context learning) for all kinds of LLMs or fine-tuning (\ie instruction following) for open-source LLMs only~\cite{zhao2023survey}. 

Figure~\ref{fig:trend} compares the accuracy of PLM-based (blue dots) and LLM-based (green dots) \nlsql models on  Spider leaderboard~\cite{DBLP:conf/emnlp/YuZYYWLMLYRZR18}.
It shows that LLM-based \nlsql models started in Feb 2023 (DIN-SQL $+$ CodeX) with comparable accuracy to PLM-based models. However, with the fast evolution of LLMs, the performance gap between LLM- and PLM-based models has been widening, highlighting the advantages of LLM-based approaches.

\stitle{Q2: Are LLM-based Models the Clear Winner?}
Based on Figure~\ref{fig:trend}, \textit{can we conclude that LLM-based models are ``the choice'' for any \nlsql application?}
In other words, is selecting the model ranked at the top of the leaderboard always the best strategy?


Correctly answering this question is crucial in helping \textit{researchers} and \textit{practitioners} design and select the right model for different needs. 
Let's consider classical Business Intelligence (BI) use cases.

%

[{\em Various Data Domains.}]  
BI platforms like Tableau~\cite{tableau} often have various database domains (\eg movies and sports) with unique schemas and terminologies.
An ideal \nlsql model must generalize across these varied domains while adapting to each specific domain to meet ad-hoc requirements effectively.

[{\em Complex SQL operations.}]  Real-world applications often require the execution of complex \sql queries, involving advanced operations such as multiple {\tt JOIN}s, nested queries, and aggregation functions. The capability to accurately generate complex queries is an important criterion for evaluating \nlsql models. 

[{\em New Linguistic Phenomena.}]
For the same query intent, different users may pose \nlq questions with different abbreviations, synonyms, and question styles. 
Thus, the ability of an \nlsql model to accurately interpret various \nlq query variants is crucial.

\begin{example}
\label{exam:comp}
Figure~\ref{fig:leaderboard} compares the SOTA PLM- and LLM-based models from different angles on the Spider development dataset in terms of the Execution-Accuracy metric.


$[$Various Data Domains$]$
Figure~\ref{fig:leaderboard}(a) compares different models in the \textit{Competition} domain. 
\rev{The result shows that fine-tuning-based LLM/PLM methods outperform all prompt-based LLM methods. Specifically, the best PLM-based method, RESDSQL-3B$+$NatSQL~\cite{li2023resdsql}, achieves $83.9\%$ execution accuracy, which outperforms the best prompt-based LLM method, DAILSQL (with GPT-4)~\cite{gao2023text}, by $3.3\%$.
The above observations suggest that fine-tuning is a crucial strategy for enhancing the domain adaptation capabilities of \nlsql models.
}

$[$Complex SQL operations$]$
\rev{
Figure~\ref{fig:leaderboard}(b) compares different models on use cases with only \sql queries with {\tt JOIN} operators. It shows that the PLM-based method RESDSQL-3B$+$NatSQL~\cite{li2023resdsql} is ranked at the top, outperforming all LLM-based methods.

However, when we compare different methods on use cases with only nested \sql queries, as shown in Figure~\ref{fig:leaderboard}(c), we observe that the LLM-based methods generally outperform PLM-based methods. 
}


$[$New Linguistic Phenomena$]$
We also compute the average accuracy of the methods on different linguistic phenomena (\eg ``{Return all customers whose total consumption is greater than 1000}'' \vs ``{What is the list of customers who spent more than 1,000?}'').
\rev{Figure~\ref{fig:leaderboard}(d) shows that although both types of methods perform well, fine-tuned LLM and PLM for \nlsql are superior to prompting LLM for \nlsql. This is primarily because fine-tuned models better align different query variants with database schemas.} \eop
\end{example}


Example~\ref{exam:comp} shows that {\em one size does not fit all}; that is, no \nlsql model is a clear winner on different usage scenarios, even powered by currently the most powerful LLM GPT-4. In fact, real-world scenarios are much more complicated than what can be examined in public \nlsql benchmarks such as Spider and BIRD. 
\textit{Therefore, there is an urgent need for tools that can help systematically evaluate \nlsql models from different angles on a given benchmark}.



%

\stitle{Q3: Can we combine the best of both worlds and design a super NL2SQL model?}
The question following {\bf Q1} and {\bf Q2} is: if there is no single winner in different scenarios, can we design a super \nlsql model that combines the merits of both PLMs and LLMs and is robust for different scenarios.





\stitle{Contributions.}
In this paper, we present \testdb, a testbed designed to systematically evaluate PLM- and LLM-based \nlsql models across different benchmarks from multiple perspectives. \testdb can help researchers and practitioners better evaluate \nlsql models on their specific scenarios, uncover insightful experimental findings, and design a superior \nlsql model that is more robust than SOTA solutions. 
Our contributions are as follows.


\stitle{(1) \testdb: multi-angle NL2SQL evaluation.}
We design a testbed, \testdb, for fine-grained evaluation of \nlsql solutions.
\testdb can assess different \nlsql methods against established benchmarks or tailor their evaluations based on specific criteria (\eg varying data domains or \sql characteristics).
(Section~\ref{sec:method})

\stitle{(2) New experimental findings.}
\rev{We tested 13 LLM-based and 7 PLM-based \nlsql solutions on the Spider and BIRD datasets, varying in 15 different settings to analyze their performance in various usage scenarios (Section~\ref{sec:exp})}.
The key findings are as follows:




\underline{\em (i) Accuracy.} 
\rev{
Fine-tuning is crucial for enhancing performance. Specifically, LLM-based methods with fine-tuning excel in the EX metric, while PLM-based methods lead in the EM metric. 
Furthermore, PLM/LLM-based methods can be distinguished as winners in subsets of \sql with specific characteristics. For example, methods using GPT-4 perform notably better with subqueries.
}

\underline{\em (ii) NL Query Variance.} %
\rev{
For generating the same target \sql from different \nlq Queries, LLMs and PLMs fine-tuned on scenario-specific data exhibit stronger stability.
}
\note{R4 W1}%
\note{R4 D1}%
\note{R4 W2}%
\note{R4 D2}%
	

\underline{\em (iii) Domain Adaption.}
For \nlsql tasks across different domains, there is no clear winner between LLM-based and PLM-based methods. \rev{
However, in-domain data during the fine-tuning process is crucial for model performance in specific domains.
}


\underline{\em (iv) The Impact of Corpus in Pre-training.}
Our experiments reveal that after fine-tuning, LLMs pre-trained on code-specific datasets—like CodeLlama-7B, StarCoder-7B, and Deepseek-Coder-7B—outperform Llama2-7B, which is trained on general text, in \nlsql tasks. This highlights the significant impact of an LLM's pre-training data domain, or its intrinsic code capabilities, on its performance in specialized tasks such as \nlsql.

\rev{\stitle{(3) \sys: A robust NL2SQL model.}
We systematically categorize and analyze the most representative \nlsql modules based on LLMs and PLMs, highlighting their commonalities and distinct features. Building on this exploration, we propose \sys, which achieves competitive execution accuracy of \textbf{87\%} on the Spider test set and \textbf{62.66\%} on the BIRD test set.
(Section~\ref{sec:q3})}

\stitle{(4) What needs to be done next.}
Based on our experimental findings, design space exploration, and the implementation and testing of \sys, we identify three future research opportunities:
i) enhancing the trustworthiness of \nlsql methods, which includes handling ambiguous \nlq queries, diagnosing the match between the \nlq query and the predicted \sql, and interpreting the query results back to the \nlq query;
ii) developing cost-effective \nlsql solutions;
and
iii) automatically and adaptively generating training data (\nlq, \sql) based on evaluation results. (Section~\ref{sec:opp})

\section{Natural Language to SQL}
\label{sec:pre}


Let $\mathcal{N}$ be an \nlq query, $\mathcal{D}$ be a relational database with $n$ tables $\{T_1, \ldots, T_n\}$. The problem of natural language to \sql (\nlsql) is to generate an \sql query $\mathcal{Q}$ based on  $\mathcal{N}$ and the database $\mathcal{D}$.


Next, we describe related work by categorizing recent LLM-based/PLM-based \nlsql solutions into a taxonomy. We close this section by discussing the limitations of the existing works.




 \subsection{Related Works: A Bird's-Eye View}



Figure~\ref{fig:tree} illustrates an evolutionary tree of \nlsql techniques, categorized into four main branches: rule-based methods, neural network-based methods, PLM-based, and LLM-based methods.

\stitle{Rule-based Methods.} 
Early work relied on pre-defined rules or semantic parsers~\cite{DBLP:journals/corr/abs-2204-00498, 10.1145/2588555.2594519,DBLP:journals/vldb/KatsogiannisMeimarakisK23, DBLP:conf/sigmod/Katsogiannis-Meimarakis21}.
For example, NaLIR~\cite{10.1145/2588555.2594519} uses a syntactic parser and handcrafted rules to convert \nlq queries into \sql queries. 
However, these methods are limited in adaptability, scalability, and generalization.

\stitle{Neural Network-based Methods.} 
To address these limitations, researchers began using neural networks to translate \nlq queries to \sql queries.
Several large-scale benchmark datasets such as WikiSQL~\cite{zhong2017seq2sql} and Spider~\cite{DBLP:conf/emnlp/YuZYYWLMLYRZR18} were released.
Sequence-to-sequence \nlsql methods such as IRNet~\cite{mismatch} were developed. IRNet encodes \nlq queries and database schemas with an encoder and generates \sql queries with a decoder.

\stitle{PLM-based Methods.}
Around 2017, the introduction of the Transformer~\cite{vaswani2017attention} and the Spider dataset led to the rise of neural network-based methods, which soon dominated the field. Models like BERT~\cite{devlin2018bert} and T5~\cite{raffel2020exploring} heralded the era of pre-trained language models, achieving top results on benchmark datasets~\cite{li2023graphix, li2023resdsql, scholak2021picard}. For instance, RESDSQL~\cite{li2023resdsql}, a top performer on the Spider leaderboard, uses a two-stage framework: it first identifies relevant schema elements from the \nlq query, then constructs the \sql query.

\stitle{LLM-based Methods.}
The emergence of large language models like ChatGPT and GPT-4~\cite{gpt4} has revolutionized \nlsql solutions. LLM-based methods now dominate the \nlsql landscape~\cite{pourreza2023din, gao2023text, dong2023c3, codes, macsql}. For instance, DAIL-SQL~\cite{gao2023text} uses GPT-4 with prompt engineering, achieving competitive results on the Spider dataset.

Given the growth trend observed in the \nlsql evolutionary tree, we anticipate that LLM-based/PLM-based \nlsql methods will continue to dominate the field in the coming years. Therefore, it is important for us to fully understand the capabilities, limitations, and potential improvements of these \nlsql methods.

\stitle{Key Modules in NL2SQL Systems.}
Table~\ref{tab:taxonomy} categorizes state-of-the-art \nlsql methods based on backbone models and several key components.
Roughly speaking, recent competitive methods adopt language models as the backbone for \nlsql translation, either using giant and API-based large language models such as GPT-4 or tunable language models like T5 and Llama.


The schema linking module is vital to the \nlsql process, as it enhances the accuracy and relevance of generated \sql queries by incorporating database content. This underscores the importance of understanding both the schema and content of databases for \nlsql tasks. 
During \sql generation, PLM-based methods use beam search-like strategies (\eg PICARD~\cite{scholak2021picard}) to find optimal tokens within \sql syntax constraints, while LLM-based methods use greedy strategies. 
For post-processing, LLM-based methods employ heuristic prompting strategies like Self-Correction and Self-Consistency to refine outputs to better match intended \sql queries.

\begin{table*}[t!]
	\centering
	\caption{\rev{Taxonomy of PLM- and LLM-based NL2SQL Methods.}}
	\label{tab:taxonomy}
    \myvspace{-.5em}
	\resizebox{\textwidth}{!}{%
		\begin{tabular}{|cc|c|c|c|c|c|ccc|c|}
			\hline
			\multicolumn{2}{|c|}{\multirow{2}{*}{Types}} &
			\multirow{2}{*}{Methods} &
			\multirow{2}{*}{\begin{tabular}[c]{@{}c@{}}Backbone\\ Models\end{tabular}} &
			\multirow{2}{*}{\begin{tabular}[c]{@{}c@{}}Example\\ Selection \\ (Few-shot) \end{tabular}} &
			\multirow{2}{*}{\begin{tabular}[c]{@{}c@{}}Schema\\ Linking\end{tabular}} &
			\multirow{2}{*}{\begin{tabular}[c]{@{}c@{}}DB\\ Content\end{tabular}} &
			\multicolumn{3}{c|}{SQL Generation Strategy} &
			\multirow{2}{*}{\begin{tabular}[c]{@{}c@{}}Post-processing\\ Strategy\end{tabular}} \\ \cline{8-10}  
            \multicolumn{2}{|c|}{} &
			&
			&
			&
			&
            &
			\multicolumn{1}{c|}{{Multi-Step}} &
			\multicolumn{1}{c|}{\begin{tabular}[c]{@{}c@{}}Intermediate \\ Representation\end{tabular}} &
			Decoding Strategy &
			\\ \hline \hline
			\multicolumn{1}{|c|}{\multirow{6}{*}{\rotatebox[origin=c]{90}{LLM-based \hspace{1cm}}}} &
            \rev{\multirow{5}{*}{\rotatebox[origin=c]{90}{Prompting \hspace{0.8cm}}}} &
			DIN-SQL~\cite{pourreza2023din} &  GPT-4 &
			Manual &
			\ding{51} &
			\ding{55} &
			\multicolumn{1}{c|}{\begin{tabular}[c]{@{}c@{}}Classification\\ Decomposition\end{tabular}} 
			&
			\multicolumn{1}{c|}{NatSQL} &
			Greedy Search &
			Self-Correction \\ \cline{3-11} 
			\multicolumn{1}{|c|}{} &
            &
			\begin{tabular}[c]{@{}c@{}}DAIL-SQL~\cite{gao2023text}\\ (with Self-Consistency)\end{tabular}& GPT-4 &
		  Similarity-based   & 
			\ding{55} &
			\ding{55} &
			\multicolumn{1}{c|}{\ding{55}} &
			\multicolumn{1}{c|}{\ding{55}} &
			Greedy Search &
			Self-Consistency \\ \cline{3-11} 
            \multicolumn{1}{|c|}{} &
            &
			MAC-SQL~\cite{macsql} &  GPT-4 &
			    N/A &
			  \ding{51}&
			\ding{55} &
			\multicolumn{1}{c|}{
				\begin{tabular}[c]{@{}c@{}}Sub-question\\ Decomposer\end{tabular}
				 } 
			&
			\multicolumn{1}{c|}{\ding{55}} &
			Greedy Search &
			Refiner \\ \cline{3-11} 
            \multicolumn{1}{|c|}{} &
            &
			C3-SQL~\cite{dong2023c3} & GPT-3.5 &
			N/A &
			\ding{51} &
			\ding{55} &
			\multicolumn{1}{c|}{\ding{55}} &
			\multicolumn{1}{c|}{\ding{55}} &
			Greedy Search &
			Self-Consistency \\ \cline{3-11} 
            \multicolumn{1}{|c|}{} &
            &
            \rev{CodeS~\cite{codes}} &
            \rev{StarCoder} &
            \rev{Similarity-based} &
            \rev{\ding{51}} &
            \rev{\ding{51}} &
            \multicolumn{1}{c|}{\rev{\ding{55}}} &
            \multicolumn{1}{c|}{\rev{\ding{55}}} &
            \rev{Beam Search} &
            \begin{tabular}[c]{@{}c@{}}\rev{Execution-Guided}\\ \rev{SQL Selector}\end{tabular} \\ \cline{2-11} 
            \multicolumn{1}{|c|}{} &
            \rev{\multirow{10}{*}{\rotatebox{90}{Fine-tuning \hspace{0.5cm}}}} &
            \rev{SFT CodeS~\cite{codes}} &
            \rev{StarCoder} &
            \rev{N/A} &
            \rev{\ding{51}} &
            \rev{\ding{51}} &
            \multicolumn{1}{c|}{\rev{\ding{55}}} &
            \multicolumn{1}{c|}{\rev{\ding{55}}} &
            \rev{Beam Search} &
            \begin{tabular}[c]{@{}c@{}}\rev{Execution-Guided}\\ \rev{SQL Selector}\end{tabular} \\ \cline{1-1} \cline{3-11} 
			\multicolumn{1}{|c|}{\multirow{9}{*}{\rotatebox{90}{PLM-based}}} &
            &
			RESDSQL + NatSQL~\cite{li2023resdsql} & T5 &
			N/A &
			\ding{51} &
			\ding{51} &
			\multicolumn{1}{c|}{{Skeleton Parsing}} &
			\multicolumn{1}{c|}{NatSQL} &
			Beam Search &
			\begin{tabular}[c]{@{}c@{}}Execution-Guided\\ SQL Selector\end{tabular} \\ \cline{3-11} 
            \multicolumn{1}{|c|}{} &
			&
			Graphix + PICARD~\cite{li2023graphix} & T5 &
			N/A &
			\ding{51} &
			\ding{51} &
			\multicolumn{1}{c|}{\ding{55}} &
			\multicolumn{1}{c|}{\ding{55}} &
			PICARD &
			\ding{55} \\ \cline{3-11} 
            \multicolumn{1}{|c|}{} &
			&
			N-best Rerankers + PICARD~\cite{zeng2023n} & T5 &
			N/A &
			\ding{51} &
			\ding{51} &
			\multicolumn{1}{c|}{\ding{55}} &
			\multicolumn{1}{c|}{\ding{55}} &
			PICARD &
			N-best Rerankers \\ \cline{3-11} 
            \multicolumn{1}{|c|}{} &
			&
			T5 + NatSQL + Token Preprocessing~\cite{rai2023improving} & T5 &
			N/A &
			\ding{51} &
			\ding{51} &
			\multicolumn{1}{c|}{\ding{55}} &
			\multicolumn{1}{c|}{NatSQL} &
			Greedy Search &
			\ding{55} \\ \cline{3-11} 
            \multicolumn{1}{|c|}{} &
			&
			RASAT + PICARD~\cite{qi2022rasat} & T5 &
			N/A &
			\ding{51} &
			\ding{51} &
			\multicolumn{1}{c|}{\ding{55}} &
			\multicolumn{1}{c|}{\ding{55}} &
			PICARD &
			\ding{55} \\ \cline{3-11} 
            \multicolumn{1}{|c|}{} &
			&
			SHiP + PICARD~\cite{zhao2022importance} & T5 &
			N/A &
			\ding{55} &
			\ding{51} &
			\multicolumn{1}{c|}{\ding{55}} &
			\multicolumn{1}{c|}{\ding{55}} &
			PICARD &
			\ding{55} \\ \cline{3-11} 
            \multicolumn{1}{|c|}{} &
			&
			T5 + PICARD~\cite{scholak2021picard} & T5 &
			N/A &
			\ding{55} &
			\ding{51} &
			\multicolumn{1}{c|}{\ding{55}} &
			\multicolumn{1}{c|}{\ding{55}} &
			PICARD &
			\ding{55} \\ \cline{3-11} 
            \multicolumn{1}{|c|}{} &
			&
			RATSQL + GAP + NatSQL~\cite{gan2021natural}  & BART &
			N/A &
			\ding{51} &
			\ding{51} &
			\multicolumn{1}{c|}{\ding{55}} &
			\multicolumn{1}{c|}{NatSQL} &
			\ding{55} &
			\ding{55} \\ \cline{3-11} 
            \multicolumn{1}{|c|}{} &
			&
			BRIDGE v2~\cite{lin2020bridging} & BERT &
			N/A &
			\ding{55} &
			\ding{51} &
			\multicolumn{1}{c|}{\ding{55}} &
			\multicolumn{1}{c|}{\ding{55}} &
			\begin{tabular}[c]{@{}c@{}}Schema-Consistency\\ Guided Decoding\end{tabular} &
			\ding{55} \\ \hline	
		\end{tabular}%
	}
    \myvspace{-.5em}
\end{table*}

\subsection{\mbox{Existing Experiments and Their Limitations}}

\stitle{Existing Experiments.}
There are several experimental studies relevant to our research. 
For example, Gao et al.~\cite{gao2023text} evaluated the potential of open-source LLMs for \nlsql tasks through prompt engineering. 
Rajkumar et al.~\cite{DBLP:journals/corr/abs-2204-00498} explored the capabilities of the Codex language model in handling the \nlsql task under zero-shot and few-shot settings.
Gkini et al.~\cite{DBLP:conf/sigmod/GkiniBKI21} conducted an in-depth evaluation of parsing-based and keyword-based \nlsql. While the first two studies mainly focused on evaluating LLM-based \nlsql solutions, the third investigated parsing-based \nlsql methods. 

\stitle{Their Limitations.} 
Existing experiments have several limitations.

\stab(1) {\bf \em Overlook the Usage Scenarios.}
Existing evaluations typically report overall results on the entire benchmark datasets (\eg Spider). 
While this provides a broad overview, it falls short in offering detailed comparisons across specific subsets of the data (see Figure~\ref{fig:leaderboard}).
For example, we can filter the evaluated datasets based on distinct \sql characteristics or database domains, which could yield valuable insights into the relative effectiveness of different \nlsql models for particular \sql query types or domain-specific scenarios.

\stab(2) {\bf \em Lack of Direct and Comprehensive Comparisons.} 
One primary limitation is that many recent \nlsql solutions, especially those based on LLM and PLM, have not been systematically compared on well-established benchmarks and customized datasets.


\stab(3) {\bf \em Limited Exploration of the NL2SQL Design Space.} 
Current \nlsql research and practice have limited exploration of the \nlsql framework's design space using both LLM and PLM-based modules.
This restricts our understanding of how different architectural and functional modules from both LLM and PLM can be synergistically incorporated to enhance \nlsql solutions.

\section{NL2SQL360: A Testbed for NL2SQL}
\label{sec:method}





\begin{figure}[t!]
	\centering
\includegraphics[width=.7\columnwidth]{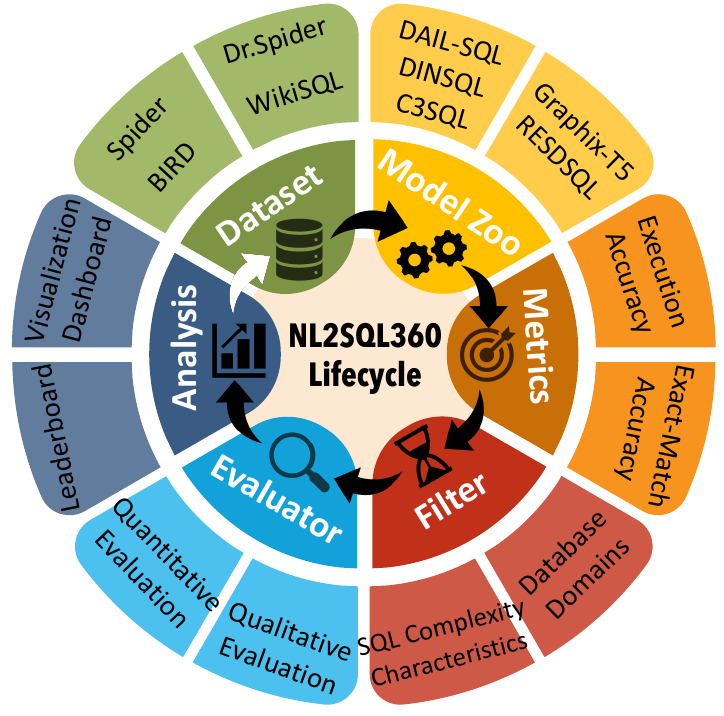}
    \myvspace{-1em}
	\caption{An Overview of \testdb.}
	\label{fig:hypersql}
    \myvspace{-2em}
\end{figure}

Figure~\ref{fig:hypersql} shows the framework of our testbed, \testdb, comprising six core components, as discussed below.

\stitle{Benchmark Datasts.}  \rev{This module maintains widely-used benchmarks: Spider~\cite{DBLP:conf/emnlp/YuZYYWLMLYRZR18}, BIRD~\cite{bird}, Spider-Realistic~\cite{spider-realistic}, Dr.Spider~\cite{DBLP:conf/iclr/Chang0DPZLLZJLA23}, KaggleDBQA~\cite{DBLP:conf/acl/LeePR20}, WikiSQL~\cite{zhong2017seq2sql}, etc.}






\stitle{Model Zoo.} This module hosts a collection of competitive and open-source \nlsql models featured on the Spider and BIRD leaderboards. It mainly includes LLM-based and PLM-based methods.

%








\stitle{Dataset Filter.} 
Traditional evaluations, which average performance across whole benchmark datasets, miss nuanced \nlsql insights for different scenarios (Section~\ref{sec:intro}). To address this, we select specific benchmark subsets, including particular databases, \nlq, and \sql queries, highlighting unique traits like query complexity, database schema diversity, and \sql features such as {\tt JOIN} operations or nested queries.

%
Therefore, we introduce a dataset filtering mechanism in our \testdb. This allows for the segregation of testing datasets into more focused subsets based on various criteria:

(1) \textit{\bf Scenario-1: SQL Complexity}. 
This scenario differentiates \sql queries by complexity, from straightforward to intricate queries with multiple clauses and conditions. The classification follows the criteria established by Spider~\cite{DBLP:conf/emnlp/YuZYYWLMLYRZR18}, aiming to evaluate how well \nlsql methods handle varying levels of \sql difficulty.

(2) \textit{\bf Scenario-2: SQL Characteristics}.
It examines \sql queries that primarily utilize specific features, such as {\tt JOIN} operations, subqueries, or aggregate functions. By categorizing queries based on these characteristics, we can evaluate an \nlsql system's ability to manage distinct \sql functionalities. \textit{For example, business intelligence platforms often handle analytic queries with nested subqueries.}

(3) \textit{\bf Scenario-3: Data Domains}.
This scenario explores the system's performance across various data domains, such as finance, healthcare, and retail. By categorizing \nlsql databases according to their data domains, we provide a structured framework for evaluating domain-specific capabilities and potential limitations.

(4) \textit{\bf Scenario-4: Query Variance Testing}.
It assesses the \nlsql system's robustness and flexibility in handling variations in natural language queries. It tests the \nlsql system's response to different phrasings and structures, measuring user-friendliness and adaptability to diverse linguistic styles. We use a variety of natural language queries from \nlsql datasets as testing samples.

\stitle{Evaluation Metrics.} \rev{We support a set of widely-accepted metrics.
Specifically, we adopt \textit{Execution Accuracy} (EX) and \textit{Exact Match Accuracy} (EM)~\cite{DBLP:conf/emnlp/YuZYYWLMLYRZR18} to assess the effectiveness of the generated SQL queries. In addition, we use the \textit{Valid Efficiency Score} (VES)~\cite{bird} to measure the efficiency of generating valid \sql queries.

To further evaluate the robustness and flexibility of \nlsql solutions in handling variations in natural language queries, we propose a new metric called \textit{Query Variance Testing}. This metric assesses how well the models can adapt to different forms of \nlq queries.}

\note{R1 D2}%
\note{R4 W3}%
\note{R4 D3}%
\rev{
    Given a \sql query $Q_i$, there typically exist multiple corresponding \nlq queries, denoted as pairs  \{($N_1$, $Q_i$), ($N_2$, $Q_i$), \ldots, ($N_m$, $Q_i$)\}. %
	In evaluating an \nlsql model, these \nlq and \sql query pairs are incorporated into the test set only if the model accurately processes at least one pair among them. This allows us to construct a specific test set for each model to compute their average accuracy.
	
	The formula for computing \textit{QVT} accuracy is defined as follows:
    
	\begin{equation}
		\small
	QVT = \frac{1}{M} \sum_{i=1}^{M} \left( \frac{\sum_{j=1}^{m_i} \mathds{1} \left( \mathcal{F}(N_{ij}) = Q_i \right)}{m_i} \right)
	\label{eq:qvt}
    \end{equation}
    
	\myvspace{-.5em}
	where:
	\begin{itemize}
		\item $M$ is the total number of \sql queries in the test set.
		\item $m_i$ is the number of natural language query variations corresponding to the \sql query $Q_i$.
		\item $\mathcal{F}(N_{ij})$ represents the \sql query generated by the \nlsql model for the $j$-th natural language query variation of $Q_i$.
  	\item $\mathds{1}(\cdot)$ is the indicator function that returns 1 if the query results inside are equal, and 0 otherwise.
	\end{itemize}
}

\stitle{Executor and Logs.} 
Users can tailor the evaluation workflow of \nlsql models, setting parameters like hyper-parameters and metrics. The testbed then automatically runs these models on benchmarks (e.g., Spider) and custom subsets (e.g., nested queries), logging every outcome. These logs offer detailed insights into each model's performance, serving as the resource for model analysis.

\stitle{Evaluator.}
Leveraging data from Logs, the \texttt{Evaluator} automatically generates quantitative assessments, presented in easily interpretable formats like tables or leaderboards. Additionally, our testbed offers visualization tools and a dashboard for interactive analysis, allowing users to compare \nlsql solutions across dimensions such as database domains and \sql characteristics.


\section{Experiments}
\label{sec:exp}

\subsection{Experimental Settings}
\label{sec:exp_settings}

\begin{table}[t!]
	\centering
	\caption{\rev{Spider \vs BIRD Dataset Statistics.}}
    \myvspace{-1em}
	\label{tab:dataset_compare}
	\resizebox{\columnwidth}{!}{%
		\begin{tabular}{|c|ccc|ccc|ccc|ccc|ccc|}
			\hline
			\multirow{2}{*}{\textbf{Dataset}} &
			\multicolumn{3}{c|}{\textbf{\#-Tables / DB}} &
			\multicolumn{3}{c|}{\textbf{\#-Columns / DB}} &
			\multicolumn{3}{c|}{\textbf{\#-Columns / Tables}} &
			\multicolumn{3}{c|}{\textbf{\#-PKs / DB}} &
			\multicolumn{3}{c|}{\textbf{\#-FKs / DB}} \\ \cline{2-16} 
			&
			\multicolumn{1}{c|}{\textbf{Min}} &
			\multicolumn{1}{c|}{\textbf{Max}} &
			\textbf{Avg} &
			\multicolumn{1}{c|}{\textbf{Min}} &
			\multicolumn{1}{c|}{\textbf{Max}} &
			\textbf{Avg} &
			\multicolumn{1}{c|}{\textbf{Min}} &
			\multicolumn{1}{c|}{\textbf{Max}} &
			\textbf{Avg} &
			\multicolumn{1}{c|}{\textbf{Min}} &
			\multicolumn{1}{c|}{\textbf{Max}} &
			\textbf{Avg} &
			\multicolumn{1}{c|}{\textbf{Min}} &
			\multicolumn{1}{c|}{\textbf{Max}} &
			\textbf{Avg} \\ \hline \hline
			\textbf{\begin{tabular}[c]{@{}c@{}}Spider\\ Train Set\end{tabular}} &
			\multicolumn{1}{c|}{2} &
			\multicolumn{1}{c|}{26} &
			5.4 &
			\multicolumn{1}{c|}{6} &
			\multicolumn{1}{c|}{352} &
			27.8 &
			\multicolumn{1}{c|}{2} &
			\multicolumn{1}{c|}{48} &
			5.1 &
			\multicolumn{1}{c|}{0} &
			\multicolumn{1}{c|}{18} &
			4.8 &
			\multicolumn{1}{c|}{0} &
			\multicolumn{1}{c|}{25} &
			5.0 \\ \hline
			\textbf{\begin{tabular}[c]{@{}c@{}}BIRD\\ Train Set\end{tabular}} &
			\multicolumn{1}{c|}{2} &
			\multicolumn{1}{c|}{65} &
			7.6 &
			\multicolumn{1}{c|}{6} &
			\multicolumn{1}{c|}{455} &
			51.3 &
			\multicolumn{1}{c|}{1} &
			\multicolumn{1}{c|}{62} &
			6.8 &
			\multicolumn{1}{c|}{0} &
			\multicolumn{1}{c|}{65} &
			6.7 &
			\multicolumn{1}{c|}{0} &
			\multicolumn{1}{c|}{61} &
			6.1 \\ \hline \hline
			\textbf{\begin{tabular}[c]{@{}c@{}}Spider\\ Dev Set\end{tabular}} &
			\multicolumn{1}{c|}{2} &
			\multicolumn{1}{c|}{11} &
			4.1 &
			\multicolumn{1}{c|}{7} &
			\multicolumn{1}{c|}{56} &
			22.1 &
			\multicolumn{1}{c|}{2} &
			\multicolumn{1}{c|}{32} &
			5.4 &
			\multicolumn{1}{c|}{1} &
			\multicolumn{1}{c|}{10} &
			3.7 &
			\multicolumn{1}{c|}{1} &
			\multicolumn{1}{c|}{11} &
			3.2 \\ \hline
			\textbf{\begin{tabular}[c]{@{}c@{}}BIRD\\ Dev Set\end{tabular}} &
			\multicolumn{1}{c|}{3} &
			\multicolumn{1}{c|}{13} &
			6.8 &
			\multicolumn{1}{c|}{11} &
			\multicolumn{1}{c|}{199} &
			72.5 &
			\multicolumn{1}{c|}{2} &
			\multicolumn{1}{c|}{115} &
			10.6 &
			\multicolumn{1}{c|}{2} &
			\multicolumn{1}{c|}{13} &
			6.5 &
			\multicolumn{1}{c|}{1} &
			\multicolumn{1}{c|}{29} &
			9.3 \\ \hline
		\end{tabular}%
	}
    \myvspace{-1em}
\end{table}

\stitle{Datasets.}
%
%
\note{R1 W2}%
\note{R1 D3}%
\note{R4 W3}%
\note{R4 D3}%
\rev{We use the development sets of Spider~\cite{DBLP:conf/emnlp/YuZYYWLMLYRZR18} and BIRD~\cite{bird} as our experimental datasets, which contain 1034 and 1534 (\nlq, \sql) samples, respectively. The \sql structure from the BIRD dataset is more complex and includes some keywords not covered by Spider, such as {\tt CASE}, {\tt IIF}, etc. This added complexity challenges the model's \nlsql ability. In addition, the databases in BIRD are more complex than those in Spider, as shown in Table~\ref{tab:dataset_compare}.}

\stitle{Methods.} We evaluate the state-of-the-art open-source LLM-based and PLM-based \nlsql methods.

\etitle{Prompt-based LLMs.} \rev{ We compare 4 prompt-based methods}:
\note{R4 W1}%
\note{R4 D1}%
\note{R4 W2}%
\note{R4 D2}%

\stab (1) DINSQL~\cite{pourreza2023din} decomposes the generation of \sql queries into different sub-problems and designs different prompts for each sub-problem to instruct GPT-4 to generate final \sql queries.

\stab (2) DAILSQL~\cite{gao2023text} encodes the question and database schema in \sql code style. It selects few-shot examples based on their structural (skeleton) similarities and query similarities. These elements are combined into an efficient prompt to guide GPT-4. 

\stab (3) DAILSQL(SC)~\cite{gao2023text}
is the version of DAILSQL with a Self-Consistency (SC) strategy for post-processing.

\stab (4) C3SQL~\cite{dong2023c3} uses schema linking filtering and a tailored calibration bias prompt with GPT-3.5 for \sql query generation, incorporating a self-consistency strategy for post-processing.
\note{R4 W1}%
\note{R4 D1}%

\etitle{Fine-tuning-based LLMs.} We evaluate 9 fine-tuning-based methods.

\rev{\stab (5-8) SFT CodeS (1B/3B/7B/15B)~\cite{codes}: CodeS is incrementally pre-trained based on StarCoder~\cite{DBLP:journals/corr/abs-2305-06161} using a large SQL-related corpus, which has demonstrated outstanding performance on many challenging \nlsql benchmarks. In the following experiments, we use SFT CodeS which is fine-tuned with Spider or BIRD datasets.
There are \textit{4} versions of SFT CodeS family models in our experiments.

\stab (9) Llama2-7B~\cite{DBLP:journals/corr/abs-2307-09288} uses an optimized Transformer as an auto-regressive language model, pre-trained on a vast corpus by Meta.

\stab (10) Llama3-8B~\cite{llama3modelcard} on over 15T tokens of data – a training dataset 7x larger than that used for Llama 2, including 4x more code.

\stab (11) StarCoder-7B~\cite{DBLP:journals/corr/abs-2305-06161} is a Code LLM that has been trained on permissively licensed data from GitHub. The data encompasses a wide range of content, including code from over 80 programming languages, Git commits, GitHub issues, and Jupyter notebooks. 

\stab (12) CodeLlama-7B~\cite{DBLP:journals/corr/abs-2308-12950} is an enhanced variant of Llama2, refined with additional training on code repository datasets.

\stab (13) Deepseek-Coder-7B~\cite{DBLP:journals/corr/abs-2401-14196} is trained on project-level code corpora and fill-in-the-blank tasks to boost code completion.}



\etitle{PLM-based NL2SQL.} We evaluate 7 the state-of-the-art methods:



\stab (1) Graphix-3B+PICARD~\cite{li2023graphix} integrates a pre-trained T5-3B transformer with graph-aware enhancements for \nlsql tasks, utilizing PICARD~\cite{scholak2021picard} to enhance performance.


\stab (2-4) RESDSQL(Base/Large/3B)~\cite{li2023resdsql}  introduces a ranking-enhanced encoding and skeleton-aware decoding to separate schema linking from skeleton parsing. 

\stab (5-7) RESDSQL(Base/Large/3B)$+$NatSQL~\cite{li2023resdsql} is the version incorporated with NatSQL~\cite{gan2021natural} for better performance. 
There are 6 versions of RESDSQL family models used in the experiments.

\stitle{Metrics.}
We evaluate different methods on Exact Match Accuracy (EM), Execution Accuracy (EX), Query Variance Testing (QVT), Valid Efficiency Socre (VES), Token Efficiency, and Latency metrics.

\stitle{Hardware and Platform.} All experiments are conducted on an Ubuntu 22.04.3 LTS server equipped with 512GB RAM and two 40-core Intel(R) Xeon(R) Platinum 8383C CPUs @ 2.70GHz. For the supervised fine-tuning of LLM experiments, we use 8 NVIDIA A800 (80GB) GPUs to fine-tune the models.

\subsection{Experiments on Evaluating Accuracy}
\label{sec:exp_accuracy}

\stitle{Exp-1: Accuracy vs. SQL Complexity.}
%
We evaluated all methods by varying \sql complexity using Spider and BIRD development sets. We computed Execution Accuracy (\textcolor{orange}{\bf EX}) and Exact Match Accuracy (\textcolor{blue}{\bf EM}) metrics. Note that we retrained RESDSQL from scratch on the BIRD train set but did not include NatSQL due to unavailable code. Additionally, We omitted DINSQL due to GPT resource constraints.

Table~\ref{tab:overall_spider} and Table~\ref{tab:overall_bird} report the results.
The state-of-the-art (SOTA) \textcolor{orange}{\bf EX} and \textcolor{blue}{\bf EM} in specific SQL complexity are marked as \textcolor{orange}{\bf orange} and \textcolor{blue}{\bf blue} in the table, respectively. 

\begin{table}[t!]
	\centering
	\caption{\rev{Accuracy \vs SQL Complexity in Spider-Dev.}}
    \myvspace{-1em}
	\label{tab:overall_spider}
	\resizebox{\columnwidth}{!}{%
		\begin{tabular}{|cc|c|c|ccccc|}
			\hline
			\multicolumn{2}{|c|}{\multirow{2}{*}{\textbf{Types}}} &
			\multirow{2}{*}{\textbf{Methods}} &
			\multirow{2}{*}{\textbf{Metrics}} &
			\multicolumn{5}{c|}{\textbf{Spider-Dev}} \\ \cline{5-9} 
			\multicolumn{2}{|c|}{} &
			&
			&
			\multicolumn{1}{c|}{\textbf{Easy}} &
			\multicolumn{1}{c|}{\textbf{Med.}} &
			\multicolumn{1}{c|}{\textbf{Hard}} &
			\multicolumn{1}{c|}{\textbf{Extra}} &
			\textbf{All} \\ \hline \hline
			\multicolumn{1}{|c|}{\multirow{16}{*}{\rotatebox[origin=c]{90}{\textbf{LLM-based}}}} &
			\multirow{8}{*}{\rotatebox[origin=c]{90}{\rev{\textbf{Prompting}}}} &
			\multirow{2}{*}{C3SQL} &
			EX &
			\multicolumn{1}{c|}{92.7} &
			\multicolumn{1}{c|}{85.2} &
			\multicolumn{1}{c|}{77.6} &
			\multicolumn{1}{c|}{62.0} &
			82.0 \\ \cline{4-9} 
			\multicolumn{1}{|c|}{} &
			&
			&
			EM &
			\multicolumn{1}{c|}{80.2} &
			\multicolumn{1}{c|}{43.5} &
			\multicolumn{1}{c|}{35.6} &
			\multicolumn{1}{c|}{18.1} &
			46.9 \\ \cline{3-9} 
			\multicolumn{1}{|c|}{} &
			&
			\multirow{2}{*}{DINSQL} &
			EX &
			\multicolumn{1}{c|}{92.3} &
			\multicolumn{1}{c|}{87.4} &
			\multicolumn{1}{c|}{76.4} &
			\multicolumn{1}{c|}{62.7} &
			82.8 \\ \cline{4-9} 
			\multicolumn{1}{|c|}{} &
			&
			&
			EM &
			\multicolumn{1}{c|}{82.7} &
			\multicolumn{1}{c|}{65.5} &
			\multicolumn{1}{c|}{42.0} &
			\multicolumn{1}{c|}{30.7} &
			60.1 \\ \cline{3-9} 
			\multicolumn{1}{|c|}{} &
			&
			\multirow{2}{*}{DAILSQL} &
			EX &
			\multicolumn{1}{c|}{91.5} &
			\multicolumn{1}{c|}{89.2} &
			\multicolumn{1}{c|}{77.0} &
			\multicolumn{1}{c|}{60.2} &
			83.1 \\ \cline{4-9} 
			\multicolumn{1}{|c|}{} &
			&
			&
			EM &
			\multicolumn{1}{c|}{89.5} &
			\multicolumn{1}{c|}{74.2} &
			\multicolumn{1}{c|}{55.5} &
			\multicolumn{1}{c|}{45.2} &
			70.0 \\ \cline{3-9} 
			\multicolumn{1}{|c|}{} &
			&
			\multirow{2}{*}{DAILSQL(SC)} &
			EX &
			\multicolumn{1}{c|}{91.5} &
			\multicolumn{1}{c|}{90.1} &
			\multicolumn{1}{c|}{75.3} &
			\multicolumn{1}{c|}{62.7} &
			83.6 \\ \cline{4-9}   \cline{4-9}
			\multicolumn{1}{|c|}{} &
			&
			&
			EM &
			\multicolumn{1}{c|}{88.3} &
			\multicolumn{1}{c|}{73.5} &
			\multicolumn{1}{c|}{54.0} &
			\multicolumn{1}{c|}{41.6} &
			68.7 \\ \cline{2-9} 
			\multicolumn{1}{|c|}{} &
			\multirow{14}{*}{\rotatebox[origin=c]{90}{\rev{\textbf{Fine-tuning}}}} &
			\multirow{2}{*}{\rev{SFT CodeS-1B}} &
			EX &
			\multicolumn{1}{c|}{92.3} &
			\multicolumn{1}{c|}{83.6} &
			\multicolumn{1}{c|}{70.1} &
			\multicolumn{1}{c|}{49.4} &
			77.9 \\ \cline{4-9} 
			\multicolumn{1}{|c|}{} &
			&
			&
			EM &
			\multicolumn{1}{c|}{91.5} &
			\multicolumn{1}{c|}{74.4} &
			\multicolumn{1}{c|}{65.5} &
			\multicolumn{1}{c|}{41.0} &
			71.7 \\ \cline{3-9} 
			\multicolumn{1}{|c|}{} &
			&
			\multirow{2}{*}{\rev{SFT CodeS-3B}} &
			EX &
			\multicolumn{1}{c|}{94.8} &
			\multicolumn{1}{c|}{88.3} &
			\multicolumn{1}{c|}{75.3} &
			\multicolumn{1}{c|}{60.8} &
			83.3 \\ \cline{4-9} 
			\multicolumn{1}{|c|}{} &
			&
			&
			EM &
			\multicolumn{1}{c|}{\textcolor{blue}{\textbf{94.4}}} &
			\multicolumn{1}{c|}{80.7} &
			\multicolumn{1}{c|}{67.8} &
			\multicolumn{1}{c|}{49.4} &
			76.8 \\ \cline{3-9} 
			\multicolumn{1}{|c|}{} &
			&
			\multirow{2}{*}{\rev{SFT CodeS-7B}} &
			EX &
			\multicolumn{1}{c|}{94.8} &
			\multicolumn{1}{c|}{\textcolor{orange}{\textbf{91.0}}} &
			\multicolumn{1}{c|}{75.3} &
			\multicolumn{1}{c|}{\textcolor{orange}{\textbf{66.9}}} &
			\textcolor{orange}{\textbf{85.4}} \\ \cline{4-9} 
			\multicolumn{1}{|c|}{} &
			&
			&
			EM &
			\multicolumn{1}{c|}{92.7} &
			\multicolumn{1}{c|}{\textcolor{blue}{\textbf{85.2}}} &
			\multicolumn{1}{c|}{67.8} &
			\multicolumn{1}{c|}{56.0} &
			79.4 \\ \cline{3-9} 
			\multicolumn{1}{|c|}{} &
			&
			\multirow{2}{*}{\rev{SFT CodeS-15B}} &
			EX &
			\multicolumn{1}{c|}{\textcolor{orange}{\textbf{95.6}}} &
			\multicolumn{1}{c|}{90.4} &
			\multicolumn{1}{c|}{\textcolor{orange}{\textbf{78.2}}} &
			\multicolumn{1}{c|}{61.4} &
			84.9 \\ \cline{4-9} 
			\multicolumn{1}{|c|}{} &
			&
			&
			EM &
			\multicolumn{1}{c|}{93.1} &
			\multicolumn{1}{c|}{83.4} &
			\multicolumn{1}{c|}{67.2} &
			\multicolumn{1}{c|}{54.2} &
			78.3 \\ \cline{1-1} \cline{3-9} 
			\multicolumn{1}{|c|}{\multirow{6}{*}{\rotatebox[origin=c]{90}{\textbf{PLM-based}}}} &
			&
			\multirow{2}{*}{RESDSQL-3B} &
			EX &
			\multicolumn{1}{c|}{94.8} &
			\multicolumn{1}{c|}{87.7} &
			\multicolumn{1}{c|}{73.0} &
			\multicolumn{1}{c|}{56.0} &
			81.8 \\ \cline{4-9} 
			\multicolumn{1}{|c|}{} &
			&
			&
			EM &
			\multicolumn{1}{c|}{94.0} &
			\multicolumn{1}{c|}{83.0} &
			\multicolumn{1}{c|}{66.7} &
			\multicolumn{1}{c|}{53.0} &
			78.0 \\ \cline{3-9} 
			\multicolumn{1}{|c|}{} &
			&
			\multirow{2}{*}{\begin{tabular}[c]{@{}c@{}}RESDSQL-3B \\ + NatSQL\end{tabular}} &
			EX &
			\multicolumn{1}{c|}{94.4} &
			\multicolumn{1}{c|}{87.9} &
			\multicolumn{1}{c|}{77.0} &
			\multicolumn{1}{c|}{66.3} &
			84.1 \\ \cline{4-9} 
			\multicolumn{1}{|c|}{} &
			&
			&
			EM &
			\multicolumn{1}{c|}{93.1} &
			\multicolumn{1}{c|}{83.0} &
			\multicolumn{1}{c|}{\textcolor{blue}{\textbf{70.1}}} &
			\multicolumn{1}{c|}{\textcolor{blue}{\textbf{65.7}}} &
			\textcolor{blue}{\textbf{80.5}} \\ \cline{3-9} 
			\multicolumn{1}{|c|}{} &
			&
			\multirow{2}{*}{\begin{tabular}[c]{@{}c@{}}Graphix-3B \\ + PICARD\end{tabular}} &
			EX &
			\multicolumn{1}{c|}{92.3} &
			\multicolumn{1}{c|}{86.3} &
			\multicolumn{1}{c|}{73.6} &
			\multicolumn{1}{c|}{57.2} &
			80.9 \\ \cline{4-9} 
			\multicolumn{1}{|c|}{} &
			&
			&
			EM &
			\multicolumn{1}{c|}{91.9} &
			\multicolumn{1}{c|}{82.3} &
			\multicolumn{1}{c|}{65.5} &
			\multicolumn{1}{c|}{53.0} &
			77.1 \\ \hline \hline
			\multicolumn{2}{|c|}{\multirow{2}{*}{\textbf{Hybird}}} &
			\multirow{2}{*}{\textbf{\sys}} &
			EX &
			\multicolumn{1}{c|}{94.4} &
			\multicolumn{1}{c|}{\begin{tabular}[c]{@{}c@{}} \textcolor{green}{\textbf{91.3}} \\ \textcolor{green}{(\textbf{0.3 $\uparrow$})} \end{tabular}} &
			\multicolumn{1}{c|}{\begin{tabular}[c]{@{}c@{}} \textcolor{green}{\textbf{83.3}} \\ \textcolor{green}{(\textbf{5.1 $\uparrow$})} \end{tabular}} &
			\multicolumn{1}{c|}{\begin{tabular}[c]{@{}c@{}} \textcolor{green}{\textbf{68.7}} \\ \textcolor{green}{(\textbf{1.8 $\uparrow$})} \end{tabular}} &
			\multicolumn{1}{c|}{\begin{tabular}[c]{@{}c@{}} \textcolor{green}{\textbf{87.0}} \\ \textcolor{green}{(\textbf{1.6 $\uparrow$})} \end{tabular}} \\ \cline{4-9} 
			\multicolumn{2}{|c|}{} &
			&
			EM &
			\multicolumn{1}{c|}{90.3} &
			\multicolumn{1}{c|}{76.7} &
			\multicolumn{1}{c|}{61.5} &
			\multicolumn{1}{c|}{44.0} &
			72.1 \\ \hline
		\end{tabular}%
	}
    \myvspace{-1em}
\end{table}

\begin{table}[t!]
\centering
\caption{\rev{Accuracy \vs SQL Complexity in BIRD-Dev.}}
\myvspace{-1em}
\label{tab:overall_bird}
\renewcommand\arraystretch{1.2} 
\resizebox{\columnwidth}{!}{%
\begin{tabular}{|cc|c|c|cccc|}
\hline
\multicolumn{2}{|c|}{\multirow{2}{*}{\textbf{Types}}} &
  \multirow{2}{*}{\textbf{Methods}} &
  \multirow{2}{*}{\textbf{Metrics}} &
  \multicolumn{4}{c|}{\textbf{BIRD-Dev}} \\ \cline{5-8} 
\multicolumn{2}{|c|}{} &
   &
   &
  \multicolumn{1}{c|}{\textbf{Simple}} &
  \multicolumn{1}{c|}{\textbf{Moderate}} &
  \multicolumn{1}{c|}{\textbf{Challenging}} &
  \textbf{All} \\ \hline \hline
\multicolumn{1}{|c|}{\multirow{7}{*}{\rotatebox[origin=c]{90}{\textbf{LLM-based}}}} &
  \multirow{3}{*}{\rotatebox[origin=c]{90}{  \begin{tabular}[c]{@{}c@{}} \rev{\textbf{Prompt-}} \\ \rev{\textbf{ing}} \end{tabular}  }} &
  C3SQL &
  EX &
  \multicolumn{1}{c|}{58.9} &
  \multicolumn{1}{c|}{38.5} &
  \multicolumn{1}{c|}{31.9} &
  50.2 \\ \cline{3-8} 
\multicolumn{1}{|c|}{} &
   &
  DAILSQL &
  EX &
  \multicolumn{1}{c|}{62.5} &
  \multicolumn{1}{c|}{43.2} &
  \multicolumn{1}{c|}{37.5} &
  54.3 \\ \cline{3-8} 
\multicolumn{1}{|c|}{} &
   &
  DAILSQL(SC) &
  EX &
  \multicolumn{1}{c|}{63.0} &
  \multicolumn{1}{c|}{45.6} &
  \multicolumn{1}{c|}{\textcolor{orange}{\textbf{43.1}}} &
  55.9 \\ \cline{2-8} 
\multicolumn{1}{|c|}{} &
  \multirow{7}{*}{\rotatebox[origin=c]{90}{\rev{\textbf{Fine-tuning}}}} &
  \rev{SFT CodeS-1B} &
  EX &
  \multicolumn{1}{c|}{58.7} &
  \multicolumn{1}{c|}{37.6} &
  \multicolumn{1}{c|}{36.8} &
  50.3 \\ \cline{3-8} 
\multicolumn{1}{|c|}{} &
   &
  \rev{SFT CodeS-3B} &
  EX &
  \multicolumn{1}{c|}{62.8} &
  \multicolumn{1}{c|}{44.3} &
  \multicolumn{1}{c|}{38.2} &
  54.9 \\ \cline{3-8} 
\multicolumn{1}{|c|}{} &
   &
  \rev{SFT CodeS-7B} &
  EX &
  \multicolumn{1}{c|}{64.6} &
  \multicolumn{1}{c|}{46.9} &
  \multicolumn{1}{c|}{40.3} &
  57.0 \\ \cline{3-8} 
\multicolumn{1}{|c|}{} &
   &
  \rev{SFT CodeS-15B} &
  EX &
  \multicolumn{1}{c|}{\textcolor{orange}{\textbf{65.8}}} &
  \multicolumn{1}{c|}{\textcolor{orange}{\textbf{48.8}}} &
  \multicolumn{1}{c|}{42.4} &
  \textcolor{orange}{\textbf{58.5}} \\ \cline{1-1} \cline{3-8} 
\multicolumn{1}{|c|}{\multirow{3}{*}{\rotatebox[origin=c]{90}{
\begin{tabular}[c]{@{}c@{}} \textbf{PLM-} \\ \textbf{based} \end{tabular}
}}} &
   &
  RESDSQL-Base &
  EX &
  \multicolumn{1}{c|}{42.3} &
  \multicolumn{1}{c|}{20.2} &
  \multicolumn{1}{c|}{16.0} &
  33.1 \\ \cline{3-8} 
\multicolumn{1}{|c|}{} &
   &
  RESDSQL-Large &
  EX &
  \multicolumn{1}{c|}{46.5} &
  \multicolumn{1}{c|}{27.7} &
  \multicolumn{1}{c|}{22.9} &
  38.6 \\ \cline{3-8} 
\multicolumn{1}{|c|}{} &
   &
  RESDSQL-3B &
  EX &
  \multicolumn{1}{c|}{53.5} &
  \multicolumn{1}{c|}{33.3} &
  \multicolumn{1}{c|}{16.7} &
  43.9 \\ \hline \hline
\multicolumn{2}{|c|}{\textbf{Hybird}} &
  \rev{\sys} &
  EX &
  \multicolumn{1}{c|}{\textcolor{green}{\begin{tabular}[c]{@{}c@{}}\textbf{66.9} \\ \textbf{(1.1$\uparrow$) }\end{tabular}}} &
  \multicolumn{1}{c|}{46.5} &
  \multicolumn{1}{c|}{\textcolor{green}{\begin{tabular}[c]{@{}c@{}}\textbf{43.8} \\ \textbf{(0.7$\uparrow$) }\end{tabular}}} &
  \textcolor{green}{\textbf{58.5}} \\ \hline
\end{tabular}%
}
\myvspace{-1em}
\end{table}

\begin{figure*}[t!]
	\centering
	\includegraphics[width=1.0\textwidth]{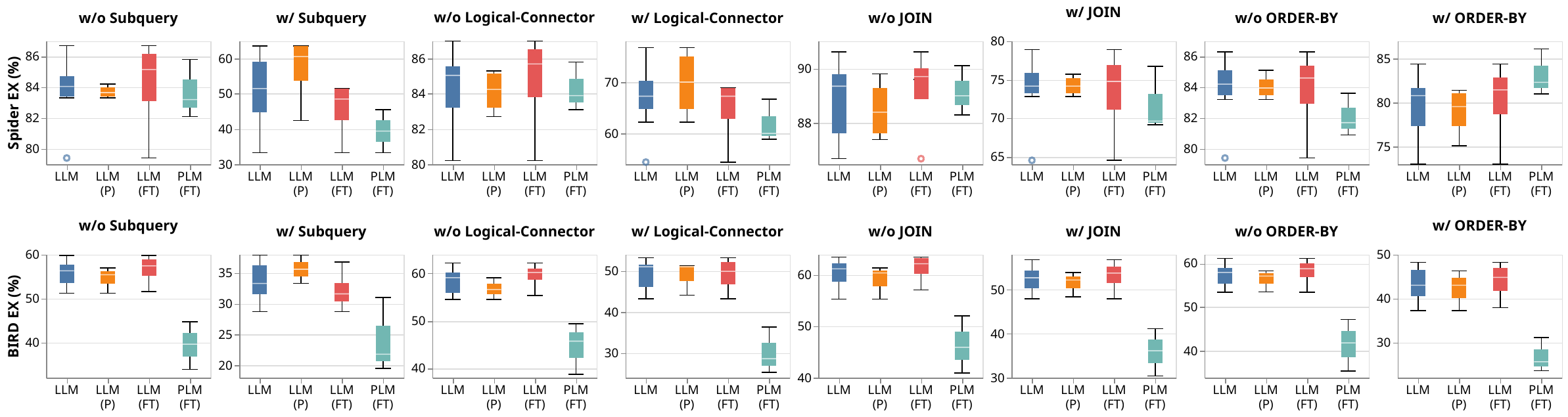}
    \myvspace{-2em}
	\caption{\rev{EX \vs SQL Characteristics. {\small (LLM (P): Prompt-based LLMs, LLM (FT): Fine-tuned LLMs, PLM (FT): Fine-tuned PLMs.)}}}
	\label{fig:ex_boxplot}
    \myvspace{-1em}
\end{figure*}

\begin{figure*}[t!]
    \begin{center}
        \begin{minipage}{0.49\textwidth}
            \centering
        	\includegraphics[width=0.97\linewidth]{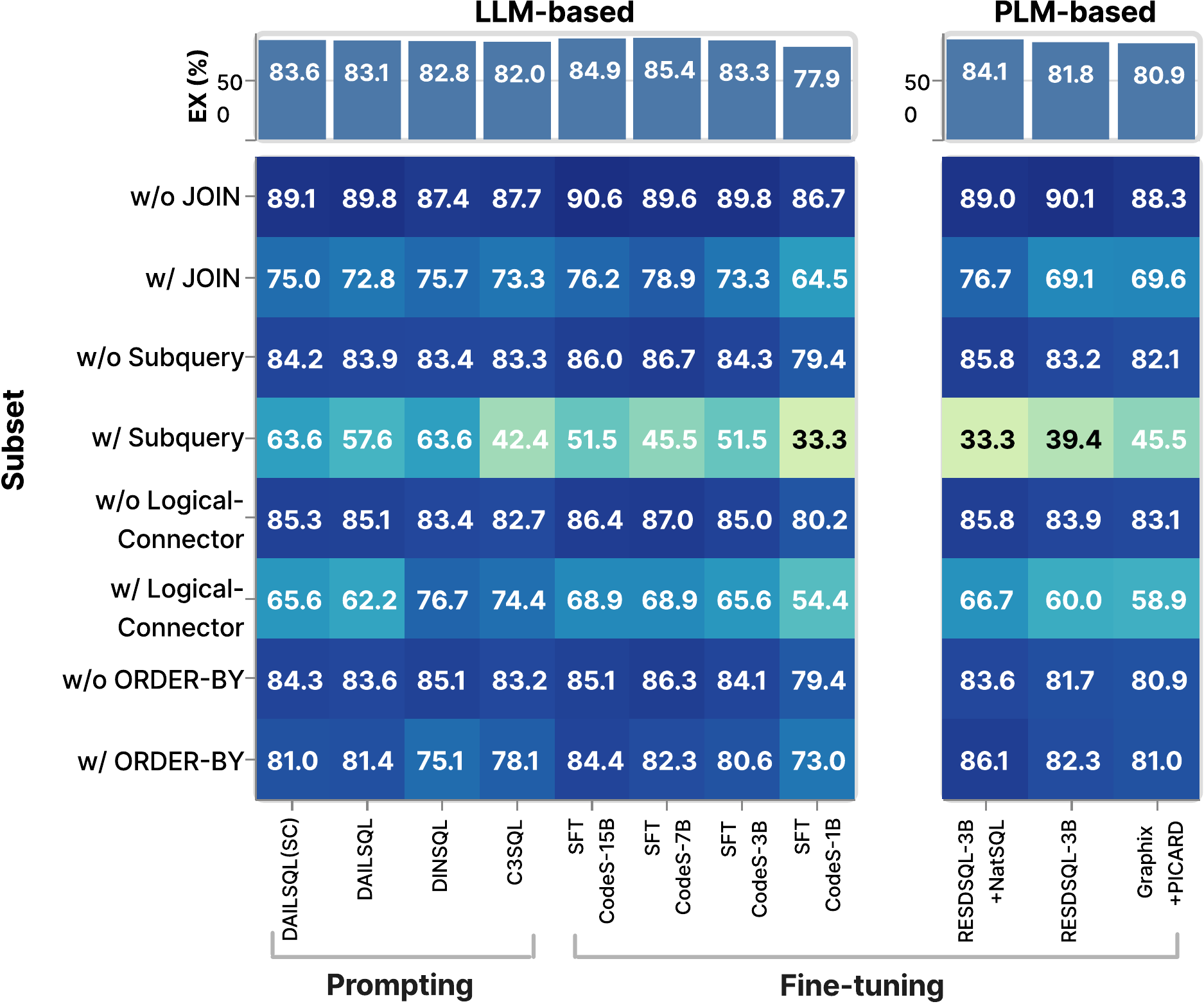}
            \myvspace{-1em}
            \caption{\rev{EX \vs SQL Characteristics on Spider.}}
            \label{fig:heatmap_spider}
        \end{minipage}
        \begin{minipage}{0.49\textwidth}
            \centering
        	\includegraphics[width=0.97\linewidth]{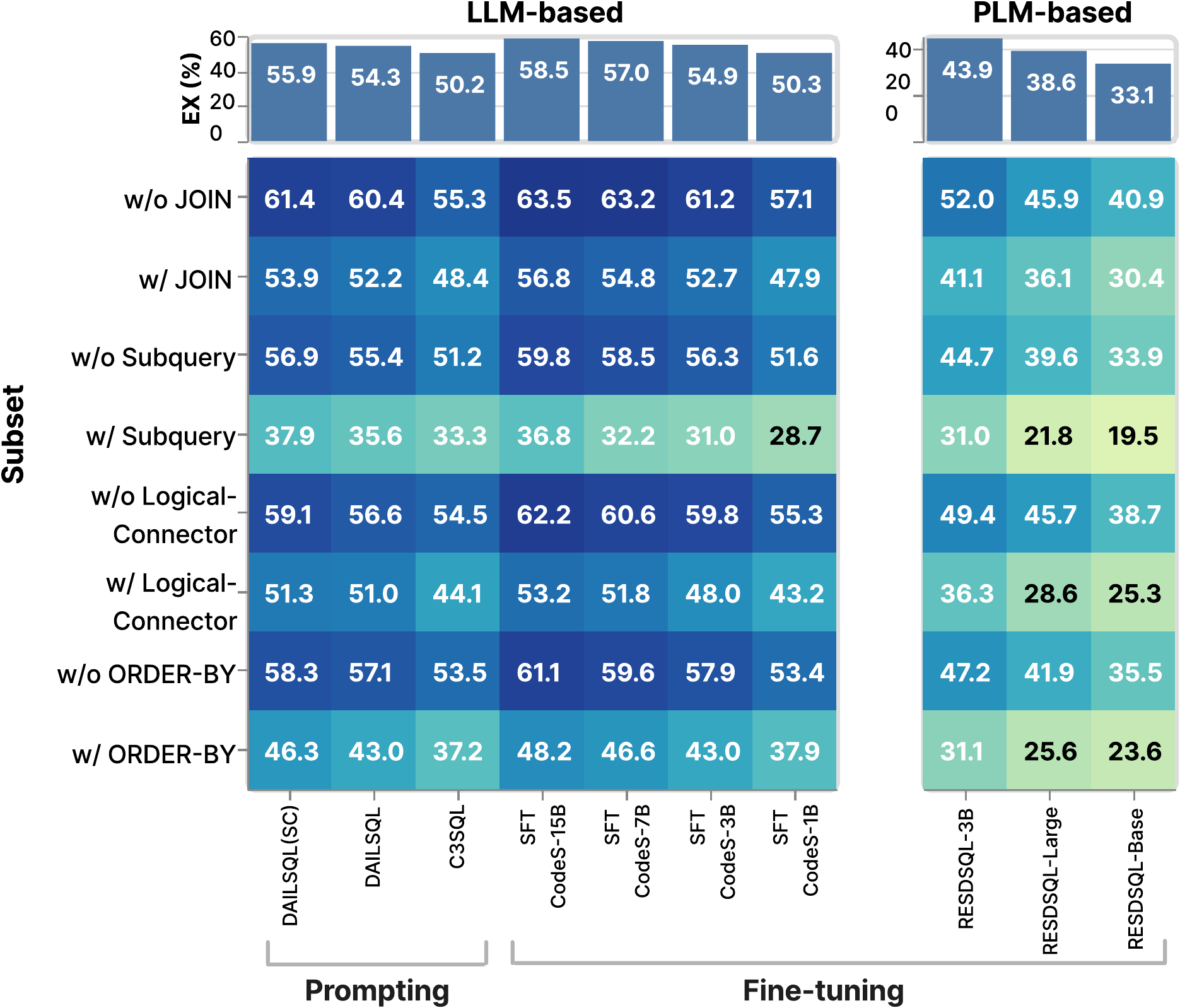}
            \myvspace{-1em}
            \caption{\rev{EX \vs SQL Characteristics on BIRD.}}
            \label{fig:heatmap_bird}
        \end{minipage}
    \end{center}
    \myvspace{-1.5em}
\end{figure*}

\etitle{Insights based on the \textcolor{orange}{\bf EX} metric.} 
\note{R1 W1}%
\note{R1 D1}%
\note{R3 W1}%
\note{R3 D1}%
\rev{
As shown in Table~\ref{tab:overall_spider} and Table~\ref{tab:overall_bird}, we find that the EX of the LLM-based method exceeded the PLM-based method in different difficulty subsets. Particularly, in Table~\ref{tab:overall_bird}, DAILSQL(SC) outperforms LLM-based SOTA method SFT CodeS-15B on the Challenging subset, which may benefit from GPT-4's stronger reasoning capabilities.
}

\etitle{Insights based on the \textcolor{blue}{\bf EM} metric.}   
\note{R1 W1}%
\note{R1 D1}%
\note{R3 W1}%
\note{R3 D1}%
\note{R4 W1}%
\note{R4 D1}%
\rev{In Table~\ref{tab:overall_spider}, we find that LLM-based methods after supervised fine-tuning generally have higher EM performance than prompt-based LLM methods. After fine-tuning, both the LLM- and PLM-based model's output aligns more closely with the specific dataset's data distribution, leading it to predict \sql structures similar to those in that dataset.}

\vspace{.5em}
\begin{finding}
Finding 1. \textit{
Fine-tuning is an essential strategy to improve performance. Specifically, LLM-based methods with fine-tuning achieve the best overall results on the EX metric, while PLM-based methods perform best on the EM metric overall.
}
\end{finding}

\stitle{Exp-2: Accuracy vs. SQL Characteristics.}
Real-world applications often require generating \sql queries involving advanced operations like subqueries, logical connectors, {\tt ORDER BY}, and multiple {\tt JOIN}s. Therefore, we will evaluate the capability of \nlsql models to accurately generate \sql queries with varying characteristics.

\note{R4 W3}%
\note{R4 D3}%
\rev{To this end, we classify \sql queries based on four criteria: 
	(1) the presence of subqueries,
	(2) the number of logical connectors, 
	(3) the use of {\tt ORDER BY}, and
	(4) the number of {\tt JOIN}s.  Note that our \testdb supports \sql query filtering based on individual \sql clauses, their combinations, or user-defined conditions. However, due to space constraints, we demonstrate only four representative aspects.
We run all methods on these four subsets of \sql queries and compute their EX metrics. }


We classify LLM-based methods into prompt-based and fine-tuning-based. Figure~\ref{fig:ex_boxplot} shows the EX performance distribution across subsets of the Spider and BIRD datasets. Figure~\ref{fig:heatmap_spider} and ~\ref{fig:heatmap_bird} provide detailed results. The bar chart presents the overall EX for each method, with methods on the x-axis and subsets on the y-axis.

%

\stab \textbf{Exp-2.1: \#-Subquery.}
%
%
\note{R1 W1}%
\note{R1 D1}%
\note{R3 W1}%
\note{R3 D1}%
\note{R4 W1}%
\note{R4 D1}%
\note{R4 W2}%
\note{R4 D2}%
\note{R4 W3}%
\note{R4 D3}%
\rev{As shown in Figure~\ref{fig:heatmap_spider} and Figure~\ref{fig:heatmap_bird}, all methods perform worst in cases with subqueries, indicating that reasoning through subqueries is a challenging task.
Figure~\ref{fig:ex_boxplot} shows that in scenarios without subqueries, the LLM-based methods slightly outperform the PLM-based methods on Spider and significantly outperform them on BIRD on average.
In scenarios with subqueries, the LLM-based methods excel on both datasets. 

This is because generating \sql with subqueries requires the model to first consider the subquery and then generate the entire \sql, demanding strong reasoning abilities.
We find that all LLM-based methods, especially those prompted by GPT-4, perform better in subquery, surpassing both fine-tuned LLM-based methods and PLM-based methods. This suggests that the model's inherent reasoning ability is crucial for processing \sql with subqueries.}

\vspace{.5em}
\begin{finding}
Finding 2. \textit{\rev{In scenarios involving subqueries, LLM-based methods outperform PLM-based methods overall, with methods using GPT-4 (\ie prompt-based LLM) showing particularly better performance. The inherent reasoning ability of these models is likely crucial for success in predicting the subqueries.}}
\end{finding}

\stab \textbf{Exp-2.2: \#-Logical Connector.}
Logical Connectors (\eg {\tt AND}, {\tt OR}) are used to link conditions, filter query results, and perform other operations, making it essential to understand the model's performance with respect to logical connectors. 



\note{R1 W1}%
\note{R1 D1}%
\note{R3 W1}%
\note{R3 D1}%
\note{R4 W3}%
\note{R4 D3}%

Without Logical Connectors (Figure~\ref{fig:ex_boxplot}), LLM-based methods do not outperform PLM-based methods on the Spider dataset. However, on the more complex BIRD dataset (Table~\ref{tab:dataset_compare}), LLM-based methods excel. When Logical Connectors are needed, LLM-based methods consistently outperform PLM-based methods on both datasets.

\vspace{.5em}
\begin{finding}
Finding 3. \textit{
    \rev{
        In scenarios where Logical Connectors are required, the LLM-based methods are better than the PLM-based methods.
    }
}
\end{finding}

\stab \textbf{Exp-2.3: \#-JOIN.}
In many usage scenarios, we need to generate \sql queries with {\tt JOIN}s across multiple tables. This challenges the model's ability to correctly understand complex database schemas.
%

\etitle{SQL without JOIN.} %
\note{R1 W1}%
\note{R1 D1}%
\note{R3 W1}%
\note{R3 D1}%
\note{R4 W3}%
\note{R4 D3}%
\rev{As shown in Figure~\ref{fig:ex_boxplot}, in scenarios without {\tt JOIN} operations, LLM-based and PLM-based methods show inconsistent performance on Spider and BIRD, with no clear winner. Figure~\ref{fig:heatmap_spider} and Figure~\ref{fig:heatmap_bird} provide similar insights.}

\etitle{SQL with JOIN.} \rev{However, for scenarios requiring {\tt JOIN} operations, LLM-based methods outperform PLM-based methods in both datasets. This could be due to the {\tt JOIN} operation's need for understanding complex database schemas, where LLMs typically excel due to their superior context-understanding capabilities.} 

\etitle{Impact of NatSQL.} \rev{In Figure~\ref{fig:heatmap_spider}, for \sql queries with {\tt JOIN}, DINSQL works best in prompt-based methods, while RESDSQL-3B+NatSQL is the best among PLM-based methods. Both utilize NatSQL~\cite{gan2021natural} as an intermediate representation, likely benefiting from its streamlined form that omits {\tt JOIN} keywords and reduces schema item prediction, thus easing \sql prediction in {\tt JOIN} scenarios.}

\vspace{.5em}
\begin{finding}
Finding 4. \textit{
    \rev{In scenarios involving {\tt JOIN} operations, LLM-based methods outperform PLM-based methods. Taking NatSQL as an intermediate representation reduces the complexity of predicting JOIN operations and potentially enhances the model performance.}
}
\end{finding}

\stab \textbf{Exp-2.4: \#-ORDER BY.}
%
\rev{As shown in Figure~\ref{fig:ex_boxplot}, we observed that without the {\tt ORDER BY} clause, LLM-based methods outperform PLM-based methods on both the Spider and BIRD datasets. However, with the {\tt ORDER BY} clause, LLM-based methods underperform compared to PLM-based methods on the Spider dataset, while they outperform PLM-based methods on the BIRD dataset. This difference might be because the BIRD dataset is more complex than the Spider dataset.
}
\note{R1 W1}%
\note{R1 D1}%
\note{R3 W1}%
\note{R3 D1}%

\vspace{.5em}
\begin{finding}
Finding 5. \textit{
    \rev{In scenarios with the {\tt ORDER BY} clause, the performance of PLM/LLM-based methods varies across different datasets. Generally, LLM-based methods demonstrate stronger generalization capability.}
}
\end{finding}

\begin{figure}[t!]
	\centering
	\includegraphics[width=0.9\columnwidth]{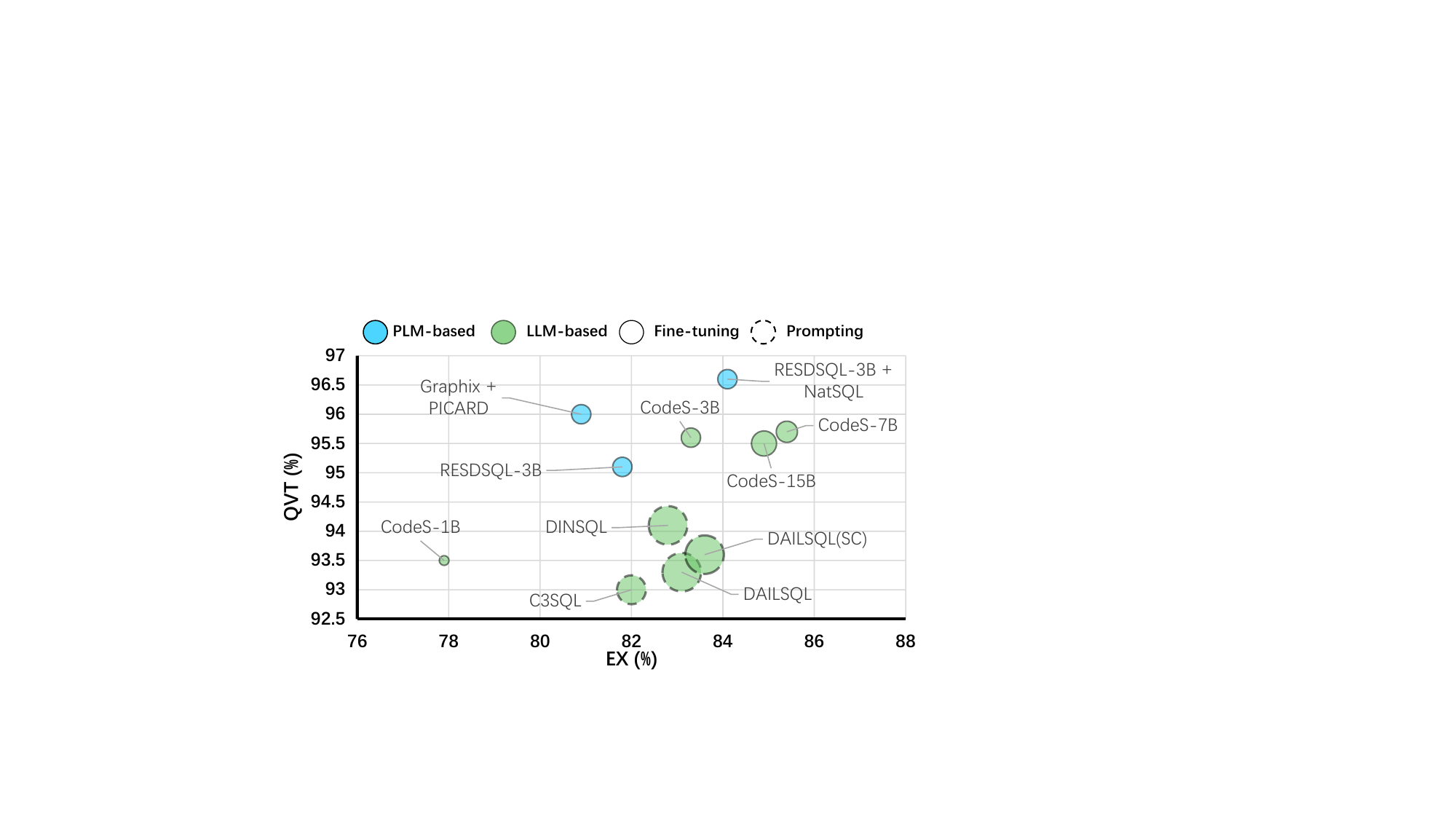}
    \myvspace{-1em}
	\caption{\rev{QVT \vs Execution Accuracy (EX).}}
	\label{fig:acc_sql_qua}
    \myvspace{-0.5em}
\end{figure}

\begin{figure}[t!]
    \begin{minipage}[t]{\columnwidth}
        \centering
        \includegraphics[width=\columnwidth]{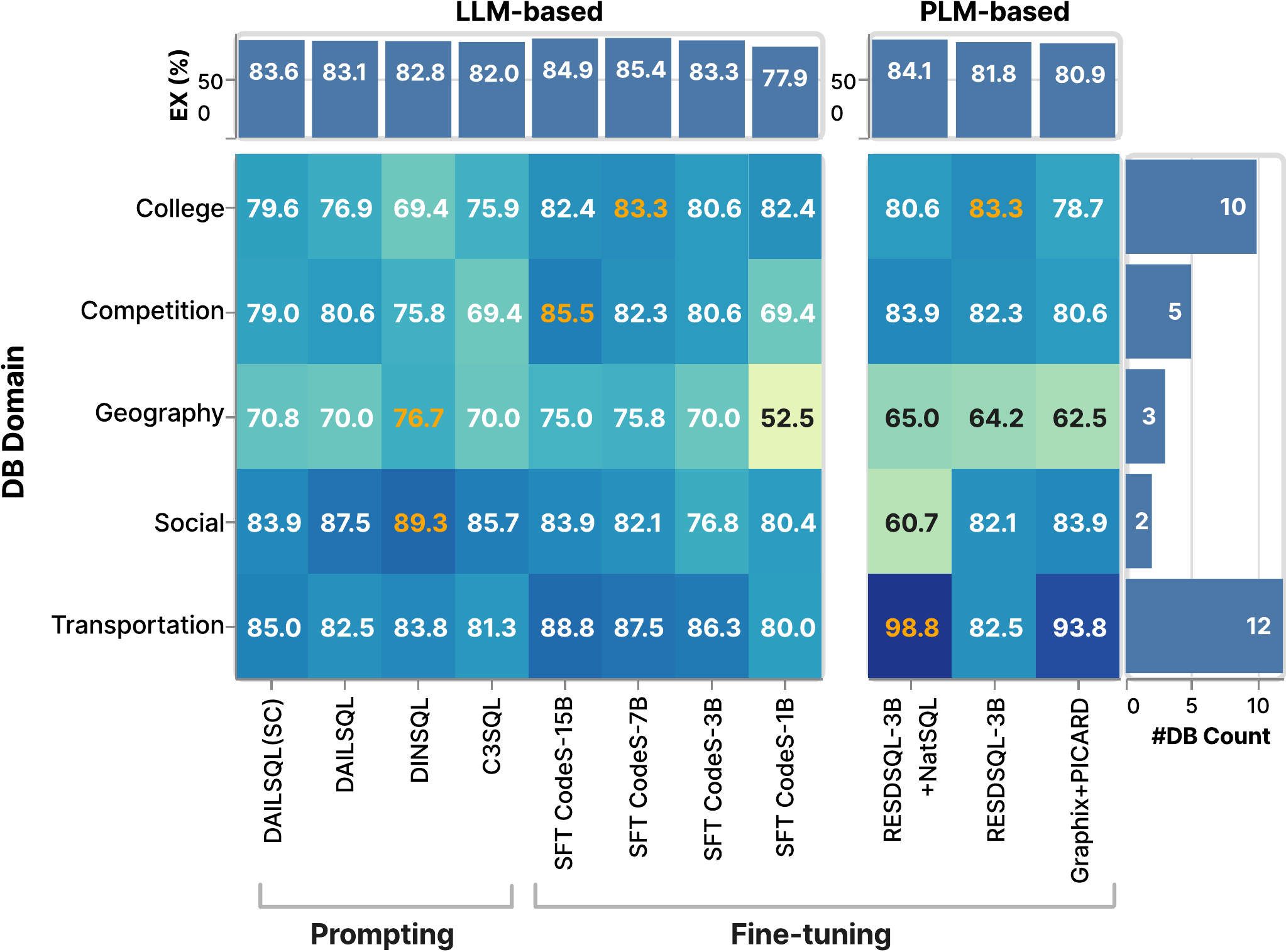}
        \subcaption*{\rev{(a) Detailed Results}}
        \label{fig:spider_dev_domain_a} 
    \end{minipage}
    \begin{minipage}[t]{\columnwidth}
        \centering
        \includegraphics[width=\columnwidth]{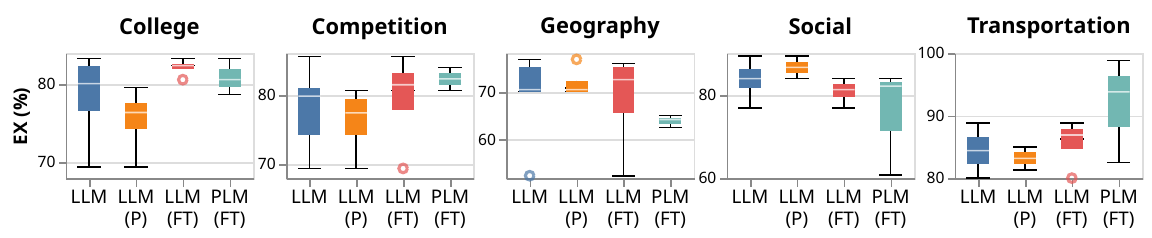}
        \subcaption*{\rev{(b) Overall Results}}
        \label{fig:spider_dev_domain_b}
    \end{minipage}
    \myvspace{-1em}
	\caption{\rev{EX \vs Different Domains on Spider.}}
	\label{fig:spider_dev_domain}
    \myvspace{-1em}
\end{figure}

\stitle{Exp-3: Query Variance Testing.}
\note{R1 D2}%
\note{R4 W3}%
\note{R4 D3}%
We evaluate the \nlsql system's adaptability to diverse natural language phrasings and structures, reflecting the variety expected in practical applications.
\rev{Note that there are seldom \sql queries with multiple corresponding \nlq queries in the BIRD dataset.  Thus, we build the QVT dataset using Spider Dev set, as it contains 469 \sqls corresponding to more than two different \nlq queries, aligning with QVT's purpose.
We compute the QVT scores based on the Equation~\eqref{eq:qvt}.}
\note{R1 W1}%
\note{R1 D1}%
\note{R3 W1}%
\note{R3 D1}%
\note{R4 W1}%
\note{R4 D1}%
\note{R4 W2}%
\note{R4 D2}%

\rev{As shown in Figure~\ref{fig:acc_sql_qua}, there is no clear winner between LLM-based methods and PLM-based methods in terms of QVT. However, Fine-tuned LLMs generally exhibit higher QVT than prompting LLMs. This improvement may result from the alignment of model input with specific data distributions after fine-tuning, reducing the impact of \nlq changes on performance. Notably, although the Graphix+PICARD method underperforms in overall EX compared to all prompt-based methods, it surpasses them in QVT.
}

\vspace{.5em}
\begin{finding}
Finding 6. \textit{
    \rev{
        There is no clear winner between LLM-based and PLM-based methods in QVT. Fine-tuning the model with task-specific datasets may help stabilize its performance against \nlq variations.
    }
}
\end{finding}

\stitle{Exp-4: Database Domain Adaption.}
\rev{In practical \nlsql applications, scenarios usually involve domain-specific databases, such as movies or sports, each with unique schema designs and terminologies. Assessing the detailed performance of methods across different domains is crucial for effective model application. We classified the 140 databases in the Spider training set and the 20 databases in the development set into 33 domains. All fine-tuning-based LLMs and PLMs are tuned using the training set.}
\note{R1 W1}%
\note{R1 D1}%
\note{R3 W1}%
\note{R3 D1}%
\note{R4 W1}%
\note{R4 D1}%
\note{R4 W2}%
\note{R4 D2}%
\rev{Figure~\ref{fig:spider_dev_domain}(a) shows the EX performance across diverse database domains in the Spider dataset. Figure~\ref{fig:spider_dev_domain}(b) shows the overall performance.}

\rev{As shown in Figure~\ref{fig:spider_dev_domain}(a), we discovered that different \nlsql methods exhibit varying biases towards different domains and there is no clear winner between LLM-based and PLM-based methods.}

\rev{
    However, in Figure~\ref{fig:spider_dev_domain}(b), we observe that fine-tuning-based methods outperform in domains with more training databases (College, Competition, Transportation). Conversely, in domains with fewer training databases, prompt-based methods excel. This suggests that in-domain training data during the fine-tuning process is crucial for enhancing model performance in specific domains.
}

\vspace{.5em}
\begin{finding}
Finding 7. \textit{
Different methods exhibit varying biases towards different domains, and there is no clear winner between LLM-based and PLM-based methods.
However, in-domain training data during fine-tuning is crucial for model performance in specific domains.
}
\end{finding}

\begin{figure}[t!]
	\centering
\includegraphics[width=1\columnwidth]{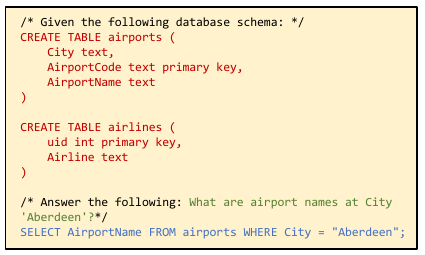}
    \myvspace{-2em}
	\caption{An Example of SQL-style Prompt.}
	\label{fig:prompt_example}
\end{figure}

\begin{figure}[t!]
	\centering
\includegraphics[width=\columnwidth]{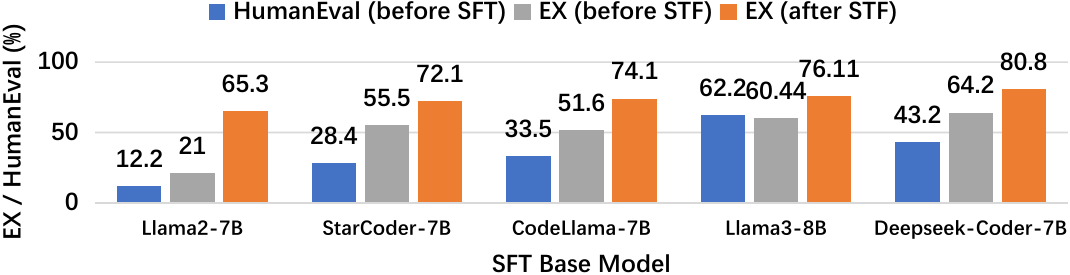}
    \myvspace{-2em}
	\caption{\rev{EX / HumanEval \vs SFT Base Models.}}
	\label{fig:sft_llm}
    \myvspace{-1em}
\end{figure}

\stitle{Exp-5: Supervised Fine-tuning of LLM-based Methods.}
\rev{We investigated Supervised Fine-tuning (SFT) of \textit{open-source} LLMs for the \nlsql task. DAILSQL~\cite{gao2023text} examines the impact of varying shot and prompt representation during SFT but does not address \textit{which open-source LLMs are best suited for SFT in the \nlsql task}. DAILSQL found that SQL-style prompts were beneficial, so we adopted a similar prompt approach in a zero-shot setting, as shown in Figure~\ref{fig:prompt_example}. Given that \nlsql is a code-related task, we selected five open-source LLMs with varying code abilities, evaluated using the HumanEval (Pass@1) metric~\cite{humaneval}. To ensure a fair comparison and account for hardware limitations, all chosen LLMs have similar parameters. The suffix in the model name, such as \textbf{7B}, indicates the model has 7 billion parameters.} 

\etitle{Settings.} \rev{We compare 5 fine-tuning-based LLMs introduced in Section~\ref{sec:exp_settings}. We use an instruction-tuning approach, \ie Alpaca~\cite{alpaca}.} We use the Adam optimizer with a learning rate of 1e-5 and no weight decay. The learning rate follows a cosine decay to zero by the end of training. We train with a global batch size of 16 for a single epoch to mitigate over-fitting risks. \rev{After SFT, LLMs are evaluated on the Spider Dev set using the EX metric.}

\etitle{Results.} \rev{As shown in Figure~\ref{fig:sft_llm}, after SFT, the performance (EX) improves but varies significantly across different base models. Importantly, a positive correlation is observed between these performance variations and the models' intrinsic coding abilities (HumanEval) before SFT. This suggests that \textit{selecting base LLMs with advanced coding capabilities is beneficial for adaptation in the \nlsql task.}}

\vspace{.5em}
\begin{finding}
Finding 8. \textit{After Supervised Fine-tuning (SFT) on open-source LLMs for the \nlsql task, we found a positive correlation between performance after SFT and the model's inherent coding ability prior to SFT. This indicates that base LLMs with advanced coding abilities are important for adapting to the \nlsql task.}
\end{finding}

\subsection{Experiments on Evaluating Efficiency}
\label{sec:exp_efficiency}

\stitle{Exp-6: Economy of LLM-based Methods.}
\rev{Prompt-based LLM methods} utilize commercial GPT API interfaces to accomplish the \nlsql task. As of June 2024, compared to GPT-3.5-turbo, the API interface of GPT-4 is 60 times more expensive for input tokens and 40 times more expensive for output tokens. In practical applications, our concern extends beyond the performance of \nlsql methods to include cost considerations. %
\note{R4 W3}%
\note{R4 D3}%
\rev{In this experiment, we compute several metrics for each prompt-based method based on the development set of Spider and BIRD}. \rev{These include the number of tokens and the cost (in dollars) per \nlsql task}. As shown in Table~\ref{tab:llm-economy}, we also calculate the ratio of EX to Average Cost, which indicates the cost-effectiveness of the \nlsql method to some extent.

\note{R1 W1}%
\note{R1 D1}%
\note{R3 W1}%
\note{R3 D1}%
\rev{Although C3SQL scores lowest in EX on both datasets, its EX to average cost ratio is the highest, benefiting from the lower cost of the GPT-3.5-turbo interface compared with GPT-4. Among methods using GPT-4, DINSQL is the least cost-effective, whereas DAILSQL emerges as the most cost-efficient. Although DAILSQL(SC) outperforms DAILSQL on both datasets, it introduces higher costs.}

\vspace{.5em}
\begin{finding}
Finding 9. \textit{Based on the ratio of Execution Accuracy (EX) to the Average Cost, we observe that \rev{prompt-based LLM methods} calling GPT-3.5-turbo offer higher cost-effectiveness. \rev{Although DAILSQL(SC) shows EX improvements over DAILSQL on Spider and BIRD datasets, it introduces higher costs reducing its cost-effectiveness.}}
\end{finding}

\begin{table}[t!]
	\centering
	\caption{\rev{Accuracy vs. LLM Economy on Spider/BIRD Dev Set.}}
	\label{tab:llm-economy}
    \myvspace{-1em}
	\resizebox{\columnwidth}{!}{%
		\begin{tabular}{|c|c|cc|cc|cc|cc|}
			\hline
			\multirow{2}{*}{\textbf{Methods}} &
			\multirow{2}{*}{\textbf{LLMs}} &
			\multicolumn{2}{c|}{\textbf{Avg. Tokens / Query}} &
			\multicolumn{2}{c|}{\textbf{Avg. Cost / Query}} &
			\multicolumn{2}{c|}{\textbf{EX(\%)}} &
			\multicolumn{2}{c|}{\textbf{EX / Avg. Cost}} \\ \cline{3-10} 
			&
			&
			\multicolumn{1}{c|}{\textbf{Spider}} &
			\textbf{BIRD} &
			\multicolumn{1}{c|}{\textbf{Spider}} &
			\textbf{BIRD} &
			\multicolumn{1}{c|}{\textbf{Spider}} &
			\textbf{BIRD} &
			\multicolumn{1}{c|}{\textbf{Spider}} &
			\textbf{BIRD} \\ \hline \hline
			C3SQL &
			GPT-3.5 &
			\multicolumn{1}{c|}{5702} &
			5890 &
			\multicolumn{1}{c|}{0.0103} &
			0.0104 &
			\multicolumn{1}{c|}{82.0} &
			50.2 &
			\multicolumn{1}{c|}{7961} &
			4825 \\ \hline
			DINSQL &
			GPT-4 &
			\multicolumn{1}{c|}{9571} &
			- &
			\multicolumn{1}{c|}{0.2988} &
			- &
			\multicolumn{1}{c|}{82.8} &
			- &
			\multicolumn{1}{c|}{277} &
			- \\ \hline
			DAILSQL &
			GPT-4 &
			\multicolumn{1}{c|}{930} &
			1559 &
			\multicolumn{1}{c|}{0.0288} &
			0.0486 &
			\multicolumn{1}{c|}{83.1} &
			54.3 &
			\multicolumn{1}{c|}{2885} &
			1117 \\ \hline
			DAILSQL(SC) &
			GPT-4 &
			\multicolumn{1}{c|}{1063} &
			1886 &
			\multicolumn{1}{c|}{0.0377} &
			0.0683 &
			\multicolumn{1}{c|}{83.6} &
			55.9 &
			\multicolumn{1}{c|}{2218} &
			819 \\ \hline \hline
			\sys &
			\textbf{GPT-4} &
			\multicolumn{1}{c|}{\textbf{942}} &
			\textbf{1412} &
			\multicolumn{1}{c|}{\textbf{0.0354}} &
			\textbf{0.0555} &
			\multicolumn{1}{c|}{\textbf{87.0}} &
			\textbf{58.5} &
			\multicolumn{1}{c|}{\textbf{2458}} &
			\textbf{1053} \\ \hline
		\end{tabular}%
	}
 \vspace{-1em}
\end{table}

\begin{table}[t!]
	\centering
	\caption{The Efficiency of PLM-based Methods.}
	\label{tab:plm_efficiency}
 \vspace{-1em}
	\resizebox{\columnwidth}{!}{%
		\begin{tabular}{|c|c|c|c|c|}
			\hline
			\textbf{Methods} &
			\textbf{Parameters} &
			\textbf{EX (\%)} &
			\textbf{\begin{tabular}[c]{@{}c@{}}Latency \\ Per Sample (sec)\end{tabular}} &
			\textbf{\begin{tabular}[c]{@{}c@{}}GPU Memory Used\\ (GiB)\end{tabular}} \\ \hline \hline
			RESDSQL-Base           & 220M & 77.9 & 1.10 & 3.87  \\ \hline
			RESDSQL-Base + NatSQL  & 220M & 80.2 & 1.01 & 3.59  \\ \hline
			RESDSQL-Large          & 770M & 80.1 & 1.71 & 7.55  \\ \hline
			RESDSQL-Large + NatSQL & 770M & 81.9 & 1.57 & 6.83  \\ \hline
			RESDSQL-3B             & 3B   & 81.8 & 1.91 & 24.66 \\ \hline
			RESDSQL-3B + NatSQL    & 3B   & 84.1 & 1.97 & 21.59 \\ \hline
		\end{tabular}%
	}
    \myvspace{-1em}
\end{table}

\begin{table}[t!]
	\centering
	\caption{\rev{The Valid Efficiency Score Results.}}
    \myvspace{-1em}
	\label{tab:sql-ves}
	\begin{subtable}{\columnwidth}
		\centering
		\caption{\rev{The Valid Efficiency Score in Spider-Dev.}}
		\label{tab:sql-ves-spider}
		\renewcommand\arraystretch{1.5} 
		\resizebox{\columnwidth}{!}{%
			\begin{tabular}{|cc|c|ccccc|}
				\hline
				\multicolumn{2}{|c|}{\multirow{2}{*}{\textbf{Types}}} &
				\multirow{2}{*}{\textbf{Methods}} &
				\multicolumn{5}{c|}{\textbf{Spider-Dev}} \\ \cline{4-8} 
				\multicolumn{2}{|c|}{} &
				&
				\multicolumn{1}{c|}{\textbf{Easy}} &
				\multicolumn{1}{c|}{\textbf{Medium}} &
				\multicolumn{1}{c|}{\textbf{Hard}} &
				\multicolumn{1}{c|}{\textbf{Extra}} &
				\textbf{All} \\ \hline \hline
				\multicolumn{1}{|c|}{\multirow{8}{*}{\rotatebox[origin=c]{90}{\textbf{LLM-based}}}} &
				\multirow{4}{*}{\rotatebox[origin=c]{90}{\textbf{\rev{Prompting}}}} &
				C3SQL &
				\multicolumn{1}{c|}{104.68} &
				\multicolumn{1}{c|}{96.04} &
				\multicolumn{1}{c|}{84.55} &
				\multicolumn{1}{c|}{69.63} &
				91.94 \\ \cline{3-8} 
				\multicolumn{1}{|c|}{} &
				&
				DINSQL &
				\multicolumn{1}{c|}{102.99} &
				\multicolumn{1}{c|}{97.49} &
				\multicolumn{1}{c|}{84.05} &
				\multicolumn{1}{c|}{67.81} &
				91.78 \\ \cline{3-8} 
				\multicolumn{1}{|c|}{} &
				&
				DAILSQL &
				\multicolumn{1}{c|}{102.73} &
				\multicolumn{1}{c|}{100.36} &
				\multicolumn{1}{c|}{86.15} &
				\multicolumn{1}{c|}{66.10} &
				93.04 \\ \cline{3-8} 
				\multicolumn{1}{|c|}{} &
				&
				DAILSQL(SC) &
				\multicolumn{1}{c|}{103.86} &
				\multicolumn{1}{c|}{102.73} &
				\multicolumn{1}{c|}{86.40} &
				\multicolumn{1}{c|}{71.59} &
				95.25 \\ \cline{2-8} 
				\multicolumn{1}{|c|}{} &
				\multirow{7}{*}{\rotatebox[origin=c]{90}{\textbf{\rev{Fine-tuning}}}} &
				\rev{SFT CodeS-1B} &
				\multicolumn{1}{c|}{103.23} &
				\multicolumn{1}{c|}{94.13} &
				\multicolumn{1}{c|}{80.37} &
				\multicolumn{1}{c|}{55.02} &
				87.72 \\ \cline{3-8} 
				\multicolumn{1}{|c|}{} &
				&
				\rev{SFT CodeS-3B} &
				\multicolumn{1}{c|}{106.17} &
				\multicolumn{1}{c|}{99.72} &
				\multicolumn{1}{c|}{80.80} &
				\multicolumn{1}{c|}{68.10} &
				93.01 \\ \cline{3-8} 
				\multicolumn{1}{|c|}{} &
				&
				\rev{SFT CodeS-7B} &
				\multicolumn{1}{c|}{108.77} &
				\multicolumn{1}{c|}{102.90} &
				\multicolumn{1}{c|}{84.05} &
				\multicolumn{1}{c|}{73.42} &
				\textcolor{orange}{\textbf{96.41}} \\ \cline{3-8} 
				\multicolumn{1}{|c|}{} &
				&
				\rev{SFT CodeS-15B} &
				\multicolumn{1}{c|}{107.91} &
				\multicolumn{1}{c|}{\textcolor{orange}{\textbf{103.02}}} &
				\multicolumn{1}{c|}{\textcolor{orange}{\textbf{87.10}}} &
				\multicolumn{1}{c|}{68.92} &
				96.04 \\ \cline{1-1} \cline{3-8} 
				\multicolumn{1}{|c|}{\multirow{3}{*}{\rotatebox[origin=c]{90}{\textbf{PLM-based}}}} &
				&
				RESDSQL-3B &
				\multicolumn{1}{c|}{106.22} &
				\multicolumn{1}{c|}{98.61} &
				\multicolumn{1}{c|}{83.06} &
				\multicolumn{1}{c|}{61.60} &
				91.88 \\ \cline{3-8} 
				\multicolumn{1}{|c|}{} &
				&
				RESDSQL-3B + NatSQL &
				\multicolumn{1}{c|}{106.91} &
				\multicolumn{1}{c|}{97.98} &
				\multicolumn{1}{c|}{86.78} &
				\multicolumn{1}{c|}{\textcolor{orange}{\textbf{73.83}}} &
				94.36 \\ \cline{3-8} 
				\multicolumn{1}{|c|}{} &
				&
				Graphix + PICARD &
				\multicolumn{1}{c|}{\textcolor{orange}{\textbf{108.92}}} &
				\multicolumn{1}{c|}{102.71} &
				\multicolumn{1}{c|}{83.64} &
				\multicolumn{1}{c|}{68.61} &
				95.51 \\ \hline \hline
				\multicolumn{2}{|c|}{\textbf{Hybird}} &
				\sys &
				\multicolumn{1}{c|}{107.54} &
				\multicolumn{1}{c|}{ \textcolor{green}{\begin{tabular}[c]{@{}c@{}}\textbf{104.32} \\ \textbf{(1.30$\uparrow$) }\end{tabular}} } &
				\multicolumn{1}{c|}{ \textcolor{green}{\begin{tabular}[c]{@{}c@{}}\textbf{96.98} \\ \textbf{(9.88$\uparrow$) }\end{tabular}} } &
				\multicolumn{1}{c|}{ \textcolor{green}{\begin{tabular}[c]{@{}c@{}}\textbf{75.18} \\ \textbf{(1.35$\uparrow$) }\end{tabular}} } &
				\textcolor{green}{\begin{tabular}[c]{@{}c@{}}\textbf{99.18} \\ \textbf{(2.77$\uparrow$) }\end{tabular}} \\ \hline
			\end{tabular}%
		}
	\end{subtable}
	
	\vspace{.5em}
	\begin{subtable}{\columnwidth}
		\centering
		\caption{\rev{The Valid Efficiency Score in BIRD-Dev.}}
		\label{tab:sql-ves-bird}
		\renewcommand\arraystretch{1.2} 
		\resizebox{\columnwidth}{!}{%
			\begin{tabular}{|cc|c|cccc|}
				\hline
				\multicolumn{2}{|c|}{\multirow{2}{*}{\textbf{Types}}} &
				\multirow{2}{*}{\textbf{Methods}} &
				\multicolumn{4}{c|}{\textbf{BIRD-Dev}} \\ \cline{4-7} 
				\multicolumn{2}{|c|}{} &
				&
				\multicolumn{1}{c|}{\textbf{Simple}} &
				\multicolumn{1}{c|}{\textbf{Moderate}} &
				\multicolumn{1}{c|}{\textbf{Challenging}} &
				\textbf{All} \\ \hline \hline
				\multicolumn{1}{|c|}{\multirow{7}{*}{\rotatebox[origin=c]{90}{\textbf{LLM-based}}}} &
				\multirow{3}{*}{\rotatebox[origin=c]{90}{  \begin{tabular}[c]{@{}c@{}} \textbf{Prompt-} \\ \textbf{ing} \end{tabular}  }} &
				C3SQL &
				\multicolumn{1}{c|}{59.82} &
				\multicolumn{1}{c|}{41.68} &
				\multicolumn{1}{c|}{31.93} &
				51.70 \\ \cline{3-7} 
				\multicolumn{1}{|c|}{} &
				&
				DAILSQL &
				\multicolumn{1}{c|}{65.04} &
				\multicolumn{1}{c|}{43.35} &
				\multicolumn{1}{c|}{39.33} &
				56.05 \\ \cline{3-7} 
				\multicolumn{1}{|c|}{} &
				&
				DAILSQL(SC) &
				\multicolumn{1}{c|}{66.54} &
				\multicolumn{1}{c|}{46.14} &
				\multicolumn{1}{c|}{45.18} &
				58.35 \\ \cline{2-7} 
				\multicolumn{1}{|c|}{} &
				\multirow{7}{*}{\rotatebox[origin=c]{90}{\textbf{Fine-tuning}}} &
				SFT CodeS-1B &
				\multicolumn{1}{c|}{61.11} &
				\multicolumn{1}{c|}{39.89} &
				\multicolumn{1}{c|}{37.38} &
				52.45 \\ \cline{3-7} 
				\multicolumn{1}{|c|}{} &
				&
				SFT CodeS-3B &
				\multicolumn{1}{c|}{64.96} &
				\multicolumn{1}{c|}{50.98} &
				\multicolumn{1}{c|}{38.99} &
				58.28 \\ \cline{3-7} 
				\multicolumn{1}{|c|}{} &
				&
				SFT CodeS-7B &
				\multicolumn{1}{c|}{66.88} &
				\multicolumn{1}{c|}{49.53} &
				\multicolumn{1}{c|}{\textcolor{orange}{\textbf{58.42}}} &
				60.83 \\ \cline{3-7} 
				\multicolumn{1}{|c|}{} &
				&
				SFT CodeS-15B &
				\multicolumn{1}{c|}{\textcolor{orange}{\textbf{67.87}}} &
				\multicolumn{1}{c|}{\textcolor{orange}{\textbf{51.69}}} &
				\multicolumn{1}{c|}{52.71} &
				\textcolor{orange}{\textbf{61.54}} \\ \cline{1-1} \cline{3-7} 
				\multicolumn{1}{|c|}{\multirow{3}{*}{\rotatebox[origin=c]{90}{\begin{tabular}[c]{@{}c@{}} \textbf{PLM-} \\ \textbf{based} \end{tabular}}}} &
				&
				RESDSQL-Base &
				\multicolumn{1}{c|}{42.75} &
				\multicolumn{1}{c|}{22.16} &
				\multicolumn{1}{c|}{16.54} &
				34.05 \\ \cline{3-7} 
				\multicolumn{1}{|c|}{} &
				&
				RESDSQL-Large &
				\multicolumn{1}{c|}{47.21} &
				\multicolumn{1}{c|}{30.00} &
				\multicolumn{1}{c|}{34.67} &
				40.81 \\ \cline{3-7} 
				\multicolumn{1}{|c|}{} &
				&
				RESDSQL-3B &
				\multicolumn{1}{c|}{53.35} &
				\multicolumn{1}{c|}{35.49} &
				\multicolumn{1}{c|}{28.84} &
				45.64 \\ \hline \hline
				\multicolumn{2}{|c|}{\textbf{Hybird}} &
				\sys &
				\multicolumn{1}{c|}{\textcolor{green}{\begin{tabular}[c]{@{}c@{}}\textbf{69.75} \\ \textbf{(1.88$\uparrow$) }\end{tabular}}} &
				\multicolumn{1}{c|}{50.55} &
				\multicolumn{1}{c|}{49.08} &
				\textcolor{green}{\begin{tabular}[c]{@{}c@{}}\textbf{61.99} \\ \textbf{(0.45$\uparrow$) }\end{tabular}} \\ \hline
			\end{tabular}%
		}
	\end{subtable}
    \myvspace{-1em}
\end{table}

\stitle{Exp-7: Efficiency of PLM-based Methods.}
In practical applications, considering both the performance and efficiency of \nlsql methods is essential, especially latency per sample. Different methods have varying GPU memory requirements, increasing with model size. Selecting the appropriate method based on available hardware and latency requirements is a common challenge.
We assess three metrics across six models: RESDSQL-Base/Large/3B and RESDSQL-Base/Large/3B + NatSQL, focusing on Execution Accuracy (EX), Latency Per Sample, and GPU Memory Used, using the Spider development set. Model efficiency is dataset-agnostic, so we omit the BIRD dataset due to space constraints.

Table~\ref{tab:plm_efficiency} shows that as model size increases, so do GPU memory and latency. However, RESDSQL-Base+NatSQL (220M parameters) and RESDSQL-Large (770M parameters) achieve similar EX (80.2\% and 80.1\%), with the former having lower latency and memory usage. Similarly, RESDSQL-Large+NatSQL and RESDSQL-3B have comparable EX but differ in latency and hardware needs. Therefore, model selection should consider latency and hardware resources.


\vspace{.5em}
\begin{finding}
Finding 10. \textit{For the same method, as model parameters increase in size, there is a corresponding rise in the latency and hardware resource requirements. \rev{Furthermore, models with similar performance can differ in latency and hardware resource requirements.}}
\end{finding}

\stitle{Exp-8: SQL Efficiency - Valid Efficiency Score.}
In practical scenarios, it's crucial not only to focus on the correctness of the \sql queries generated by models but also on their execution efficiency.
\rev{BIRD~\cite{bird} introduces the Valid Efficiency Score (VES) to evaluate the execution efficiency of correctly generated \sql queries. The VES score is determined by dividing the execution time of the ground truth \sql query by the execution time of the predicted \sql query.}
\note{R4 W3}%
\note{R4 D3}%
\rev{We evaluate different methods on the development set of Spider and BIRD using the VES metric to compare the execution efficiency of \sql generated by different methods.} 

Table~\ref{tab:sql-ves} reports experimental results. The highest VES score is highlighted in \textcolor{orange}{orange} in the table.
The methods with the highest VES on subsets of varying difficulty are inconsistent, with no clear advantage for LLM-based or PLM-based approaches. Generally, a method shows lower VES on harder subsets, likely due to increased complexity and longer execution time.

\vspace{.5em}
\begin{finding}
Finding 11. \rev{Based on the VES metric, there is no clear winner between LLM-based and PLM-based methods. For the same method, it tends to have lower VES on more difficult subsets.}
\end{finding}

\begin{figure}[t!]
	\centering	\includegraphics[width=1\columnwidth]{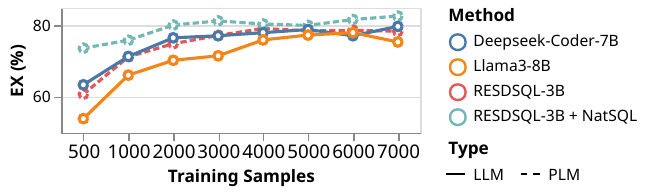}
    \myvspace{-2em}
	\caption{\rev{EX \vs \#-Training Samples on Spider.}}
	\label{fig:train_data}
    \myvspace{-1em}
\end{figure}

\stitle{\rev{Exp-9: The Impact of the \#-Training Samples.}}
In real-world scenarios, limited in-domain data often hinders performance. We experiment on the Spider training set, sampling subsets with size increments of 1000 and a smaller subset of 500. We train various methods on these subsets and evaluate their EX performance on the Spider development set. The training hyper-parameters for RESDSQL-3B and RESDSQL-3B+NatSQL match those in ~\cite{li2023resdsql}, while other methods follow \textbf{Exp-5.}

The results in Figure~\ref{fig:train_data} show that both PLM-based and fine-tuned LLM methods improve with more \nlsql training data and achieve acceptable performance with 4000 training samples. However, the EX performance gains decrease as dataset size increases.

\vspace{.5em}
\begin{finding}
Finding 12. \textit{Both PLM-based and LLM-based methods improve with more \nlsql training data. However, the EX performance gains decrease as dataset size increases. If data privacy is a concern or sufficient labeled data is available, fine-tuning LLM/PLM is promising.}
\end{finding}

\section{\mbox{Combining the Best of both Worlds}}
\label{sec:q3}

\subsection{A Design Space Exploration}

\begin{figure}[h!]
	\centering
\includegraphics[width=\columnwidth]{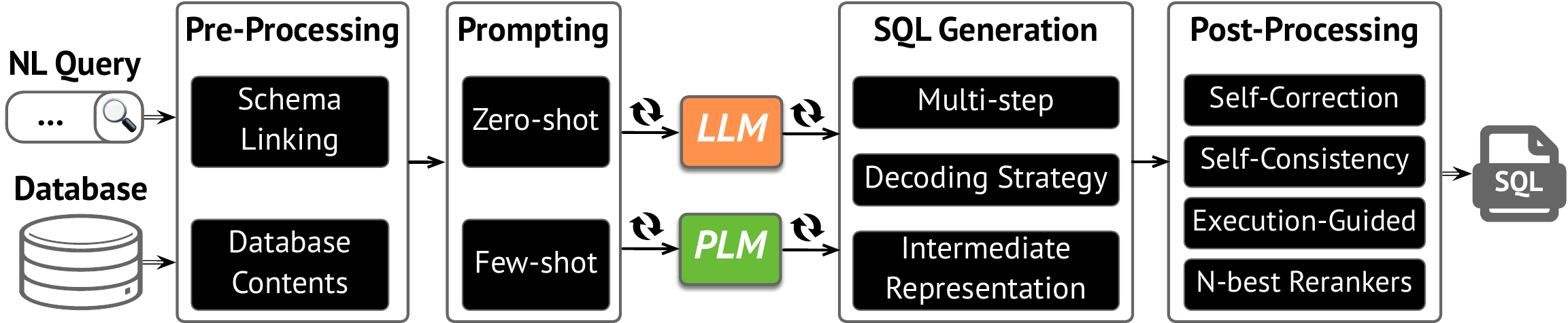}
    \myvspace{-1.5em}
	\caption{\rev{The Design Space of the NL2SQL Solution.}}
	\label{fig:nl2sql_overview}
    \myvspace{-.5em}
\end{figure}

\rev{We explore the design space of \nlsql solution powered by language models, as shown in Figure~\ref{fig:nl2sql_overview}.}

\stab (1) \textbf{Pre-Processing}: 
The Pre-Processing module comprises schema linking and DB contents. Schema linking maps \nlq query references to database schema elements (tables, columns), enhancing cross-domain generalizability and complex query generation~\cite{lei-etal-2020-examining}. This approach is adopted by leading LLM-based~\cite{dong2023c3, pourreza2023din} and PLM-based methods~\cite{li2023resdsql, li2023graphix}.
Additionally, the DB content module aligns query conditions with database content, often enriching column details via string matching~\cite{lin2020bridging}. As detailed in Table~\ref{tab:taxonomy}, while prevalent in PLM-based methods, it's seldom utilized in LLM-based approaches.

\stab (2) \textbf{Prompting Strategy}: 
Prompting strategies fall into zero-shot, where no \nlsql examples are included in the model input, and few-shot, which incorporates such examples, denoted as ``3-shot'', ``5-shot'', etc., depending on the number of examples used. Table~\ref{tab:taxonomy} shows PLM-based methods typically use zero-shot, while LLM-based methods vary: C3SQL~\cite{dong2023c3} employs zero-shot, whereas DAILSQL\cite{gao2023text} and DINSQL\cite{pourreza2023din} use few-shot.
\note{R3 W2}
\note{R3 D2}
\rev{The few-shot examples for DINSQL are manually designed and fixed, whereas those for DAILSQL are dynamically selected based on the similarity between the target question and training set examples.} 


\stab (3) \textbf{SQL Generation Strategy}:
Language models employ various strategies for generating \sql, categorized into three key aspects: Multi-Step, Decoding Strategy, and Intermediate Representation.

(a) \textit{Multi-Step}
akin to the Chain-of-Thought (COT) process, involves generating \sql queries in stages, particularly useful for complex queries~\cite{cot}. %
\rev{
We include two types of multi-step strategies: ``\sql skeleton - \sql'' from PLM-based RESDSQL~\cite{li2023resdsql} and ``Subquery - \sql'' from DINSQL~\cite{pourreza2023din}.
}
\note{R3 W2}
\note{R3 D2}


(b) \textit{Decoding Strategy} 
 involves the model's decoding process to ensure output validity. The PLM-based PICARD~\cite{scholak2021picard} enforces \sql syntax compliance in its output, whereas LLM-based methods, utilizing OpenAI's API, lack this decoding-level restriction.
 

(c) \textit{Intermediate Representation} strategy explores if a model employs an intermediary query form to address the \nlq to \sql translation's \textit{mismatch problem}, where \sql's design for relational databases doesn't align with natural language semantics. Various solutions like ~\cite{mismatch} and NatSQL~\cite{gan2021natural} have been introduced. LLM-based DINSQL~\cite{pourreza2023din} and several PLM-based methods~\cite{li2023resdsql, rai2023improving, gan2021natural} adopt NatSQL. \rev{In our setting, we only include NatSQL for simplification.}

\stab (4) \textbf{Post-Processing}: we consider the following strategies.

\rev{
(a) \textit{Self-Correction} is proposed in DINSQL~\cite{pourreza2023din}. It provides the generated \sql to the model for fixing potential issues.
}
\note{R3 W2}%
\note{R3 D2}%

\rev{
(b) \textit{Self-Consistency}
involves executing various valid \sql queries for a single \nlq query, using a voting mechanism on the outcomes to determine the most consistent \sql as the final choice.  It is used in C3SQL~\cite{dong2023c3} and DAILSQL~\cite{gao2023text}.
}

(c) \textit{Execution-Guided SQL Selector}
is a module~\cite{li2023resdsql} that sequentially executes model-generated SQL queries, identifying the first error-free execution as the valid SQL.

(d) \textit{N-best Rerankers} rank multiple candidate \sql queries to select the most probable one as the final query~\cite{zeng2023n}.

\subsection{NL2SQL360 Facilitates Better NL2SQL}
After categorizing different LLM- and PLM-based methods into a unified modular framework, 
a question arises: \textit{Can automatically exploring and combining different modules from various \nlsql solutions achieve stronger performance?}

To address this question, inspired by the Neural Architecture Search (NAS) algorithm~\cite{nas}, we designed an \nlsql Automated Architecture Search algorithm (\textbf{NL2SQL360-AAS}) within our NL2SQL360 framework. The key intuition behind \textbf{NL2SQL360-AAS} is to automatically explore the predefined design space (\ie predefined search space) of the \nlsql solution. Therefore, we adopt the standard Genetic Algorithm (GA)~\cite{ga} to achieve this goal.


\rev{There are some key concepts relevant to our \textbf{\testdb-AAS}.

 \underline{(1) \textit{Search Space.}} This includes various modules used in \nlsql, such as \sql generation strategies, post-processing modules, and prompting techniques, as shown in Figure~\ref{fig:nl2sql_overview}.

 \underline{(2) \textit{Individual}}. A valid combination of different modules in the search space, \ie a valid \nlsql solution, is an \textit{individual}.

\underline{(3) \textit{Target Metrics.}} We aim to select better individuals based on target metrics like Execution Accuracy (EX), Exact-Match Accuracy (EM), and Valid Efficiency Score (VES) on a specified dataset.}


\begin{figure}[t!]
	\centering	\includegraphics[width=\columnwidth]{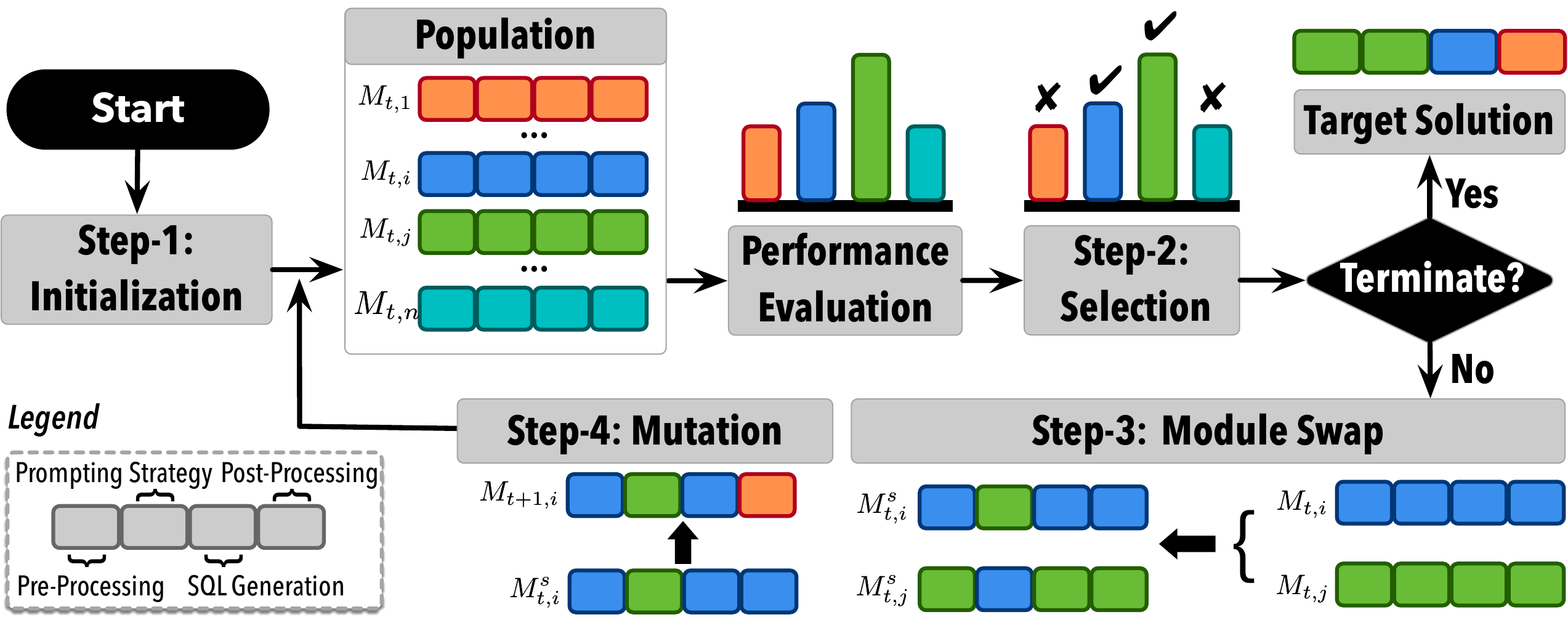}
    \myvspace{-1.5em}
	\caption{NL2SQL360-AAS Algorithm Overview.}
	\label{fig:nl2sql360_aas_overview}
    \myvspace{-1em}
\end{figure}

\stitle{NL2SQL360-AAS: An Overview.}
\rev{As shown in Figure~\ref{fig:nl2sql360_aas_overview}, our algorithm consists of four main steps, \ie Initialization, Individual Selection, \nlsql Module Swap, and \nlsql Module Mutation. Note that, $M_{t,i}$ is the $i$-th individual in the $t$-th generation population.}
\note{R3 W2}%
\note{R3 D2}%

\etitle{Step-1: Initialization.}
\rev{We initialize $N$ randomized \nlsql system individuals $\{M_{0, n}\}^N_{n=1}$ that are composed of random modules as shown in Figure~\ref{fig:nl2sql_overview}, resulting in $0$-th generation population.}

\etitle{Step-2: Individual Selection.} %
\rev{
We evaluate the population of $N$ individuals on the specified dataset (\eg Spider) using the target metric (\eg EX). We implement a {\em Russian Roulette Process}~\cite{nas} for individual selection. This process probabilistically samples individuals based on their target metric distribution, ensuring that individuals with higher target metrics have a greater likelihood of being selected, while consistently eliminating the lowest performers.
}



\etitle{Step-3: \nlsql Module Swap.} %
\rev{Two individuals selected from the previous step will exchange their \nlsql modules based on the module swap probability $p_s$. For example, if individual $M_{t,i}$ has a Self-Correction module and individual $M_{t,j}$ has a Self-Consistency module in the Post-Processing Layer before the swap, these two modules could be exchanged. In Figure~\ref{fig:nl2sql360_aas_overview}, the individuals after the module swap are labeled as $M^{s}_{t,i}$ and $M^{s}_{t,j}$, respectively.}

\etitle{Step-4: \nlsql Module Mutation.} %
\rev{
Next, the individual $M^{s}_{t,i}$ (similarly $M^{s}_{t,j}$) will undergo module mutation in each layer (\eg Pre-Processing Layer) based on the module mutation probability $p_m$. For example, if the Pre-Processing Layer of $M^{s}_{t,i}$ does not use the DB Contents module, a successful mutation will result in the inclusion of this module. After mutation, the individual is labeled as $M_{t+1,i}$ and will enter the next generation population. 

We repeat \textit{Step} 2--4 until we obtain the complete next generation population $\{M_{t+1, n}\}^N_{n=1}$, marking one entire population iteration.
}

\subsection{SuperSQL: A Superior NL2SQL Solution Suggested by NL2SQL360-AAS}
\label{sec:exp_nlsql360_aas}

\begin{figure}[t!]
	\centering
\includegraphics[width=.95\columnwidth]{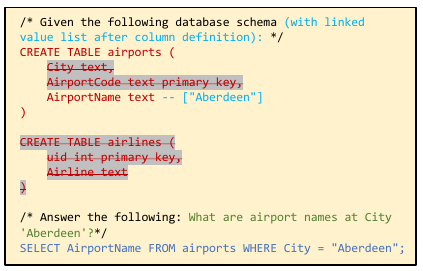}
    \myvspace{-1.em}
	\caption{Clear Schema with DB Content Prompt.}
\label{fig:clear_prompt_with_content}
    \myvspace{-1.5em}
\end{figure}

In this section, we validate the effectiveness of the \textbf{\testdb-AAS} algorithm. The search space is defined as shown in Figure~\ref{fig:nl2sql_overview}. %
For simplification, we utilize only the Few-shot module from DAILSQL in the Prompting Strategy.
Since we use GPT as our backbone, we cannot adjust its decoding strategy and thus we can only use Greedy Search in the Decoding phase. To save costs, we use GPT-3.5-turbo in the algorithm.
The population size $N$ is set to 10, the number of population generations $T$ is 20, and the probabilities for \nlsql module swap and \nlsql module mutation, $p_s$ and $p_m$, are set to 0.5 and 0.2, respectively. We run the algorithm on the Spider development set with the EX as the target metric.

\stitle{\sys.} We select the individual with the highest Execution Accuracy from the final generation produced by the NL2SQL360-AAS as our final searched \nlsql solution, namely \sys. 

The composition of \sys is as follows: %
(1) Pre-Processing: Schema Linking from RESDSQL and DB Contents from BRIDGE v2; %
(2) Prompting: DAILSQL's Few-shot module selects in-context examples by similarity; %
(3) SQL Generation: OpenAI's Greedy-decoding strategy, without Multi-step or NatSQL; %
(4) Post-Processing: Self-Consistency from DAILSQL. %
We explored the organization of the prompt for this composition, as illustrated in Figure~\ref{fig:clear_prompt_with_content}. Under this combination, the RESDSQL schema linking module is used to filter out irrelevant schema items. Furthermore, it incorporates the DB content module from the BRIDGE v2 method. 
The relevant content is added as comments next to the columns in the prompt, enhancing column information. We then replace the backbone model with GPT-4 for improved performance.

\stitle{The Effectiveness of \sys.} %
We evaluate \sys on the Spider development set, achieving 87.0\% in EX and outperforming other methods (Table~\ref{tab:overall_spider}). \sys excels in Medium, Hard, and Extra subsets, proving its effectiveness. On the BIRD development set, \sys also shows competitive performance (Table~\ref{tab:overall_bird}).

We also evaluate \sys on the Spider and BIRD test sets, achieving 87.0\% EX on Spider (2nd place) and 62.66\% EX on BIRD (9th place).
Note that \sys surpasses all baselines within its design space. Specifically, \sys outperforms the strongest \textit{baseline}—DAILSQL(SC)—by 5.25\% on the BIRD test set. This improvement is primarily due to our \textbf{NL2SQL360-AAS}, which effectively searches for superior module combinations based on different \textit{baselines} in the design space. 
Adding more powerful \textit{baselines} is expected to further improve \sys through \textbf{NL2SQL360-AAS}.

%



\stitle{The Efficiency of \sys.} 
\rev{
We calculate the VES metric to evaluate \sql efficiency on the development set of Spider and BIRD. According to Table~\ref{tab:sql-ves}, \sys attains overall VES scores of 99.18 and 61.99, respectively, outperforming other methods.
}

\stitle{The Economy of \sys.} Furthermore, we consider the economy of our method, and the results are shown in \rev{Table~\ref{tab:llm-economy}}. Compared to other GPT-4 based methods, our method uses fewer tokens and lower costs, while achieving better performance in EX metrics.

\section{Research Opportunities}
\label{sec:opp}

We discuss research opportunities based on experimental findings.

\stitle{Make NL2SQL Methods Trustworthy.}
Current methods may generate incorrect \sql results, which can be attributed to: 1) ambiguous and underspecified \nlq queries, 2) ambiguous database schemas and dirty contents, and 3) inadequate capabilities in schema linking.

\etitle{Handling ambiguous and underspecified \nlq queries.} 
We can explore the following strategies to alleviate these issues.
{\em (i) Query Rewriter} aims to automatically refine given \nlq queries and ensure their clarity. 
{\em (ii) Query Auto-completion} helps formulate user queries by suggesting candidate tokens that are well-aligned with the database.

\etitle{Interpreting NL2SQL Solution.}
 {\em (i) NL2SQL Debugger} detects incorrect \sql queries, enables users to step through the \sql generation process, and helps identify errors or mismatches.
 {\em (ii) SQL and Query Results Interpretation} method helps users assess if the generated \sql and query results meet their requirements.

\stitle{Develop Cost-effective NL2SQL Methods.}
LLM-based \nlsql methods are promising but costly in terms of token consumption, impacting both costs and inference times. Exploring ways to enhance accuracy while minimizing token use is crucial. Specifically, the potential benefits of modularized \nlsql solutions and multi-agent frameworks are becoming clear. Incorporating LLMs with these methods has the potential to optimize both accuracy and efficiency, particularly for complex queries, while conserving tokens.

\stitle{Adaptive Training Data Generation.}
The effectiveness of \nlsql methods depends greatly on the quality and coverage of training data. These methods often struggle with adapting to unseen databases.
A promising research direction is dynamically generating (\nlq, \sql) pairs using model evaluation feedback. This approach tackles domain adaptation and ensures diverse, high-quality training data by leveraging insights from \nlsql performance.

\section{Conclusion}
\label{sec:conclusion}

We proposed a multi-angle testbed, named \testdb, for evaluating \nlsql methods from different perspectives, such as the ability to handle various characteristics of \sql and database domains, in a fine-grained manner.
\rev{We utilized our \testdb to evaluate 13 LLM-based and 7 PLM-based \nlsql methods on 2 widely-used benchmarks, varying 15 settings and deriving a set of new findings.}
Furthermore, we employed our \testdb to analyze the design space for \nlsql solutions and automatically search for one of the best solutions, named \sys, tailored to user-specific needs.
Our new \sys, which interleaves LLM-based and PLM-based modules, achieves \textbf{87\%} and \textbf{62.66\%} execution accuracy on the Spider and BIRD test sets, respectively.

\begin{acks}
This paper was supported by Guangdong Basic and Applied Basic Research Foundation (2023A1515110545), National Key R\&D Program of China (2023YFB4503600), NSF of China (61925205, 62232009, 62102215),  Zhongguancun Lab, Huawei, TAL education, Beijing National Research Center for Information Science and Technology (BNRist), CCF-Huawei Populus Grove Fund (CCF-HuaweiDB202306), and the Fundamental Research Funds for the Central Universities.
\end{acks}

\newpage

\clearpage
\balance
\bibliographystyle{ACM-Reference-Format}
\bibliography{main}


\begin{thebibliography}{62}


\ifx \showCODEN    \undefined \def \showCODEN     #1{\unskip}     \fi
\ifx \showDOI      \undefined \def \showDOI       #1{#1}\fi
\ifx \showISBNx    \undefined \def \showISBNx     #1{\unskip}     \fi
\ifx \showISBNxiii \undefined \def \showISBNxiii  #1{\unskip}     \fi
\ifx \showISSN     \undefined \def \showISSN      #1{\unskip}     \fi
\ifx \showLCCN     \undefined \def \showLCCN      #1{\unskip}     \fi
\ifx \shownote     \undefined \def \shownote      #1{#1}          \fi
\ifx \showarticletitle \undefined \def \showarticletitle #1{#1}   \fi
\ifx \showURL      \undefined \def \showURL       {\relax}        \fi
\providecommand\bibfield[2]{#2}
\providecommand\bibinfo[2]{#2}
\providecommand\natexlab[1]{#1}
\providecommand\showeprint[2][]{arXiv:#2}

\bibitem[\protect\citeauthoryear{??}{tab}{2024}]%
        {tableau}
 \bibinfo{year}{2024}\natexlab{}.
\newblock \bibinfo{title}{{TABLEAU SOFTWARE, LLC, A SALESFORCE COMPANY}}.
\newblock \bibinfo{howpublished}{\url{https://www.tableau.com/}}.
\newblock
\newblock
\shownote{Accessed: 2024-2-22.}


\bibitem[\protect\citeauthoryear{AI@Meta}{AI@Meta}{2024}]%
        {llama3modelcard}
\bibfield{author}{\bibinfo{person}{AI@Meta}.} \bibinfo{year}{2024}\natexlab{}.
\newblock \showarticletitle{Llama 3 Model Card}.
\newblock  (\bibinfo{year}{2024}).
\newblock
\urldef\tempurl%
\url{https://github.com/meta-llama/llama3/blob/main/MODEL_CARD.md}
\showURL{%
\tempurl}


\bibitem[\protect\citeauthoryear{Alam, Qamar, Dixit, and Benaida}{Alam et~al\mbox{.}}{2020}]%
        {ga}
\bibfield{author}{\bibinfo{person}{Tanweer Alam}, \bibinfo{person}{Shamimul Qamar}, \bibinfo{person}{Amit Dixit}, {and} \bibinfo{person}{Mohamed Benaida}.} \bibinfo{year}{2020}\natexlab{}.
\newblock \showarticletitle{Genetic Algorithm: Reviews, Implementations, and Applications}.
\newblock \bibinfo{journal}{\emph{CoRR}}  \bibinfo{volume}{abs/2007.12673} (\bibinfo{year}{2020}).
\newblock
\showeprint[arXiv]{2007.12673}
\urldef\tempurl%
\url{https://arxiv.org/abs/2007.12673}
\showURL{%
\tempurl}


\bibitem[\protect\citeauthoryear{Chang, Wang, Dong, Pan, Zhu, Li, Lan, Zhang, Jiang, Lilien, Ash, Wang, Wang, Castelli, Ng, and Xiang}{Chang et~al\mbox{.}}{2023}]%
        {DBLP:conf/iclr/Chang0DPZLLZJLA23}
\bibfield{author}{\bibinfo{person}{Shuaichen Chang}, \bibinfo{person}{Jun Wang}, \bibinfo{person}{Mingwen Dong}, \bibinfo{person}{Lin Pan}, \bibinfo{person}{Henghui Zhu}, \bibinfo{person}{Alexander~Hanbo Li}, \bibinfo{person}{Wuwei Lan}, \bibinfo{person}{Sheng Zhang}, \bibinfo{person}{Jiarong Jiang}, \bibinfo{person}{Joseph Lilien}, \bibinfo{person}{Steve Ash}, \bibinfo{person}{William~Yang Wang}, \bibinfo{person}{Zhiguo Wang}, \bibinfo{person}{Vittorio Castelli}, \bibinfo{person}{Patrick Ng}, {and} \bibinfo{person}{Bing Xiang}.} \bibinfo{year}{2023}\natexlab{}.
\newblock \showarticletitle{Dr.Spider: {A} Diagnostic Evaluation Benchmark towards Text-to-SQL Robustness}. In \bibinfo{booktitle}{\emph{{ICLR}}}. \bibinfo{publisher}{OpenReview.net}.
\newblock


\bibitem[\protect\citeauthoryear{Chen, Tworek, Jun, Yuan, de~Oliveira~Pinto, Kaplan, Edwards, Burda, Joseph, Brockman, Ray, Puri, Krueger, Petrov, Khlaaf, Sastry, Mishkin, Chan, Gray, Ryder, Pavlov, Power, Kaiser, Bavarian, Winter, Tillet, Such, Cummings, Plappert, Chantzis, Barnes, Herbert{-}Voss, Guss, Nichol, Paino, Tezak, Tang, Babuschkin, Balaji, Jain, Saunders, Hesse, Carr, Leike, Achiam, Misra, Morikawa, Radford, Knight, Brundage, Murati, Mayer, Welinder, McGrew, Amodei, McCandlish, Sutskever, and Zaremba}{Chen et~al\mbox{.}}{2021}]%
        {humaneval}
\bibfield{author}{\bibinfo{person}{Mark Chen}, \bibinfo{person}{Jerry Tworek}, \bibinfo{person}{Heewoo Jun}, \bibinfo{person}{Qiming Yuan}, \bibinfo{person}{Henrique~Pond{\'{e}} de Oliveira~Pinto}, \bibinfo{person}{Jared Kaplan}, \bibinfo{person}{Harrison Edwards}, \bibinfo{person}{Yuri Burda}, \bibinfo{person}{Nicholas Joseph}, \bibinfo{person}{Greg Brockman}, \bibinfo{person}{Alex Ray}, \bibinfo{person}{Raul Puri}, \bibinfo{person}{Gretchen Krueger}, \bibinfo{person}{Michael Petrov}, \bibinfo{person}{Heidy Khlaaf}, \bibinfo{person}{Girish Sastry}, \bibinfo{person}{Pamela Mishkin}, \bibinfo{person}{Brooke Chan}, \bibinfo{person}{Scott Gray}, \bibinfo{person}{Nick Ryder}, \bibinfo{person}{Mikhail Pavlov}, \bibinfo{person}{Alethea Power}, \bibinfo{person}{Lukasz Kaiser}, \bibinfo{person}{Mohammad Bavarian}, \bibinfo{person}{Clemens Winter}, \bibinfo{person}{Philippe Tillet}, \bibinfo{person}{Felipe~Petroski Such}, \bibinfo{person}{Dave Cummings}, \bibinfo{person}{Matthias Plappert}, \bibinfo{person}{Fotios
  Chantzis}, \bibinfo{person}{Elizabeth Barnes}, \bibinfo{person}{Ariel Herbert{-}Voss}, \bibinfo{person}{William~Hebgen Guss}, \bibinfo{person}{Alex Nichol}, \bibinfo{person}{Alex Paino}, \bibinfo{person}{Nikolas Tezak}, \bibinfo{person}{Jie Tang}, \bibinfo{person}{Igor Babuschkin}, \bibinfo{person}{Suchir Balaji}, \bibinfo{person}{Shantanu Jain}, \bibinfo{person}{William Saunders}, \bibinfo{person}{Christopher Hesse}, \bibinfo{person}{Andrew~N. Carr}, \bibinfo{person}{Jan Leike}, \bibinfo{person}{Joshua Achiam}, \bibinfo{person}{Vedant Misra}, \bibinfo{person}{Evan Morikawa}, \bibinfo{person}{Alec Radford}, \bibinfo{person}{Matthew Knight}, \bibinfo{person}{Miles Brundage}, \bibinfo{person}{Mira Murati}, \bibinfo{person}{Katie Mayer}, \bibinfo{person}{Peter Welinder}, \bibinfo{person}{Bob McGrew}, \bibinfo{person}{Dario Amodei}, \bibinfo{person}{Sam McCandlish}, \bibinfo{person}{Ilya Sutskever}, {and} \bibinfo{person}{Wojciech Zaremba}.} \bibinfo{year}{2021}\natexlab{}.
\newblock \showarticletitle{Evaluating Large Language Models Trained on Code}.
\newblock \bibinfo{journal}{\emph{CoRR}}  \bibinfo{volume}{abs/2107.03374} (\bibinfo{year}{2021}).
\newblock
\showeprint[arXiv]{2107.03374}
\urldef\tempurl%
\url{https://arxiv.org/abs/2107.03374}
\showURL{%
\tempurl}


\bibitem[\protect\citeauthoryear{Chen, Chen, White, Mooney, Payani, Srinivasa, Su, and Sun}{Chen et~al\mbox{.}}{2023}]%
        {DBLP:conf/acl/ChenCWMPSS023}
\bibfield{author}{\bibinfo{person}{Ziru Chen}, \bibinfo{person}{Shijie Chen}, \bibinfo{person}{Michael White}, \bibinfo{person}{Raymond~J. Mooney}, \bibinfo{person}{Ali Payani}, \bibinfo{person}{Jayanth Srinivasa}, \bibinfo{person}{Yu Su}, {and} \bibinfo{person}{Huan Sun}.} \bibinfo{year}{2023}\natexlab{}.
\newblock \showarticletitle{Text-to-SQL Error Correction with Language Models of Code}. In \bibinfo{booktitle}{\emph{{ACL} {(2)}}}. \bibinfo{pages}{1359--1372}.
\newblock


\bibitem[\protect\citeauthoryear{Deng, Awadallah, Meek, Polozov, Sun, and Richardson}{Deng et~al\mbox{.}}{2021}]%
        {spider-realistic}
\bibfield{author}{\bibinfo{person}{Xiang Deng}, \bibinfo{person}{Ahmed~Hassan Awadallah}, \bibinfo{person}{Christopher Meek}, \bibinfo{person}{Oleksandr Polozov}, \bibinfo{person}{Huan Sun}, {and} \bibinfo{person}{Matthew Richardson}.} \bibinfo{year}{2021}\natexlab{}.
\newblock \showarticletitle{Structure-Grounded Pretraining for Text-to-SQL}. In \bibinfo{booktitle}{\emph{Proceedings of the 2021 Conference of the North American Chapter of the Association for Computational Linguistics: Human Language Technologies, {NAACL-HLT} 2021, Online, June 6-11, 2021}}, \bibfield{editor}{\bibinfo{person}{Kristina Toutanova}, \bibinfo{person}{Anna Rumshisky}, \bibinfo{person}{Luke Zettlemoyer}, \bibinfo{person}{Dilek Hakkani{-}T{\"{u}}r}, \bibinfo{person}{Iz~Beltagy}, \bibinfo{person}{Steven Bethard}, \bibinfo{person}{Ryan Cotterell}, \bibinfo{person}{Tanmoy Chakraborty}, {and} \bibinfo{person}{Yichao Zhou}} (Eds.). \bibinfo{publisher}{Association for Computational Linguistics}, \bibinfo{pages}{1337--1350}.
\newblock
\urldef\tempurl%
\url{https://doi.org/10.18653/V1/2021.NAACL-MAIN.105}
\showDOI{\tempurl}


\bibitem[\protect\citeauthoryear{Devlin, Chang, Lee, and Toutanova}{Devlin et~al\mbox{.}}{2019}]%
        {devlin2018bert}
\bibfield{author}{\bibinfo{person}{Jacob Devlin}, \bibinfo{person}{Ming{-}Wei Chang}, \bibinfo{person}{Kenton Lee}, {and} \bibinfo{person}{Kristina Toutanova}.} \bibinfo{year}{2019}\natexlab{}.
\newblock \showarticletitle{{BERT:} Pre-training of Deep Bidirectional Transformers for Language Understanding}. In \bibinfo{booktitle}{\emph{Proceedings of the 2019 Conference of the North American Chapter of the Association for Computational Linguistics: Human Language Technologies, {NAACL-HLT} 2019, Minneapolis, MN, USA, June 2-7, 2019, Volume 1 (Long and Short Papers)}}, \bibfield{editor}{\bibinfo{person}{Jill Burstein}, \bibinfo{person}{Christy Doran}, {and} \bibinfo{person}{Thamar Solorio}} (Eds.). \bibinfo{publisher}{Association for Computational Linguistics}, \bibinfo{pages}{4171--4186}.
\newblock
\urldef\tempurl%
\url{https://doi.org/10.18653/V1/N19-1423}
\showDOI{\tempurl}


\bibitem[\protect\citeauthoryear{Dong, Zhang, Ge, Mao, Gao, Chen, Lin, and Lou}{Dong et~al\mbox{.}}{2023}]%
        {dong2023c3}
\bibfield{author}{\bibinfo{person}{Xuemei Dong}, \bibinfo{person}{Chao Zhang}, \bibinfo{person}{Yuhang Ge}, \bibinfo{person}{Yuren Mao}, \bibinfo{person}{Yunjun Gao}, \bibinfo{person}{Lu Chen}, \bibinfo{person}{Jinshu Lin}, {and} \bibinfo{person}{Dongfang Lou}.} \bibinfo{year}{2023}\natexlab{}.
\newblock \showarticletitle{{C3:} Zero-shot Text-to-SQL with ChatGPT}.
\newblock \bibinfo{journal}{\emph{CoRR}}  \bibinfo{volume}{abs/2307.07306} (\bibinfo{year}{2023}).
\newblock
\urldef\tempurl%
\url{https://doi.org/10.48550/ARXIV.2307.07306}
\showDOI{\tempurl}
\showeprint[arXiv]{2307.07306}


\bibitem[\protect\citeauthoryear{Gan, Chen, Xie, Purver, Woodward, Drake, and Zhang}{Gan et~al\mbox{.}}{2021}]%
        {gan2021natural}
\bibfield{author}{\bibinfo{person}{Yujian Gan}, \bibinfo{person}{Xinyun Chen}, \bibinfo{person}{Jinxia Xie}, \bibinfo{person}{Matthew Purver}, \bibinfo{person}{John~R. Woodward}, \bibinfo{person}{John~H. Drake}, {and} \bibinfo{person}{Qiaofu Zhang}.} \bibinfo{year}{2021}\natexlab{}.
\newblock \showarticletitle{Natural {SQL:} Making {SQL} Easier to Infer from Natural Language Specifications}. In \bibinfo{booktitle}{\emph{Findings of the Association for Computational Linguistics: {EMNLP} 2021, Virtual Event / Punta Cana, Dominican Republic, 16-20 November, 2021}}, \bibfield{editor}{\bibinfo{person}{Marie{-}Francine Moens}, \bibinfo{person}{Xuanjing Huang}, \bibinfo{person}{Lucia Specia}, {and} \bibinfo{person}{Scott~Wen{-}tau Yih}} (Eds.). \bibinfo{publisher}{Association for Computational Linguistics}, \bibinfo{pages}{2030--2042}.
\newblock
\urldef\tempurl%
\url{https://doi.org/10.18653/V1/2021.FINDINGS-EMNLP.174}
\showDOI{\tempurl}


\bibitem[\protect\citeauthoryear{Gao, Wang, Li, Sun, Qian, Ding, and Zhou}{Gao et~al\mbox{.}}{2024}]%
        {gao2023text}
\bibfield{author}{\bibinfo{person}{Dawei Gao}, \bibinfo{person}{Haibin Wang}, \bibinfo{person}{Yaliang Li}, \bibinfo{person}{Xiuyu Sun}, \bibinfo{person}{Yichen Qian}, \bibinfo{person}{Bolin Ding}, {and} \bibinfo{person}{Jingren Zhou}.} \bibinfo{year}{2024}\natexlab{}.
\newblock \showarticletitle{Text-to-SQL Empowered by Large Language Models: {A} Benchmark Evaluation}.
\newblock \bibinfo{journal}{\emph{Proc. {VLDB} Endow.}} \bibinfo{volume}{17}, \bibinfo{number}{5} (\bibinfo{year}{2024}), \bibinfo{pages}{1132--1145}.
\newblock
\urldef\tempurl%
\url{https://www.vldb.org/pvldb/vol17/p1132-gao.pdf}
\showURL{%
\tempurl}


\bibitem[\protect\citeauthoryear{Gkini, Belmpas, Koutrika, and Ioannidis}{Gkini et~al\mbox{.}}{2021}]%
        {DBLP:conf/sigmod/GkiniBKI21}
\bibfield{author}{\bibinfo{person}{Orest Gkini}, \bibinfo{person}{Theofilos Belmpas}, \bibinfo{person}{Georgia Koutrika}, {and} \bibinfo{person}{Yannis~E. Ioannidis}.} \bibinfo{year}{2021}\natexlab{}.
\newblock \showarticletitle{An In-Depth Benchmarking of Text-to-SQL Systems}. In \bibinfo{booktitle}{\emph{{SIGMOD} Conference}}. \bibinfo{publisher}{{ACM}}, \bibinfo{pages}{632--644}.
\newblock


\bibitem[\protect\citeauthoryear{Gu, Fan, Tang, Cao, Jia, Madden, and Du}{Gu et~al\mbox{.}}{2023}]%
        {DBLP:journals/pacmmod/GuF00JM023}
\bibfield{author}{\bibinfo{person}{Zihui Gu}, \bibinfo{person}{Ju Fan}, \bibinfo{person}{Nan Tang}, \bibinfo{person}{Lei Cao}, \bibinfo{person}{Bowen Jia}, \bibinfo{person}{Sam Madden}, {and} \bibinfo{person}{Xiaoyong Du}.} \bibinfo{year}{2023}\natexlab{}.
\newblock \showarticletitle{Few-shot Text-to-SQL Translation using Structure and Content Prompt Learning}.
\newblock \bibinfo{journal}{\emph{Proc. {ACM} Manag. Data}} \bibinfo{volume}{1}, \bibinfo{number}{2} (\bibinfo{year}{2023}), \bibinfo{pages}{147:1--147:28}.
\newblock


\bibitem[\protect\citeauthoryear{Guo, Zhu, Yang, Xie, Dong, Zhang, Chen, Bi, Wu, Li, Luo, Xiong, and Liang}{Guo et~al\mbox{.}}{2024}]%
        {DBLP:journals/corr/abs-2401-14196}
\bibfield{author}{\bibinfo{person}{Daya Guo}, \bibinfo{person}{Qihao Zhu}, \bibinfo{person}{Dejian Yang}, \bibinfo{person}{Zhenda Xie}, \bibinfo{person}{Kai Dong}, \bibinfo{person}{Wentao Zhang}, \bibinfo{person}{Guanting Chen}, \bibinfo{person}{Xiao Bi}, \bibinfo{person}{Y. Wu}, \bibinfo{person}{Y.~K. Li}, \bibinfo{person}{Fuli Luo}, \bibinfo{person}{Yingfei Xiong}, {and} \bibinfo{person}{Wenfeng Liang}.} \bibinfo{year}{2024}\natexlab{}.
\newblock \showarticletitle{DeepSeek-Coder: When the Large Language Model Meets Programming - The Rise of Code Intelligence}.
\newblock \bibinfo{journal}{\emph{CoRR}}  \bibinfo{volume}{abs/2401.14196} (\bibinfo{year}{2024}).
\newblock
\urldef\tempurl%
\url{https://doi.org/10.48550/ARXIV.2401.14196}
\showDOI{\tempurl}
\showeprint[arXiv]{2401.14196}


\bibitem[\protect\citeauthoryear{Guo, Zhan, Gao, Xiao, Lou, Liu, and Zhang}{Guo et~al\mbox{.}}{2019}]%
        {mismatch}
\bibfield{author}{\bibinfo{person}{Jiaqi Guo}, \bibinfo{person}{Zecheng Zhan}, \bibinfo{person}{Yan Gao}, \bibinfo{person}{Yan Xiao}, \bibinfo{person}{Jian{-}Guang Lou}, \bibinfo{person}{Ting Liu}, {and} \bibinfo{person}{Dongmei Zhang}.} \bibinfo{year}{2019}\natexlab{}.
\newblock \showarticletitle{Towards Complex Text-to-SQL in Cross-Domain Database with Intermediate Representation}.
\newblock \bibinfo{journal}{\emph{CoRR}}  \bibinfo{volume}{abs/1905.08205} (\bibinfo{year}{2019}).
\newblock
\showeprint[arXiv]{1905.08205}
\urldef\tempurl%
\url{http://arxiv.org/abs/1905.08205}
\showURL{%
\tempurl}


\bibitem[\protect\citeauthoryear{Hu, Zhao, Jiang, Lan, Zhu, Chauhan, Li, Pan, Wang, Hang, Zhang, Guo, Dong, Lilien, Ng, Wang, Castelli, and Xiang}{Hu et~al\mbox{.}}{2023}]%
        {zhao2022importance}
\bibfield{author}{\bibinfo{person}{Yiqun Hu}, \bibinfo{person}{Yiyun Zhao}, \bibinfo{person}{Jiarong Jiang}, \bibinfo{person}{Wuwei Lan}, \bibinfo{person}{Henghui Zhu}, \bibinfo{person}{Anuj Chauhan}, \bibinfo{person}{Alexander~Hanbo Li}, \bibinfo{person}{Lin Pan}, \bibinfo{person}{Jun Wang}, \bibinfo{person}{Chung{-}Wei Hang}, \bibinfo{person}{Sheng Zhang}, \bibinfo{person}{Jiang Guo}, \bibinfo{person}{Mingwen Dong}, \bibinfo{person}{Joseph Lilien}, \bibinfo{person}{Patrick Ng}, \bibinfo{person}{Zhiguo Wang}, \bibinfo{person}{Vittorio Castelli}, {and} \bibinfo{person}{Bing Xiang}.} \bibinfo{year}{2023}\natexlab{}.
\newblock \showarticletitle{Importance of Synthesizing High-quality Data for Text-to-SQL Parsing}. In \bibinfo{booktitle}{\emph{Findings of the Association for Computational Linguistics: {ACL} 2023, Toronto, Canada, July 9-14, 2023}}, \bibfield{editor}{\bibinfo{person}{Anna Rogers}, \bibinfo{person}{Jordan~L. Boyd{-}Graber}, {and} \bibinfo{person}{Naoaki Okazaki}} (Eds.). \bibinfo{publisher}{Association for Computational Linguistics}, \bibinfo{pages}{1327--1343}.
\newblock
\urldef\tempurl%
\url{https://doi.org/10.18653/V1/2023.FINDINGS-ACL.86}
\showDOI{\tempurl}


\bibitem[\protect\citeauthoryear{Katsogiannis{-}Meimarakis and Koutrika}{Katsogiannis{-}Meimarakis and Koutrika}{2021}]%
        {DBLP:conf/sigmod/Katsogiannis-Meimarakis21}
\bibfield{author}{\bibinfo{person}{George Katsogiannis{-}Meimarakis} {and} \bibinfo{person}{Georgia Koutrika}.} \bibinfo{year}{2021}\natexlab{}.
\newblock \showarticletitle{A Deep Dive into Deep Learning Approaches for Text-to-SQL Systems}. In \bibinfo{booktitle}{\emph{{SIGMOD} Conference}}. \bibinfo{publisher}{{ACM}}, \bibinfo{pages}{2846--2851}.
\newblock


\bibitem[\protect\citeauthoryear{Katsogiannis{-}Meimarakis and Koutrika}{Katsogiannis{-}Meimarakis and Koutrika}{2023}]%
        {DBLP:journals/vldb/KatsogiannisMeimarakisK23}
\bibfield{author}{\bibinfo{person}{George Katsogiannis{-}Meimarakis} {and} \bibinfo{person}{Georgia Koutrika}.} \bibinfo{year}{2023}\natexlab{}.
\newblock \showarticletitle{A survey on deep learning approaches for text-to-SQL}.
\newblock \bibinfo{journal}{\emph{{VLDB} J.}} \bibinfo{volume}{32}, \bibinfo{number}{4} (\bibinfo{year}{2023}), \bibinfo{pages}{905--936}.
\newblock


\bibitem[\protect\citeauthoryear{Lee, Polozov, and Richardson}{Lee et~al\mbox{.}}{2021}]%
        {DBLP:conf/acl/LeePR20}
\bibfield{author}{\bibinfo{person}{Chia{-}Hsuan Lee}, \bibinfo{person}{Oleksandr Polozov}, {and} \bibinfo{person}{Matthew Richardson}.} \bibinfo{year}{2021}\natexlab{}.
\newblock \showarticletitle{KaggleDBQA: Realistic Evaluation of Text-to-SQL Parsers}. In \bibinfo{booktitle}{\emph{Proceedings of the 59th Annual Meeting of the Association for Computational Linguistics and the 11th International Joint Conference on Natural Language Processing, {ACL/IJCNLP} 2021, (Volume 1: Long Papers), Virtual Event, August 1-6, 2021}}, \bibfield{editor}{\bibinfo{person}{Chengqing Zong}, \bibinfo{person}{Fei Xia}, \bibinfo{person}{Wenjie Li}, {and} \bibinfo{person}{Roberto Navigli}} (Eds.). \bibinfo{publisher}{Association for Computational Linguistics}, \bibinfo{pages}{2261--2273}.
\newblock
\urldef\tempurl%
\url{https://doi.org/10.18653/V1/2021.ACL-LONG.176}
\showDOI{\tempurl}


\bibitem[\protect\citeauthoryear{Lei, Wang, Ma, Gan, Lu, Kan, and Chua}{Lei et~al\mbox{.}}{2020}]%
        {lei-etal-2020-examining}
\bibfield{author}{\bibinfo{person}{Wenqiang Lei}, \bibinfo{person}{Weixin Wang}, \bibinfo{person}{Zhixin Ma}, \bibinfo{person}{Tian Gan}, \bibinfo{person}{Wei Lu}, \bibinfo{person}{Min-Yen Kan}, {and} \bibinfo{person}{Tat-Seng Chua}.} \bibinfo{year}{2020}\natexlab{}.
\newblock \showarticletitle{Re-examining the Role of Schema Linking in Text-to-{SQL}}. In \bibinfo{booktitle}{\emph{Proceedings of the 2020 Conference on Empirical Methods in Natural Language Processing (EMNLP)}}, \bibfield{editor}{\bibinfo{person}{Bonnie Webber}, \bibinfo{person}{Trevor Cohn}, \bibinfo{person}{Yulan He}, {and} \bibinfo{person}{Yang Liu}} (Eds.). \bibinfo{publisher}{Association for Computational Linguistics}, \bibinfo{address}{Online}, \bibinfo{pages}{6943--6954}.
\newblock
\urldef\tempurl%
\url{https://doi.org/10.18653/v1/2020.emnlp-main.564}
\showDOI{\tempurl}


\bibitem[\protect\citeauthoryear{Lewis, Liu, Goyal, Ghazvininejad, Mohamed, Levy, Stoyanov, and Zettlemoyer}{Lewis et~al\mbox{.}}{2020}]%
        {lewis2019bart}
\bibfield{author}{\bibinfo{person}{Mike Lewis}, \bibinfo{person}{Yinhan Liu}, \bibinfo{person}{Naman Goyal}, \bibinfo{person}{Marjan Ghazvininejad}, \bibinfo{person}{Abdelrahman Mohamed}, \bibinfo{person}{Omer Levy}, \bibinfo{person}{Veselin Stoyanov}, {and} \bibinfo{person}{Luke Zettlemoyer}.} \bibinfo{year}{2020}\natexlab{}.
\newblock \showarticletitle{{BART:} Denoising Sequence-to-Sequence Pre-training for Natural Language Generation, Translation, and Comprehension}. In \bibinfo{booktitle}{\emph{Proceedings of the 58th Annual Meeting of the Association for Computational Linguistics, {ACL} 2020, Online, July 5-10, 2020}}. \bibinfo{pages}{7871--7880}.
\newblock


\bibitem[\protect\citeauthoryear{Li and Jagadish}{Li and Jagadish}{2014}]%
        {10.1145/2588555.2594519}
\bibfield{author}{\bibinfo{person}{Fei Li} {and} \bibinfo{person}{Hosagrahar~V Jagadish}.} \bibinfo{year}{2014}\natexlab{}.
\newblock \showarticletitle{NaLIR: an interactive natural language interface for querying relational databases}. In \bibinfo{booktitle}{\emph{Proceedings of the 2014 ACM SIGMOD International Conference on Management of Data}} (Snowbird, Utah, USA) \emph{(\bibinfo{series}{SIGMOD '14})}. \bibinfo{publisher}{Association for Computing Machinery}, \bibinfo{address}{New York, NY, USA}, \bibinfo{pages}{709–712}.
\newblock
\showISBNx{9781450323765}
\urldef\tempurl%
\url{https://doi.org/10.1145/2588555.2594519}
\showDOI{\tempurl}


\bibitem[\protect\citeauthoryear{Li, Zhang, Li, and Chen}{Li et~al\mbox{.}}{2023d}]%
        {li2023resdsql}
\bibfield{author}{\bibinfo{person}{Haoyang Li}, \bibinfo{person}{Jing Zhang}, \bibinfo{person}{Cuiping Li}, {and} \bibinfo{person}{Hong Chen}.} \bibinfo{year}{2023}\natexlab{d}.
\newblock \showarticletitle{{RESDSQL:} Decoupling Schema Linking and Skeleton Parsing for Text-to-SQL}. In \bibinfo{booktitle}{\emph{Thirty-Seventh {AAAI} Conference on Artificial Intelligence, {AAAI} 2023, Thirty-Fifth Conference on Innovative Applications of Artificial Intelligence, {IAAI} 2023, Thirteenth Symposium on Educational Advances in Artificial Intelligence, {EAAI} 2023, Washington, DC, USA, February 7-14, 2023}}, \bibfield{editor}{\bibinfo{person}{Brian Williams}, \bibinfo{person}{Yiling Chen}, {and} \bibinfo{person}{Jennifer Neville}} (Eds.). \bibinfo{publisher}{{AAAI} Press}, \bibinfo{pages}{13067--13075}.
\newblock
\urldef\tempurl%
\url{https://doi.org/10.1609/AAAI.V37I11.26535}
\showDOI{\tempurl}


\bibitem[\protect\citeauthoryear{Li, Zhang, Liu, Fan, Zhang, Zhu, Wei, Pan, Li, and Chen}{Li et~al\mbox{.}}{2024}]%
        {codes}
\bibfield{author}{\bibinfo{person}{Haoyang Li}, \bibinfo{person}{Jing Zhang}, \bibinfo{person}{Hanbing Liu}, \bibinfo{person}{Ju Fan}, \bibinfo{person}{Xiaokang Zhang}, \bibinfo{person}{Jun Zhu}, \bibinfo{person}{Renjie Wei}, \bibinfo{person}{Hongyan Pan}, \bibinfo{person}{Cuiping Li}, {and} \bibinfo{person}{Hong Chen}.} \bibinfo{year}{2024}\natexlab{}.
\newblock \showarticletitle{CodeS: Towards Building Open-source Language Models for Text-to-SQL}.
\newblock \bibinfo{journal}{\emph{CoRR}}  \bibinfo{volume}{abs/2402.16347} (\bibinfo{year}{2024}).
\newblock
\urldef\tempurl%
\url{https://doi.org/10.48550/ARXIV.2402.16347}
\showDOI{\tempurl}
\showeprint[arXiv]{2402.16347}


\bibitem[\protect\citeauthoryear{Li, Hui, Cheng, Qin, Ma, Huo, Huang, Du, Si, and Li}{Li et~al\mbox{.}}{2023b}]%
        {li2023graphix}
\bibfield{author}{\bibinfo{person}{Jinyang Li}, \bibinfo{person}{Binyuan Hui}, \bibinfo{person}{Reynold Cheng}, \bibinfo{person}{Bowen Qin}, \bibinfo{person}{Chenhao Ma}, \bibinfo{person}{Nan Huo}, \bibinfo{person}{Fei Huang}, \bibinfo{person}{Wenyu Du}, \bibinfo{person}{Luo Si}, {and} \bibinfo{person}{Yongbin Li}.} \bibinfo{year}{2023}\natexlab{b}.
\newblock \showarticletitle{Graphix-T5: Mixing Pre-trained Transformers with Graph-Aware Layers for Text-to-SQL Parsing}. In \bibinfo{booktitle}{\emph{Thirty-Seventh {AAAI} Conference on Artificial Intelligence, {AAAI} 2023, Thirty-Fifth Conference on Innovative Applications of Artificial Intelligence, {IAAI} 2023, Thirteenth Symposium on Educational Advances in Artificial Intelligence, {EAAI} 2023, Washington, DC, USA, February 7-14, 2023}}, \bibfield{editor}{\bibinfo{person}{Brian Williams}, \bibinfo{person}{Yiling Chen}, {and} \bibinfo{person}{Jennifer Neville}} (Eds.). \bibinfo{publisher}{{AAAI} Press}, \bibinfo{pages}{13076--13084}.
\newblock
\urldef\tempurl%
\url{https://doi.org/10.1609/AAAI.V37I11.26536}
\showDOI{\tempurl}


\bibitem[\protect\citeauthoryear{Li, Hui, Qu, Li, Yang, Li, Wang, Qin, Cao, Geng, Huo, Zhou, Ma, Li, Chang, Huang, Cheng, and Li}{Li et~al\mbox{.}}{2023c}]%
        {bird}
\bibfield{author}{\bibinfo{person}{Jinyang Li}, \bibinfo{person}{Binyuan Hui}, \bibinfo{person}{Ge Qu}, \bibinfo{person}{Binhua Li}, \bibinfo{person}{Jiaxi Yang}, \bibinfo{person}{Bowen Li}, \bibinfo{person}{Bailin Wang}, \bibinfo{person}{Bowen Qin}, \bibinfo{person}{Rongyu Cao}, \bibinfo{person}{Ruiying Geng}, \bibinfo{person}{Nan Huo}, \bibinfo{person}{Xuanhe Zhou}, \bibinfo{person}{Chenhao Ma}, \bibinfo{person}{Guoliang Li}, \bibinfo{person}{Kevin~Chen{-}Chuan Chang}, \bibinfo{person}{Fei Huang}, \bibinfo{person}{Reynold Cheng}, {and} \bibinfo{person}{Yongbin Li}.} \bibinfo{year}{2023}\natexlab{c}.
\newblock \showarticletitle{Can {LLM} Already Serve as {A} Database Interface? {A} BIg Bench for Large-Scale Database Grounded Text-to-SQLs}.
\newblock \bibinfo{journal}{\emph{CoRR}}  \bibinfo{volume}{abs/2305.03111} (\bibinfo{year}{2023}).
\newblock
\urldef\tempurl%
\url{https://doi.org/10.48550/ARXIV.2305.03111}
\showDOI{\tempurl}
\showeprint[arXiv]{2305.03111}


\bibitem[\protect\citeauthoryear{Li, Allal, Zi, Muennighoff, Kocetkov, Mou, Marone, Akiki, Li, Chim, Liu, Zheltonozhskii, Zhuo, Wang, Dehaene, Davaadorj, Lamy{-}Poirier, Monteiro, Shliazhko, Gontier, Meade, Zebaze, Yee, Umapathi, Zhu, Lipkin, Oblokulov, Wang, V, Stillerman, Patel, Abulkhanov, Zocca, Dey, Zhang, Moustafa{-}Fahmy, Bhattacharyya, Yu, Singh, Luccioni, Villegas, Kunakov, Zhdanov, Romero, Lee, Timor, Ding, Schlesinger, Schoelkopf, Ebert, Dao, Mishra, Gu, Robinson, Anderson, Dolan{-}Gavitt, Contractor, Reddy, Fried, Bahdanau, Jernite, Ferrandis, Hughes, Wolf, Guha, von Werra, and de~Vries}{Li et~al\mbox{.}}{2023a}]%
        {DBLP:journals/corr/abs-2305-06161}
\bibfield{author}{\bibinfo{person}{Raymond Li}, \bibinfo{person}{Loubna~Ben Allal}, \bibinfo{person}{Yangtian Zi}, \bibinfo{person}{Niklas Muennighoff}, \bibinfo{person}{Denis Kocetkov}, \bibinfo{person}{Chenghao Mou}, \bibinfo{person}{Marc Marone}, \bibinfo{person}{Christopher Akiki}, \bibinfo{person}{Jia Li}, \bibinfo{person}{Jenny Chim}, \bibinfo{person}{Qian Liu}, \bibinfo{person}{Evgenii Zheltonozhskii}, \bibinfo{person}{Terry~Yue Zhuo}, \bibinfo{person}{Thomas Wang}, \bibinfo{person}{Olivier Dehaene}, \bibinfo{person}{Mishig Davaadorj}, \bibinfo{person}{Joel Lamy{-}Poirier}, \bibinfo{person}{Jo{\~{a}}o Monteiro}, \bibinfo{person}{Oleh Shliazhko}, \bibinfo{person}{Nicolas Gontier}, \bibinfo{person}{Nicholas Meade}, \bibinfo{person}{Armel Zebaze}, \bibinfo{person}{Ming{-}Ho Yee}, \bibinfo{person}{Logesh~Kumar Umapathi}, \bibinfo{person}{Jian Zhu}, \bibinfo{person}{Benjamin Lipkin}, \bibinfo{person}{Muhtasham Oblokulov}, \bibinfo{person}{Zhiruo Wang}, \bibinfo{person}{Rudra~Murthy V},
  \bibinfo{person}{Jason Stillerman}, \bibinfo{person}{Siva~Sankalp Patel}, \bibinfo{person}{Dmitry Abulkhanov}, \bibinfo{person}{Marco Zocca}, \bibinfo{person}{Manan Dey}, \bibinfo{person}{Zhihan Zhang}, \bibinfo{person}{Nour Moustafa{-}Fahmy}, \bibinfo{person}{Urvashi Bhattacharyya}, \bibinfo{person}{Wenhao Yu}, \bibinfo{person}{Swayam Singh}, \bibinfo{person}{Sasha Luccioni}, \bibinfo{person}{Paulo Villegas}, \bibinfo{person}{Maxim Kunakov}, \bibinfo{person}{Fedor Zhdanov}, \bibinfo{person}{Manuel Romero}, \bibinfo{person}{Tony Lee}, \bibinfo{person}{Nadav Timor}, \bibinfo{person}{Jennifer Ding}, \bibinfo{person}{Claire Schlesinger}, \bibinfo{person}{Hailey Schoelkopf}, \bibinfo{person}{Jan Ebert}, \bibinfo{person}{Tri Dao}, \bibinfo{person}{Mayank Mishra}, \bibinfo{person}{Alex Gu}, \bibinfo{person}{Jennifer Robinson}, \bibinfo{person}{Carolyn~Jane Anderson}, \bibinfo{person}{Brendan Dolan{-}Gavitt}, \bibinfo{person}{Danish Contractor}, \bibinfo{person}{Siva Reddy}, \bibinfo{person}{Daniel Fried},
  \bibinfo{person}{Dzmitry Bahdanau}, \bibinfo{person}{Yacine Jernite}, \bibinfo{person}{Carlos~Mu{\~{n}}oz Ferrandis}, \bibinfo{person}{Sean Hughes}, \bibinfo{person}{Thomas Wolf}, \bibinfo{person}{Arjun Guha}, \bibinfo{person}{Leandro von Werra}, {and} \bibinfo{person}{Harm de Vries}.} \bibinfo{year}{2023}\natexlab{a}.
\newblock \showarticletitle{StarCoder: may the source be with you!}
\newblock \bibinfo{journal}{\emph{CoRR}}  \bibinfo{volume}{abs/2305.06161} (\bibinfo{year}{2023}).
\newblock
\urldef\tempurl%
\url{https://doi.org/10.48550/ARXIV.2305.06161}
\showDOI{\tempurl}
\showeprint[arXiv]{2305.06161}


\bibitem[\protect\citeauthoryear{Lin, Socher, and Xiong}{Lin et~al\mbox{.}}{2020}]%
        {lin2020bridging}
\bibfield{author}{\bibinfo{person}{Xi~Victoria Lin}, \bibinfo{person}{Richard Socher}, {and} \bibinfo{person}{Caiming Xiong}.} \bibinfo{year}{2020}\natexlab{}.
\newblock \showarticletitle{Bridging Textual and Tabular Data for Cross-Domain Text-to-SQL Semantic Parsing}. In \bibinfo{booktitle}{\emph{Findings of the Association for Computational Linguistics: {EMNLP} 2020, Online Event, 16-20 November 2020}} \emph{(\bibinfo{series}{Findings of {ACL}})}, \bibfield{editor}{\bibinfo{person}{Trevor Cohn}, \bibinfo{person}{Yulan He}, {and} \bibinfo{person}{Yang Liu}} (Eds.), Vol.~\bibinfo{volume}{{EMNLP} 2020}. \bibinfo{publisher}{Association for Computational Linguistics}, \bibinfo{pages}{4870--4888}.
\newblock
\urldef\tempurl%
\url{https://doi.org/10.18653/V1/2020.FINDINGS-EMNLP.438}
\showDOI{\tempurl}


\bibitem[\protect\citeauthoryear{Liu, Hu, Lin, and Wen}{Liu et~al\mbox{.}}{2022}]%
        {DBLP:conf/kdd/LiuH0W22}
\bibfield{author}{\bibinfo{person}{Aiwei Liu}, \bibinfo{person}{Xuming Hu}, \bibinfo{person}{Li Lin}, {and} \bibinfo{person}{Lijie Wen}.} \bibinfo{year}{2022}\natexlab{}.
\newblock \showarticletitle{Semantic Enhanced Text-to-SQL Parsing via Iteratively Learning Schema Linking Graph}. In \bibinfo{booktitle}{\emph{{KDD}}}. \bibinfo{publisher}{{ACM}}, \bibinfo{pages}{1021--1030}.
\newblock


\bibitem[\protect\citeauthoryear{Liu, Shen, Li, Ma, Jiang, Zhang, Fan, Li, Luo, and Tang}{Liu et~al\mbox{.}}{2024}]%
        {nl2sqlsurvey}
\bibfield{author}{\bibinfo{person}{Xinyu Liu}, \bibinfo{person}{Shuyu Shen}, \bibinfo{person}{Boyan Li}, \bibinfo{person}{Peixian Ma}, \bibinfo{person}{Runzhi Jiang}, \bibinfo{person}{Yuxin Zhang}, \bibinfo{person}{Ju Fan}, \bibinfo{person}{Guoliang Li}, \bibinfo{person}{Yuyu Luo}, {and} \bibinfo{person}{Nan Tang}.} \bibinfo{year}{2024}\natexlab{}.
\newblock \showarticletitle{{A Survey of NL2SQL with Large Language Models:} Where are we, and where are we going?}
\newblock  (\bibinfo{year}{2024}).
\newblock
\urldef\tempurl%
\url{https://github.com/HKUSTDial/NL2SQL_Handbook}
\showURL{%
\tempurl}


\bibitem[\protect\citeauthoryear{Luo, Qin, Chai, Tang, Li, and Li}{Luo et~al\mbox{.}}{2022a}]%
        {DBLP:journals/tkde/LuoQCTLL22}
\bibfield{author}{\bibinfo{person}{Yuyu Luo}, \bibinfo{person}{Xuedi Qin}, \bibinfo{person}{Chengliang Chai}, \bibinfo{person}{Nan Tang}, \bibinfo{person}{Guoliang Li}, {and} \bibinfo{person}{Wenbo Li}.} \bibinfo{year}{2022}\natexlab{a}.
\newblock \showarticletitle{Steerable Self-Driving Data Visualization}.
\newblock \bibinfo{journal}{\emph{{IEEE} Trans. Knowl. Data Eng.}} \bibinfo{volume}{34}, \bibinfo{number}{1} (\bibinfo{year}{2022}), \bibinfo{pages}{475--490}.
\newblock


\bibitem[\protect\citeauthoryear{Luo, Qin, Tang, and Li}{Luo et~al\mbox{.}}{2018a}]%
        {DBLP:conf/icde/LuoQ0018}
\bibfield{author}{\bibinfo{person}{Yuyu Luo}, \bibinfo{person}{Xuedi Qin}, \bibinfo{person}{Nan Tang}, {and} \bibinfo{person}{Guoliang Li}.} \bibinfo{year}{2018}\natexlab{a}.
\newblock \showarticletitle{DeepEye: Towards Automatic Data Visualization}. In \bibinfo{booktitle}{\emph{{ICDE}}}. \bibinfo{publisher}{{IEEE} Computer Society}, \bibinfo{pages}{101--112}.
\newblock


\bibitem[\protect\citeauthoryear{Luo, Qin, Tang, Li, and Wang}{Luo et~al\mbox{.}}{2018b}]%
        {DBLP:conf/sigmod/LuoQ00W18}
\bibfield{author}{\bibinfo{person}{Yuyu Luo}, \bibinfo{person}{Xuedi Qin}, \bibinfo{person}{Nan Tang}, \bibinfo{person}{Guoliang Li}, {and} \bibinfo{person}{Xinran Wang}.} \bibinfo{year}{2018}\natexlab{b}.
\newblock \showarticletitle{DeepEye: Creating Good Data Visualizations by Keyword Search}. In \bibinfo{booktitle}{\emph{{SIGMOD} Conference}}. \bibinfo{publisher}{{ACM}}, \bibinfo{pages}{1733--1736}.
\newblock


\bibitem[\protect\citeauthoryear{Luo, Tang, Li, Chai, Li, and Qin}{Luo et~al\mbox{.}}{2021}]%
        {DBLP:conf/sigmod/Luo00CLQ21}
\bibfield{author}{\bibinfo{person}{Yuyu Luo}, \bibinfo{person}{Nan Tang}, \bibinfo{person}{Guoliang Li}, \bibinfo{person}{Chengliang Chai}, \bibinfo{person}{Wenbo Li}, {and} \bibinfo{person}{Xuedi Qin}.} \bibinfo{year}{2021}\natexlab{}.
\newblock \showarticletitle{Synthesizing Natural Language to Visualization {(NL2VIS)} Benchmarks from {NL2SQL} Benchmarks}. In \bibinfo{booktitle}{\emph{{SIGMOD} Conference}}. \bibinfo{publisher}{{ACM}}, \bibinfo{pages}{1235--1247}.
\newblock


\bibitem[\protect\citeauthoryear{Luo, Tang, Li, Tang, Chai, and Qin}{Luo et~al\mbox{.}}{2022b}]%
        {DBLP:journals/tvcg/LuoTLTCQ22}
\bibfield{author}{\bibinfo{person}{Yuyu Luo}, \bibinfo{person}{Nan Tang}, \bibinfo{person}{Guoliang Li}, \bibinfo{person}{Jiawei Tang}, \bibinfo{person}{Chengliang Chai}, {and} \bibinfo{person}{Xuedi Qin}.} \bibinfo{year}{2022}\natexlab{b}.
\newblock \showarticletitle{Natural Language to Visualization by Neural Machine Translation}.
\newblock \bibinfo{journal}{\emph{{IEEE} Trans. Vis. Comput. Graph.}} \bibinfo{volume}{28}, \bibinfo{number}{1} (\bibinfo{year}{2022}), \bibinfo{pages}{217--226}.
\newblock


\bibitem[\protect\citeauthoryear{Minaee, Mikolov, Nikzad, Chenaghlu, Socher, Amatriain, and Gao}{Minaee et~al\mbox{.}}{2024}]%
        {minaee2024large}
\bibfield{author}{\bibinfo{person}{Shervin Minaee}, \bibinfo{person}{Tomas Mikolov}, \bibinfo{person}{Narjes Nikzad}, \bibinfo{person}{Meysam Chenaghlu}, \bibinfo{person}{Richard Socher}, \bibinfo{person}{Xavier Amatriain}, {and} \bibinfo{person}{Jianfeng Gao}.} \bibinfo{year}{2024}\natexlab{}.
\newblock \showarticletitle{Large Language Models: {A} Survey}.
\newblock \bibinfo{journal}{\emph{CoRR}}  \bibinfo{volume}{abs/2402.06196} (\bibinfo{year}{2024}).
\newblock
\urldef\tempurl%
\url{https://doi.org/10.48550/ARXIV.2402.06196}
\showDOI{\tempurl}
\showeprint[arXiv]{2402.06196}


\bibitem[\protect\citeauthoryear{OpenAI}{OpenAI}{2023}]%
        {gpt4}
\bibfield{author}{\bibinfo{person}{OpenAI}.} \bibinfo{year}{2023}\natexlab{}.
\newblock \showarticletitle{{GPT-4} Technical Report}.
\newblock \bibinfo{journal}{\emph{CoRR}}  \bibinfo{volume}{abs/2303.08774} (\bibinfo{year}{2023}).
\newblock
\urldef\tempurl%
\url{https://doi.org/10.48550/ARXIV.2303.08774}
\showDOI{\tempurl}
\showeprint[arXiv]{2303.08774}


\bibitem[\protect\citeauthoryear{Pourreza and Rafiei}{Pourreza and Rafiei}{2023}]%
        {pourreza2023din}
\bibfield{author}{\bibinfo{person}{Mohammadreza Pourreza} {and} \bibinfo{person}{Davood Rafiei}.} \bibinfo{year}{2023}\natexlab{}.
\newblock \showarticletitle{{DIN-SQL:} Decomposed In-Context Learning of Text-to-SQL with Self-Correction}. In \bibinfo{booktitle}{\emph{Advances in Neural Information Processing Systems 36: Annual Conference on Neural Information Processing Systems 2023, NeurIPS 2023, New Orleans, LA, USA, December 10 - 16, 2023}}, \bibfield{editor}{\bibinfo{person}{Alice Oh}, \bibinfo{person}{Tristan Naumann}, \bibinfo{person}{Amir Globerson}, \bibinfo{person}{Kate Saenko}, \bibinfo{person}{Moritz Hardt}, {and} \bibinfo{person}{Sergey Levine}} (Eds.).
\newblock
\urldef\tempurl%
\url{http://papers.nips.cc/paper\_files/paper/2023/hash/72223cc66f63ca1aa59edaec1b3670e6-Abstract-Conference.html}
\showURL{%
\tempurl}


\bibitem[\protect\citeauthoryear{Qi, Tang, He, Wan, Cheng, Zhou, Wang, Zhang, and Lin}{Qi et~al\mbox{.}}{2022}]%
        {qi2022rasat}
\bibfield{author}{\bibinfo{person}{Jiexing Qi}, \bibinfo{person}{Jingyao Tang}, \bibinfo{person}{Ziwei He}, \bibinfo{person}{Xiangpeng Wan}, \bibinfo{person}{Yu Cheng}, \bibinfo{person}{Chenghu Zhou}, \bibinfo{person}{Xinbing Wang}, \bibinfo{person}{Quanshi Zhang}, {and} \bibinfo{person}{Zhouhan Lin}.} \bibinfo{year}{2022}\natexlab{}.
\newblock \showarticletitle{{RASAT:} Integrating Relational Structures into Pretrained Seq2Seq Model for Text-to-SQL}. In \bibinfo{booktitle}{\emph{Proceedings of the 2022 Conference on Empirical Methods in Natural Language Processing, {EMNLP} 2022, Abu Dhabi, United Arab Emirates, December 7-11, 2022}}, \bibfield{editor}{\bibinfo{person}{Yoav Goldberg}, \bibinfo{person}{Zornitsa Kozareva}, {and} \bibinfo{person}{Yue Zhang}} (Eds.). \bibinfo{publisher}{Association for Computational Linguistics}, \bibinfo{pages}{3215--3229}.
\newblock
\urldef\tempurl%
\url{https://doi.org/10.18653/V1/2022.EMNLP-MAIN.211}
\showDOI{\tempurl}


\bibitem[\protect\citeauthoryear{Radford, Wu, Child, Luan, Amodei, Sutskever, et~al\mbox{.}}{Radford et~al\mbox{.}}{2019}]%
        {gpt2}
\bibfield{author}{\bibinfo{person}{Alec Radford}, \bibinfo{person}{Jeffrey Wu}, \bibinfo{person}{Rewon Child}, \bibinfo{person}{David Luan}, \bibinfo{person}{Dario Amodei}, \bibinfo{person}{Ilya Sutskever}, {et~al\mbox{.}}} \bibinfo{year}{2019}\natexlab{}.
\newblock \showarticletitle{Language models are unsupervised multitask learners}.
\newblock \bibinfo{journal}{\emph{OpenAI blog}} \bibinfo{volume}{1}, \bibinfo{number}{8} (\bibinfo{year}{2019}), \bibinfo{pages}{9}.
\newblock


\bibitem[\protect\citeauthoryear{Raffel, Shazeer, Roberts, Lee, Narang, Matena, Zhou, Li, and Liu}{Raffel et~al\mbox{.}}{2020}]%
        {raffel2020exploring}
\bibfield{author}{\bibinfo{person}{Colin Raffel}, \bibinfo{person}{Noam Shazeer}, \bibinfo{person}{Adam Roberts}, \bibinfo{person}{Katherine Lee}, \bibinfo{person}{Sharan Narang}, \bibinfo{person}{Michael Matena}, \bibinfo{person}{Yanqi Zhou}, \bibinfo{person}{Wei Li}, {and} \bibinfo{person}{Peter~J Liu}.} \bibinfo{year}{2020}\natexlab{}.
\newblock \showarticletitle{Exploring the limits of transfer learning with a unified text-to-text transformer}.
\newblock \bibinfo{journal}{\emph{The Journal of Machine Learning Research}} \bibinfo{volume}{21}, \bibinfo{number}{1} (\bibinfo{year}{2020}), \bibinfo{pages}{5485--5551}.
\newblock


\bibitem[\protect\citeauthoryear{Rai, Wang, Zhou, and Yao}{Rai et~al\mbox{.}}{2023}]%
        {rai2023improving}
\bibfield{author}{\bibinfo{person}{Daking Rai}, \bibinfo{person}{Bailin Wang}, \bibinfo{person}{Yilun Zhou}, {and} \bibinfo{person}{Ziyu Yao}.} \bibinfo{year}{2023}\natexlab{}.
\newblock \showarticletitle{Improving Generalization in Language Model-based Text-to-SQL Semantic Parsing: Two Simple Semantic Boundary-based Techniques}. In \bibinfo{booktitle}{\emph{Proceedings of the 61st Annual Meeting of the Association for Computational Linguistics (Volume 2: Short Papers), {ACL} 2023, Toronto, Canada, July 9-14, 2023}}, \bibfield{editor}{\bibinfo{person}{Anna Rogers}, \bibinfo{person}{Jordan~L. Boyd{-}Graber}, {and} \bibinfo{person}{Naoaki Okazaki}} (Eds.). \bibinfo{publisher}{Association for Computational Linguistics}, \bibinfo{pages}{150--160}.
\newblock
\urldef\tempurl%
\url{https://doi.org/10.18653/V1/2023.ACL-SHORT.15}
\showDOI{\tempurl}


\bibitem[\protect\citeauthoryear{Rajkumar, Li, and Bahdanau}{Rajkumar et~al\mbox{.}}{2022}]%
        {DBLP:journals/corr/abs-2204-00498}
\bibfield{author}{\bibinfo{person}{Nitarshan Rajkumar}, \bibinfo{person}{Raymond Li}, {and} \bibinfo{person}{Dzmitry Bahdanau}.} \bibinfo{year}{2022}\natexlab{}.
\newblock \showarticletitle{Evaluating the Text-to-SQL Capabilities of Large Language Models}.
\newblock \bibinfo{journal}{\emph{CoRR}}  \bibinfo{volume}{abs/2204.00498} (\bibinfo{year}{2022}).
\newblock


\bibitem[\protect\citeauthoryear{Rozi{\`{e}}re, Gehring, Gloeckle, Sootla, Gat, Tan, Adi, Liu, Remez, Rapin, Kozhevnikov, Evtimov, Bitton, Bhatt, Canton{-}Ferrer, Grattafiori, Xiong, D{\'{e}}fossez, Copet, Azhar, Touvron, Martin, Usunier, Scialom, and Synnaeve}{Rozi{\`{e}}re et~al\mbox{.}}{2023}]%
        {DBLP:journals/corr/abs-2308-12950}
\bibfield{author}{\bibinfo{person}{Baptiste Rozi{\`{e}}re}, \bibinfo{person}{Jonas Gehring}, \bibinfo{person}{Fabian Gloeckle}, \bibinfo{person}{Sten Sootla}, \bibinfo{person}{Itai Gat}, \bibinfo{person}{Xiaoqing~Ellen Tan}, \bibinfo{person}{Yossi Adi}, \bibinfo{person}{Jingyu Liu}, \bibinfo{person}{Tal Remez}, \bibinfo{person}{J{\'{e}}r{\'{e}}my Rapin}, \bibinfo{person}{Artyom Kozhevnikov}, \bibinfo{person}{Ivan Evtimov}, \bibinfo{person}{Joanna Bitton}, \bibinfo{person}{Manish Bhatt}, \bibinfo{person}{Cristian Canton{-}Ferrer}, \bibinfo{person}{Aaron Grattafiori}, \bibinfo{person}{Wenhan Xiong}, \bibinfo{person}{Alexandre D{\'{e}}fossez}, \bibinfo{person}{Jade Copet}, \bibinfo{person}{Faisal Azhar}, \bibinfo{person}{Hugo Touvron}, \bibinfo{person}{Louis Martin}, \bibinfo{person}{Nicolas Usunier}, \bibinfo{person}{Thomas Scialom}, {and} \bibinfo{person}{Gabriel Synnaeve}.} \bibinfo{year}{2023}\natexlab{}.
\newblock \showarticletitle{Code Llama: Open Foundation Models for Code}.
\newblock \bibinfo{journal}{\emph{CoRR}}  \bibinfo{volume}{abs/2308.12950} (\bibinfo{year}{2023}).
\newblock
\urldef\tempurl%
\url{https://doi.org/10.48550/ARXIV.2308.12950}
\showDOI{\tempurl}
\showeprint[arXiv]{2308.12950}


\bibitem[\protect\citeauthoryear{Scholak, Schucher, and Bahdanau}{Scholak et~al\mbox{.}}{2021}]%
        {scholak2021picard}
\bibfield{author}{\bibinfo{person}{Torsten Scholak}, \bibinfo{person}{Nathan Schucher}, {and} \bibinfo{person}{Dzmitry Bahdanau}.} \bibinfo{year}{2021}\natexlab{}.
\newblock \showarticletitle{{PICARD:} Parsing Incrementally for Constrained Auto-Regressive Decoding from Language Models}. In \bibinfo{booktitle}{\emph{Proceedings of the 2021 Conference on Empirical Methods in Natural Language Processing, {EMNLP} 2021, Virtual Event / Punta Cana, Dominican Republic, 7-11 November, 2021}}, \bibfield{editor}{\bibinfo{person}{Marie{-}Francine Moens}, \bibinfo{person}{Xuanjing Huang}, \bibinfo{person}{Lucia Specia}, {and} \bibinfo{person}{Scott~Wen{-}tau Yih}} (Eds.). \bibinfo{publisher}{Association for Computational Linguistics}, \bibinfo{pages}{9895--9901}.
\newblock
\urldef\tempurl%
\url{https://doi.org/10.18653/V1/2021.EMNLP-MAIN.779}
\showDOI{\tempurl}


\bibitem[\protect\citeauthoryear{Shen, Shen, Luo, Yang, Hu, Zhang, Tai, and Wang}{Shen et~al\mbox{.}}{2021}]%
        {DBLP:journals/corr/abs-2109-03506}
\bibfield{author}{\bibinfo{person}{Leixian Shen}, \bibinfo{person}{Enya Shen}, \bibinfo{person}{Yuyu Luo}, \bibinfo{person}{Xiaocong Yang}, \bibinfo{person}{Xuming Hu}, \bibinfo{person}{Xiongshuai Zhang}, \bibinfo{person}{Zhiwei Tai}, {and} \bibinfo{person}{Jianmin Wang}.} \bibinfo{year}{2021}\natexlab{}.
\newblock \showarticletitle{Towards Natural Language Interfaces for Data Visualization: {A} Survey}.
\newblock \bibinfo{journal}{\emph{CoRR}}  \bibinfo{volume}{abs/2109.03506} (\bibinfo{year}{2021}).
\newblock


\bibitem[\protect\citeauthoryear{Tang, Luo, Ouzzani, Li, and Chen}{Tang et~al\mbox{.}}{2022}]%
        {DBLP:conf/sigmod/TangLOLC22}
\bibfield{author}{\bibinfo{person}{Jiawei Tang}, \bibinfo{person}{Yuyu Luo}, \bibinfo{person}{Mourad Ouzzani}, \bibinfo{person}{Guoliang Li}, {and} \bibinfo{person}{Hongyang Chen}.} \bibinfo{year}{2022}\natexlab{}.
\newblock \showarticletitle{Sevi: Speech-to-Visualization through Neural Machine Translation}. In \bibinfo{booktitle}{\emph{{SIGMOD} Conference}}. \bibinfo{publisher}{{ACM}}, \bibinfo{pages}{2353--2356}.
\newblock


\bibitem[\protect\citeauthoryear{Tang, Yang, Fan, Cao, Luo, and Halevy}{Tang et~al\mbox{.}}{2024}]%
        {DBLP:conf/cidr/0001YF0LH24}
\bibfield{author}{\bibinfo{person}{Nan Tang}, \bibinfo{person}{Chenyu Yang}, \bibinfo{person}{Ju Fan}, \bibinfo{person}{Lei Cao}, \bibinfo{person}{Yuyu Luo}, {and} \bibinfo{person}{Alon~Y. Halevy}.} \bibinfo{year}{2024}\natexlab{}.
\newblock \showarticletitle{VerifAI: Verified Generative {AI}}. In \bibinfo{booktitle}{\emph{{CIDR}}}. \bibinfo{publisher}{www.cidrdb.org}.
\newblock


\bibitem[\protect\citeauthoryear{Taori, Gulrajani, Zhang, Dubois, Li, Guestrin, Liang, and Hashimoto}{Taori et~al\mbox{.}}{2023}]%
        {alpaca}
\bibfield{author}{\bibinfo{person}{Rohan Taori}, \bibinfo{person}{Ishaan Gulrajani}, \bibinfo{person}{Tianyi Zhang}, \bibinfo{person}{Yann Dubois}, \bibinfo{person}{Xuechen Li}, \bibinfo{person}{Carlos Guestrin}, \bibinfo{person}{Percy Liang}, {and} \bibinfo{person}{Tatsunori~B. Hashimoto}.} \bibinfo{year}{2023}\natexlab{}.
\newblock \bibinfo{title}{Stanford Alpaca: An Instruction-following LLaMA model}.
\newblock \bibinfo{howpublished}{\url{https://github.com/tatsu-lab/stanford_alpaca}}.
\newblock


\bibitem[\protect\citeauthoryear{Touvron, Martin, Stone, Albert, Almahairi, Babaei, Bashlykov, Batra, Bhargava, Bhosale, Bikel, Blecher, Canton{-}Ferrer, Chen, Cucurull, Esiobu, Fernandes, Fu, Fu, Fuller, Gao, Goswami, Goyal, Hartshorn, Hosseini, Hou, Inan, Kardas, Kerkez, Khabsa, Kloumann, Korenev, Koura, Lachaux, Lavril, Lee, Liskovich, Lu, Mao, Martinet, Mihaylov, Mishra, Molybog, Nie, Poulton, Reizenstein, Rungta, Saladi, Schelten, Silva, Smith, Subramanian, Tan, Tang, Taylor, Williams, Kuan, Xu, Yan, Zarov, Zhang, Fan, Kambadur, Narang, Rodriguez, Stojnic, Edunov, and Scialom}{Touvron et~al\mbox{.}}{2023}]%
        {DBLP:journals/corr/abs-2307-09288}
\bibfield{author}{\bibinfo{person}{Hugo Touvron}, \bibinfo{person}{Louis Martin}, \bibinfo{person}{Kevin Stone}, \bibinfo{person}{Peter Albert}, \bibinfo{person}{Amjad Almahairi}, \bibinfo{person}{Yasmine Babaei}, \bibinfo{person}{Nikolay Bashlykov}, \bibinfo{person}{Soumya Batra}, \bibinfo{person}{Prajjwal Bhargava}, \bibinfo{person}{Shruti Bhosale}, \bibinfo{person}{Dan Bikel}, \bibinfo{person}{Lukas Blecher}, \bibinfo{person}{Cristian Canton{-}Ferrer}, \bibinfo{person}{Moya Chen}, \bibinfo{person}{Guillem Cucurull}, \bibinfo{person}{David Esiobu}, \bibinfo{person}{Jude Fernandes}, \bibinfo{person}{Jeremy Fu}, \bibinfo{person}{Wenyin Fu}, \bibinfo{person}{Brian Fuller}, \bibinfo{person}{Cynthia Gao}, \bibinfo{person}{Vedanuj Goswami}, \bibinfo{person}{Naman Goyal}, \bibinfo{person}{Anthony Hartshorn}, \bibinfo{person}{Saghar Hosseini}, \bibinfo{person}{Rui Hou}, \bibinfo{person}{Hakan Inan}, \bibinfo{person}{Marcin Kardas}, \bibinfo{person}{Viktor Kerkez}, \bibinfo{person}{Madian Khabsa},
  \bibinfo{person}{Isabel Kloumann}, \bibinfo{person}{Artem Korenev}, \bibinfo{person}{Punit~Singh Koura}, \bibinfo{person}{Marie{-}Anne Lachaux}, \bibinfo{person}{Thibaut Lavril}, \bibinfo{person}{Jenya Lee}, \bibinfo{person}{Diana Liskovich}, \bibinfo{person}{Yinghai Lu}, \bibinfo{person}{Yuning Mao}, \bibinfo{person}{Xavier Martinet}, \bibinfo{person}{Todor Mihaylov}, \bibinfo{person}{Pushkar Mishra}, \bibinfo{person}{Igor Molybog}, \bibinfo{person}{Yixin Nie}, \bibinfo{person}{Andrew Poulton}, \bibinfo{person}{Jeremy Reizenstein}, \bibinfo{person}{Rashi Rungta}, \bibinfo{person}{Kalyan Saladi}, \bibinfo{person}{Alan Schelten}, \bibinfo{person}{Ruan Silva}, \bibinfo{person}{Eric~Michael Smith}, \bibinfo{person}{Ranjan Subramanian}, \bibinfo{person}{Xiaoqing~Ellen Tan}, \bibinfo{person}{Binh Tang}, \bibinfo{person}{Ross Taylor}, \bibinfo{person}{Adina Williams}, \bibinfo{person}{Jian~Xiang Kuan}, \bibinfo{person}{Puxin Xu}, \bibinfo{person}{Zheng Yan}, \bibinfo{person}{Iliyan Zarov}, \bibinfo{person}{Yuchen
  Zhang}, \bibinfo{person}{Angela Fan}, \bibinfo{person}{Melanie Kambadur}, \bibinfo{person}{Sharan Narang}, \bibinfo{person}{Aur{\'{e}}lien Rodriguez}, \bibinfo{person}{Robert Stojnic}, \bibinfo{person}{Sergey Edunov}, {and} \bibinfo{person}{Thomas Scialom}.} \bibinfo{year}{2023}\natexlab{}.
\newblock \showarticletitle{Llama 2: Open Foundation and Fine-Tuned Chat Models}.
\newblock \bibinfo{journal}{\emph{CoRR}}  \bibinfo{volume}{abs/2307.09288} (\bibinfo{year}{2023}).
\newblock
\urldef\tempurl%
\url{https://doi.org/10.48550/ARXIV.2307.09288}
\showDOI{\tempurl}
\showeprint[arXiv]{2307.09288}


\bibitem[\protect\citeauthoryear{Vaswani, Shazeer, Parmar, Uszkoreit, Jones, Gomez, Kaiser, and Polosukhin}{Vaswani et~al\mbox{.}}{2017}]%
        {vaswani2017attention}
\bibfield{author}{\bibinfo{person}{Ashish Vaswani}, \bibinfo{person}{Noam Shazeer}, \bibinfo{person}{Niki Parmar}, \bibinfo{person}{Jakob Uszkoreit}, \bibinfo{person}{Llion Jones}, \bibinfo{person}{Aidan~N. Gomez}, \bibinfo{person}{Lukasz Kaiser}, {and} \bibinfo{person}{Illia Polosukhin}.} \bibinfo{year}{2017}\natexlab{}.
\newblock \showarticletitle{Attention is All you Need}. In \bibinfo{booktitle}{\emph{Advances in Neural Information Processing Systems 30: Annual Conference on Neural Information Processing Systems 2017, December 4-9, 2017, Long Beach, CA, {USA}}}, \bibfield{editor}{\bibinfo{person}{Isabelle Guyon}, \bibinfo{person}{Ulrike von Luxburg}, \bibinfo{person}{Samy Bengio}, \bibinfo{person}{Hanna~M. Wallach}, \bibinfo{person}{Rob Fergus}, \bibinfo{person}{S.~V.~N. Vishwanathan}, {and} \bibinfo{person}{Roman Garnett}} (Eds.). \bibinfo{pages}{5998--6008}.
\newblock
\urldef\tempurl%
\url{https://proceedings.neurips.cc/paper/2017/hash/3f5ee243547dee91fbd053c1c4a845aa-Abstract.html}
\showURL{%
\tempurl}


\bibitem[\protect\citeauthoryear{Wang, Ren, Yang, Liang, Bai, Zhang, Yan, and Li}{Wang et~al\mbox{.}}{2023}]%
        {macsql}
\bibfield{author}{\bibinfo{person}{Bing Wang}, \bibinfo{person}{Changyu Ren}, \bibinfo{person}{Jian Yang}, \bibinfo{person}{Xinnian Liang}, \bibinfo{person}{Jiaqi Bai}, \bibinfo{person}{Qian{-}Wen Zhang}, \bibinfo{person}{Zhao Yan}, {and} \bibinfo{person}{Zhoujun Li}.} \bibinfo{year}{2023}\natexlab{}.
\newblock \showarticletitle{{MAC-SQL:} {A} Multi-Agent Collaborative Framework for Text-to-SQL}.
\newblock \bibinfo{journal}{\emph{CoRR}}  \bibinfo{volume}{abs/2312.11242} (\bibinfo{year}{2023}).
\newblock
\urldef\tempurl%
\url{https://doi.org/10.48550/ARXIV.2312.11242}
\showDOI{\tempurl}
\showeprint[arXiv]{2312.11242}


\bibitem[\protect\citeauthoryear{Wang, Qin, Hui, Li, Yang, Wang, Li, Sun, Huang, Si, and Li}{Wang et~al\mbox{.}}{2022}]%
        {DBLP:conf/kdd/WangQHLYWLSHSL22}
\bibfield{author}{\bibinfo{person}{Lihan Wang}, \bibinfo{person}{Bowen Qin}, \bibinfo{person}{Binyuan Hui}, \bibinfo{person}{Bowen Li}, \bibinfo{person}{Min Yang}, \bibinfo{person}{Bailin Wang}, \bibinfo{person}{Binhua Li}, \bibinfo{person}{Jian Sun}, \bibinfo{person}{Fei Huang}, \bibinfo{person}{Luo Si}, {and} \bibinfo{person}{Yongbin Li}.} \bibinfo{year}{2022}\natexlab{}.
\newblock \showarticletitle{Proton: Probing Schema Linking Information from Pre-trained Language Models for Text-to-SQL Parsing}. In \bibinfo{booktitle}{\emph{{KDD}}}. \bibinfo{publisher}{{ACM}}, \bibinfo{pages}{1889--1898}.
\newblock


\bibitem[\protect\citeauthoryear{Wei, Wang, Schuurmans, Bosma, Chi, Le, and Zhou}{Wei et~al\mbox{.}}{2022}]%
        {cot}
\bibfield{author}{\bibinfo{person}{Jason Wei}, \bibinfo{person}{Xuezhi Wang}, \bibinfo{person}{Dale Schuurmans}, \bibinfo{person}{Maarten Bosma}, \bibinfo{person}{Ed~H. Chi}, \bibinfo{person}{Quoc Le}, {and} \bibinfo{person}{Denny Zhou}.} \bibinfo{year}{2022}\natexlab{}.
\newblock \showarticletitle{Chain of Thought Prompting Elicits Reasoning in Large Language Models}.
\newblock \bibinfo{journal}{\emph{CoRR}}  \bibinfo{volume}{abs/2201.11903} (\bibinfo{year}{2022}).
\newblock
\showeprint[arXiv]{2201.11903}
\urldef\tempurl%
\url{https://arxiv.org/abs/2201.11903}
\showURL{%
\tempurl}


\bibitem[\protect\citeauthoryear{Xie and Yuille}{Xie and Yuille}{2017}]%
        {nas}
\bibfield{author}{\bibinfo{person}{Lingxi Xie} {and} \bibinfo{person}{Alan~L. Yuille}.} \bibinfo{year}{2017}\natexlab{}.
\newblock \showarticletitle{Genetic {CNN}}.
\newblock \bibinfo{journal}{\emph{CoRR}}  \bibinfo{volume}{abs/1703.01513} (\bibinfo{year}{2017}).
\newblock
\showeprint[arXiv]{1703.01513}
\urldef\tempurl%
\url{http://arxiv.org/abs/1703.01513}
\showURL{%
\tempurl}


\bibitem[\protect\citeauthoryear{Xie, Luo, Li, and Tang}{Xie et~al\mbox{.}}{2024}]%
        {xie2024haicharthumanaipaired}
\bibfield{author}{\bibinfo{person}{Yupeng Xie}, \bibinfo{person}{Yuyu Luo}, \bibinfo{person}{Guoliang Li}, {and} \bibinfo{person}{Nan Tang}.} \bibinfo{year}{2024}\natexlab{}.
\newblock \showarticletitle{HAIChart: Human and AI Paired Visualization System}.
\newblock \bibinfo{journal}{\emph{Proc. {VLDB} Endow.}}  \bibinfo{volume}{17}.
\newblock


\bibitem[\protect\citeauthoryear{Ye, Hao, Hou, Wang, Xiao, Luo, and Zeng}{Ye et~al\mbox{.}}{2024}]%
        {YE202443}
\bibfield{author}{\bibinfo{person}{Yilin Ye}, \bibinfo{person}{Jianing Hao}, \bibinfo{person}{Yihan Hou}, \bibinfo{person}{Zhan Wang}, \bibinfo{person}{Shishi Xiao}, \bibinfo{person}{Yuyu Luo}, {and} \bibinfo{person}{Wei Zeng}.} \bibinfo{year}{2024}\natexlab{}.
\newblock \showarticletitle{Generative AI for visualization: State of the art and future directions}.
\newblock \bibinfo{journal}{\emph{Visual Informatics}} \bibinfo{volume}{8}, \bibinfo{number}{2} (\bibinfo{year}{2024}), \bibinfo{pages}{43--66}.
\newblock
\showISSN{2468-502X}
\urldef\tempurl%
\url{https://doi.org/10.1016/j.visinf.2024.04.003}
\showDOI{\tempurl}


\bibitem[\protect\citeauthoryear{Yu, Zhang, Yang, Yasunaga, Wang, Li, Ma, Li, Yao, Roman, Zhang, and Radev}{Yu et~al\mbox{.}}{2018}]%
        {DBLP:conf/emnlp/YuZYYWLMLYRZR18}
\bibfield{author}{\bibinfo{person}{Tao Yu}, \bibinfo{person}{Rui Zhang}, \bibinfo{person}{Kai Yang}, \bibinfo{person}{Michihiro Yasunaga}, \bibinfo{person}{Dongxu Wang}, \bibinfo{person}{Zifan Li}, \bibinfo{person}{James Ma}, \bibinfo{person}{Irene Li}, \bibinfo{person}{Qingning Yao}, \bibinfo{person}{Shanelle Roman}, \bibinfo{person}{Zilin Zhang}, {and} \bibinfo{person}{Dragomir~R. Radev}.} \bibinfo{year}{2018}\natexlab{}.
\newblock \showarticletitle{Spider: {A} Large-Scale Human-Labeled Dataset for Complex and Cross-Domain Semantic Parsing and Text-to-SQL Task}. In \bibinfo{booktitle}{\emph{{EMNLP}}}. \bibinfo{publisher}{Association for Computational Linguistics}, \bibinfo{pages}{3911--3921}.
\newblock


\bibitem[\protect\citeauthoryear{Zeng, Parthasarathi, and Hakkani{-}Tur}{Zeng et~al\mbox{.}}{2022}]%
        {zeng2023n}
\bibfield{author}{\bibinfo{person}{Lu Zeng}, \bibinfo{person}{Sree Hari~Krishnan Parthasarathi}, {and} \bibinfo{person}{Dilek Hakkani{-}Tur}.} \bibinfo{year}{2022}\natexlab{}.
\newblock \showarticletitle{N-Best Hypotheses Reranking for Text-to-SQL Systems}. In \bibinfo{booktitle}{\emph{{IEEE} Spoken Language Technology Workshop, {SLT} 2022, Doha, Qatar, January 9-12, 2023}}. \bibinfo{publisher}{{IEEE}}, \bibinfo{pages}{663--670}.
\newblock
\urldef\tempurl%
\url{https://doi.org/10.1109/SLT54892.2023.10023434}
\showDOI{\tempurl}


\bibitem[\protect\citeauthoryear{Zhao, Zhou, Li, Tang, Wang, Hou, Min, Zhang, Zhang, Dong, Du, Yang, Chen, Chen, Jiang, Ren, Li, Tang, Liu, Liu, Nie, and Wen}{Zhao et~al\mbox{.}}{2023}]%
        {zhao2023survey}
\bibfield{author}{\bibinfo{person}{Wayne~Xin Zhao}, \bibinfo{person}{Kun Zhou}, \bibinfo{person}{Junyi Li}, \bibinfo{person}{Tianyi Tang}, \bibinfo{person}{Xiaolei Wang}, \bibinfo{person}{Yupeng Hou}, \bibinfo{person}{Yingqian Min}, \bibinfo{person}{Beichen Zhang}, \bibinfo{person}{Junjie Zhang}, \bibinfo{person}{Zican Dong}, \bibinfo{person}{Yifan Du}, \bibinfo{person}{Chen Yang}, \bibinfo{person}{Yushuo Chen}, \bibinfo{person}{Zhipeng Chen}, \bibinfo{person}{Jinhao Jiang}, \bibinfo{person}{Ruiyang Ren}, \bibinfo{person}{Yifan Li}, \bibinfo{person}{Xinyu Tang}, \bibinfo{person}{Zikang Liu}, \bibinfo{person}{Peiyu Liu}, \bibinfo{person}{Jian{-}Yun Nie}, {and} \bibinfo{person}{Ji{-}Rong Wen}.} \bibinfo{year}{2023}\natexlab{}.
\newblock \showarticletitle{A Survey of Large Language Models}.
\newblock \bibinfo{journal}{\emph{CoRR}}  \bibinfo{volume}{abs/2303.18223} (\bibinfo{year}{2023}).
\newblock
\urldef\tempurl%
\url{https://doi.org/10.48550/ARXIV.2303.18223}
\showDOI{\tempurl}
\showeprint[arXiv]{2303.18223}


\bibitem[\protect\citeauthoryear{Zhong, Xiong, and Socher}{Zhong et~al\mbox{.}}{2017}]%
        {zhong2017seq2sql}
\bibfield{author}{\bibinfo{person}{Victor Zhong}, \bibinfo{person}{Caiming Xiong}, {and} \bibinfo{person}{Richard Socher}.} \bibinfo{year}{2017}\natexlab{}.
\newblock \showarticletitle{Seq2SQL: Generating Structured Queries from Natural Language using Reinforcement Learning}.
\newblock \bibinfo{journal}{\emph{CoRR}}  \bibinfo{volume}{abs/1709.00103} (\bibinfo{year}{2017}).
\newblock
\showeprint[arXiv]{1709.00103}
\urldef\tempurl%
\url{http://arxiv.org/abs/1709.00103}
\showURL{%
\tempurl}


\bibitem[\protect\citeauthoryear{Zhu, Du, Li, Luo, and Tang}{Zhu et~al\mbox{.}}{2024}]%
        {DBLP:journals/corr/abs-2406-07815}
\bibfield{author}{\bibinfo{person}{Yizhang Zhu}, \bibinfo{person}{Shiyin Du}, \bibinfo{person}{Boyan Li}, \bibinfo{person}{Yuyu Luo}, {and} \bibinfo{person}{Nan Tang}.} \bibinfo{year}{2024}\natexlab{}.
\newblock \showarticletitle{Are Large Language Models Good Statisticians?}
\newblock \bibinfo{journal}{\emph{CoRR}}  \bibinfo{volume}{abs/2406.07815} (\bibinfo{year}{2024}).
\newblock


\end{thebibliography}

\end{document}